\newif\iflongversion
\longversiontrue

\iflongversion
\documentclass[sigconf,9pt,screen,nonacm,balance=false]{acmart}
\else
\documentclass[sigconf,9pt,screen]{acmart}
\fi

\usepackage{bm}
\usepackage{marvosym}
\usepackage[inline]{enumitem}
\usepackage[percent]{overpic}
\usepackage{fancyvrb}
\usepackage{booktabs}
\usepackage{multirow}

\usepackage[bbsets,Dfprime,setrelation]{math}
\usepackage{manalysis}

\definecolor{vblue}{rgb}{.1,.15,.62}

\definecolor{vgreen}{rgb}{.1,.5,0}
\definecolor{vgray}{rgb}{.35,.35,.35}
\definecolor{vred}{rgb}{.7,0,0}

\usepackage{wasysym}
\usepackage[prefixflatinterpret,bracketmodalinterpret,fixformat,silentconst,sidenotecalculus,longseqcontext,seqinsist]{logic}
\usepackage[prefixflatinterpret,bracketmodalinterpret,fixformat,silentconst,differentialdL,simplenames]{dL}
\def\leftrule{L}%
\def\rightrule{R}%
\newcommand{\bebecomes}{\mathrel{::=}}
\newcommand{\alternative}{~|~}

\newcommand{\cmp}{\succcurlyeq}

\makeatletter
\newcommand{\Vast}{\bBigg@{9.2}}

\newsavebox{\Rval}%
\sbox{\Rval}{$\scriptstyle\mathbb{R}$}
\newsavebox{\Rvalext}%
\sbox{\Rvalext}{$\scriptstyle\mathbb{R}_{\exp,\sin,\cos}$}
\newsavebox{\Rvalexp}%
\sbox{\Rvalexp}{$\scriptstyle\mathbb{R}_{\exp}$}

\relax

  \newdimen\linferenceRulehskipamount%
  \newdimen\lcalculuscollectionvskipamount%
\definecolor{vblue}{rgb}{.1,.15,.62}
\definecolor{vgray}{rgb}{.35,.35,.35}
\renewcommand*{\irrulename}[1]{\text{\textcolor{vgray}{\upshape#1}}}%
\renewcommand{\linferenceRuleNameSeparation}{\hspace{5pt}}
\renewcommand{\linferenceRuleSidenoteSeparation}{\quad}
\renewcommand{\linferenceRuleRefSeparation}{, }

\relax

\renewcommand*{\lie}[3][]
{\mathcal{L}_{#2}^{\ifthenelse{\equal{#1}{}}{}{^{\left(#1\right)}}}(#3)}
\renewcommand*{\lied}[3][]{\overset{\bm .}{#3}\ifthenelse{\equal{#1}{}}{}{{}^{(#1)}}}

\newcommand{\solvar}{\varphi}

\newcommand{\tvar}{t}
\newcommand{\taug}[1]{#1, \D{\tvar}=1}

\usepackage{prettyref}
\newcommand{\rref}[2][]{\prettyref{#2}}
\newrefformat{sec}{Section\,\ref{#1}}
\newrefformat{subsec}{Section\,\ref{#1}}
\newrefformat{def}{Def.\,\ref{#1}}
\newrefformat{thm}{Theorem\,\ref{#1}}
\newrefformat{prop}{Proposition\,\ref{#1}}
\newrefformat{rem}{Remark\,\ref{#1}}
\newrefformat{lem}{Lemma\,\ref{#1}}
\newrefformat{cor}{Corollary\,\ref{#1}}
\newrefformat{ex}{Example\,\ref{#1}}
\newrefformat{cex}{Counterexample\,\ref{#1}}
\newrefformat{tab}{Table\,\ref{#1}}
\newrefformat{fig}{Fig.\,\ref{#1}}
\newrefformat{case}{case\,\ref{#1}}
\newrefformat{foot}{Footnote\,\ref{#1}}

\iflongversion
\newrefformat{app}{Appendix\,\ref{#1}}
\else
\newrefformat{app}{\cite{DBLP:journals/corr/abs-2111-01928}}
\fi

\newrefformat{itm}{{\it\ref{#1}}}

\newcommand{\bdr}[1]{\partial #1}

\newcommand{\progA}{\alpha}
\newcommand{\progB}{\beta}
\newcommand{\etermA}{e}
\newcommand{\etermB}{\tilde{e}}
\newcommand{\fvarA}{\phi}
\newcommand{\fvarB}{\psi}
\newcommand{\rfvar}{P}

\newcommand{\rcfvar}{C}

\newcommand{\rrfvar}{R}

\newcommand{\expo}[1]{e^{#1}}

\newcommand{\sigfam}{\mathcal{P}}
\newcommand{\sigfams}{\mathcal{S}}
\newcommand{\sigfamu}{\mathcal{U}}
\newcommand{\lterm}{V}
\newcommand{\ltermUpper}{W}
\newcommand{\ltermLower}{U}

\newcommand{\ptermA}{p}
\newcommand{\ptermB}{q}
\newcommand{\cofterm}{g}

\newcommand{\exholotime}{\ensuremath{\alpha_\tau}}
\newcommand{\exholotimefml}{\ensuremath{\rfvar_\tau}}

\newcommand{\exholoevent}{\ensuremath{\alpha_e}}
\newcommand{\exholoeventfml}{\ensuremath{\rfvar_e}}

\newcommand{\exholoeventfull}{\ensuremath{\alpha_{\hat{e}}}}
\newcommand{\exholoeventfullfml}{\ensuremath{\rfvar_{\hat{e}}}}

\DeclareMathOperator{\sign}{sign}

\newcommand{\exmax}{\ensuremath{\alpha_m}}
\newcommand{\exmaxfml}{\ensuremath{\rfvar_m}}

\newcommand{\keyword}[1]{\textbf{\texttt{#1}}}

\newcommand{\bluec}[1]{\textcolor{vblue}{#1}}
\newcommand{\greenc}[1]{\textcolor{vgreen}{#1}}
\newcommand{\redc}[1]{\textcolor{vred}{#1}}

\newcommand{\arbswitch}{\alpha_{\texttt{arb}}}
\newcommand{\stateswitch}{\alpha_{\texttt{state}}}
\newcommand{\guardswitch}{\alpha_{\texttt{guard}}}
\newcommand{\timeswitch}{\alpha_{\texttt{time}}}
\newcommand{\ctrlswitch}{\alpha_{\texttt{ctrl}}}

\newcommand{\invariant}{\textit{Inv}}
\newcommand{\invariants}{\textit{Inv}_s}
\newcommand{\invarianta}{\textit{Inv}_a}

\newcommand{\folr}{FOL$_{\mathbb{R}}$\xspace}

\DeclareMathOperator{\stable}{UStab}
\newcommand{\stabhp}[1]{\stable(#1)}

\DeclareMathOperator{\attractive}{UGpAttr}
\newcommand{\attrhp}[1]{\attractive(#1)}

\DeclareMathOperator{\asymstab}{UGpAS}
\newcommand{\astabhp}[1]{\asymstab(#1)}

\newcommand{\I}{\dLint[state=\omega]}
\newcommand{\It}{\dLint[state=\nu]}

\usepackage{tikz}
\usetikzlibrary{arrows}

\newcommand*\circled[1]{\tikz[baseline=(char.base)]{
            \node[shape=circle,draw, minimum size=3.5mm,inner sep=0pt] (char) {#1};}}

\newtheorem{theorem}{Theorem}
\newtheorem{lemma}[theorem]{Lemma}

\newtheorem{corollary}[theorem]{Corollary}

\theoremstyle{remark}
\newtheorem{remark}{Remark}

{\itshape}{}

\copyrightyear{2022}
\acmYear{2022}
\setcopyright{rightsretained}
\acmConference[HSCC '22]{25th ACM International Conference on Hybrid Systems: Computation and Control}{May 4--6, 2022}{Milan, Italy}
\acmBooktitle{25th ACM International Conference on Hybrid Systems: Computation and Control (HSCC '22), May 4--6, 2022, Milan, Italy}\acmDOI{10.1145/3501710.3519541}
\acmISBN{978-1-4503-9196-2/22/05}

\begin{document}

\title{Verifying Switched System Stability With Logic}
\author{Yong Kiam Tan}
\affiliation{%
  \institution{Carnegie Mellon University}
  \city{Pittsburgh}
  \state{PA}
  \country{USA}
}
\email{yongkiat@cs.cmu.edu}
\orcid{0000-0001-7033-2463}

\author{Stefan Mitsch}
\affiliation{%
  \institution{Carnegie Mellon University}
  \city{Pittsburgh}
  \state{PA}
  \country{USA}
}
\email{smitsch@cs.cmu.edu}
\orcid{0000-0002-3194-9759}

\author{Andr\'e Platzer}
\affiliation{%
  \institution{Carnegie Mellon University}
  \city{Pittsburgh}
  \state{PA}
  \country{USA}
}
\email{aplatzer@cs.cmu.edu}
\orcid{0000-0001-7238-5710}

\begin{abstract}
Switched systems are known to exhibit subtle (in)stability behaviors requiring system designers to carefully analyze the stability of closed-loop systems that arise from their proposed switching control laws.
This paper presents a formal approach for verifying switched system stability that blends classical ideas from the controls and verification literature using differential dynamic logic (\dL), a logic for deductive verification of hybrid systems.
From controls, we use standard stability notions for various classes of switching mechanisms and their corresponding Lyapunov function-based analysis techniques.
From verification, we use \dL's ability to verify quantified properties of hybrid systems and \dL models of switched systems as looping hybrid programs whose stability can be formally specified and proven by finding appropriate \emph{loop invariants}, i.e., properties that are preserved across each loop iteration.
This blend of ideas enables a trustworthy implementation of switched system stability verification in the \KeYmaeraX prover based on \dL.
For standard classes of switching mechanisms, the implementation provides fully automated stability proofs, including searching for suitable Lyapunov functions.
Moreover, the generality of the deductive approach also enables verification of switching control laws that require non-standard stability arguments through the design of loop invariants that suitably express specific intuitions behind those control laws.
This flexibility is demonstrated on three case studies: a model for longitudinal flight control by Branicky, an automatic cruise controller, and Brockett's nonholonomic integrator.
\end{abstract}



\maketitle

\section{Introduction}
\label{sec:introduction}

Switched systems provide a powerful mathematical paradigm for the design and analysis of discontinuous (or nondifferentiable) control mechanisms~\cite{DBLP:books/sp/Liberzon03,871309,SunGe,morse1995}.
Examples of such mechanisms include: bang-bang controllers that switch between on/off modes; gain schedulers that switch between a family of locally valid linear controllers; and supervisory control, where a supervisor switches between candidate controllers based on logical criteria~\cite{DBLP:books/sp/Liberzon03,morse1995}.
However, switched systems are known to exhibit subtle (in)stability behaviors, e.g., switching between stable subsystems can lead to instability~\cite{DBLP:books/sp/Liberzon03}, so it is important for system designers to adequately justify the stability of their proposed switching designs.
Verification and validation are complementary approaches for such justifications: \emph{validation} approaches, such as system simulations or lab experiments, allow designers to check that their models and controllers conform to real world behavior; \emph{verification} approaches yield formal mathematical proofs that the stability properties hold for \emph{all} possible switching decisions everywhere in the model's infinite state space, not just for finitely-many simulated trajectories.

This paper presents a logic-based, deductive approach for verifying switched system stability under various classes of switching mechanisms.
The key insight is that control-theoretic stability arguments for switching control can be formally justified by blending techniques from discrete program verification with continuous differential equations analysis using differential dynamic logic (\dL), a logic for deductive verification of hybrid systems~\cite{Platzer18,DBLP:journals/jar/Platzer17}.
Intuitively, switched systems are modeled in \dL as looping \emph{hybrid programs}~\cite{DBLP:conf/adhs/TanP21}, as in the following snippet (${\{\cdot\}}^*$ denotes repetition):
\begin{align*}
  &\{ \qquad\pumod{u}{ctrl(x)};      &&\text{// switching controller (discrete dynamics)}\\
  &\;\, \qquad\pevolve{\D{x}=f_u(x)}  &&\text{// actuate decision (continuous dynamics)}\\
  &\}^* \keyword{@invariant}(\text{ ... }) &&\text{// switching loop with invariant annotation}
\end{align*}

Accordingly, switched system stability is formally specified in \dL as first-order quantified safety properties of switching loops (\rref{subsec:backgroundstab}), and the resulting specifications can then be proved rigorously by combining fundamental ideas from verification and control, namely:
\begin{enumerate*}[label=\roman*),font=\itshape]
\item identification of \emph{loop invariants} (\keyword{@invariant} above), i.e., properties of the (discrete) loop that are preserved across all executions of the loop body,
\item \emph{compositional verification} for separately analyzing the discrete and continuous dynamics of the loop body, and
\item \emph{Lyapunov functions}, i.e., auxiliary energy functions that enable stability analysis for the continuous dynamics.
\end{enumerate*}

\rref{sec:loopinv} identifies key loop invariants underlying stability arguments for various classes of switching mechanisms and derives sound stability proof rules for those mechanisms.
Crucially, these \emph{syntactic derivations} are built from \dL's sound foundations for hybrid program reasoning~\cite{Platzer18,DBLP:journals/jar/Platzer17}, \emph{without} the need to introduce new mathematical concepts such as non-classical weak solutions or nondifferentiable Lyapunov functions~\cite{4518905,4806347}.
The remaining practical challenge is how to (automatically) find suitable Lyapunov function candidates for a given switching mechanism; the correctness of any generated candidates can be soundly checked in \dL.
\rref{sec:impl} adds support for switched systems in the \KeYmaeraX prover based on \dL~\cite{DBLP:conf/cade/FultonMQVP15}, including a modeling interface for switched systems, sum-of-squares search for Lyapunov function candidates~\cite{1243743,sostools}, and fully automatic verification of stability specifications for standard switching mechanisms.
Notably, the implementation requires \emph{no extensions} to \KeYmaeraX's soundness-critical core and thereby directly inherits all of \KeYmaeraX's correctness guarantees~\cite{DBLP:conf/cade/FultonMQVP15,DBLP:series/lncs/MitschP20}.
This trustworthiness is necessary for computer-aided verification of complex switching designs because the number of correctness conditions on their Lyapunov functions scales quadratically with the number of switching modes (\rref{subsec:controlstab}), making pen-and-paper proofs error-prone or infeasible.
\rref{sec:casestudies} further applies the deductive approach on three case studies, chosen because each require subtle twists to standard switched system stability arguments: %
\begin{itemize}
\item \emph{Longitudinal flight control}~\cite{735143}: This model is parametric (5 parameters, 2 state variables) and its stability justification due to Branicky uses a ``noncustomary'' Lyapunov function~\cite{735143,871309} with intricate arithmetic reasoning.
The proof uses \emph{ghost switching}, where virtual switching modes are introduced for the sake of stability analysis, analogous to the use of ghost variables in program verification~\cite{DBLP:journals/cacm/OwickiG76,Platzer18,DBLP:journals/jacm/PlatzerT20}.

\item \emph{Automatic cruise control}~\cite{DBLP:phd/dnb/Oehlerking11}: This hybrid automaton features switching between several modes based on specific guard conditions: standard/emergency braking, accelerating, and PI control. Lyapunov function candidates can be numerically generated~\cite{DBLP:conf/hybrid/MohlmannT13}, but must be corrected for soundness.

\item \emph{Brockett's nonholonomic integrator}~\cite{Brockett83asymptoticstability}: A large class of control systems can be transformed to the nonholonomic integrator but this system is not stabilizable by continuous feedback~\cite{Brockett83asymptoticstability,DBLP:books/sp/Liberzon03}.
The stability argument must account for an initial control mode that drives the system into a suitable region before a stabilizing control law can be applied.
\end{itemize}

These case studies are verified semi-automatically in \KeYmaeraX, with user guidance to design and prove modified loop invariants that suitably capture the specific intuitions behind their respective control laws.
The flexibility and generality of this paper's deductive approach enables such (modified) stability arguments, while ensuring that every step in the argument is rigorously justified using sound \dL logical foundations.
\iflongversion
All proofs are in Appendix~\ref{app:proofs} and~\ref{app:casestudies}.
\else
All proofs are in the supplement~\rref{app:}.
\fi

\section{Background}
\label{sec:background}
This section recalls switched systems and their hybrid program models~\citep{DBLP:conf/adhs/TanP21}.
It then explains how stability for these models is formally specified and verified using differential dynamic logic (\dL)~\citep{DBLP:journals/jar/Platzer17,Platzer18}.

\subsection{Switched Systems as Hybrid Programs}

\subsubsection{Hybrid Programs} The language of \emph{hybrid programs} is generated by the following grammar, where $x$ is a variable, $\etermA$ is a \dL term, and $\ivr$ is a formula of first-order real arithmetic~\citep{DBLP:journals/jar/Platzer17,Platzer18}.
\[
  \progA,\progB~\bebecomes~\pevolvein{\D{x}=\genDE{x}}{\ivr} \alternative \pumod{x}{e} \alternative \ptest{\ivr} \alternative \progA ; \progB \alternative \pchoice{\progA}{\progB} \alternative \prepeat{\progA}
\]

Continuous dynamics are modeled using systems of ordinary differential equations (ODEs) $\pevolvein{\D{x}=\genDE{x}}{\ivr}$ evolving within domain $\ivr$; the ODE is written as $\D{x}=\genDE{x}$ when there is no domain constraint, i.e., $\ivr \mnodefequiv \ltrue$.
Discrete dynamics are modeled using assignments ($\pumod{x}{e}$ assigns the value of term $e$ to $x$) and tests ($\ptest{\ivr}$ checks whether condition $\ivr$ is true in the current state).
The program combinators are used to piece together sub-programs to form programs with hybrid dynamics.
The combinators are: sequential composition ($\progA ; \progB$ runs $\progA$ followed by $\progB$), nondeterministic choice ($\pchoice{\progA}{\progB}$ runs $\progA$ or $\progB$ nondeterministically), and nondeterministic repetition ($\prepeat{\progA}$ repeats $\progA$ for any number of iterations).

Throughout this paper, $x = (x_1,\dots,x_n)$ denotes the vector of continuous state variables for the system under consideration.
Other variables are used for program auxiliaries, e.g., to describe memory and timing components of switching controllers.

\subsubsection{Switched systems} A \emph{switched system} is described by a finite family $\sigfam$ of ODEs $\D{x}=f_p{(x)}, p \in \sigfam$ and a set of \emph{switching signals} $\sigma : [0,\infty) \to \sigfam$ that prescribe the ODE $\D{x}=f_{\sigma(t)}{(x)}$ to follow at time $t$ along the system's evolution.
Tan and Platzer~\citep{DBLP:conf/adhs/TanP21} use hybrid programs as formal models for various classes of switching mechanisms; one example is \emph{arbitrary switching}~\cite{DBLP:books/sp/Liberzon03} where the system is allowed to follow \emph{any} switching signal in order to model real world systems whose switching behavior is uncontrolled or \emph{a priori} unknown.
The hybrid program $\arbswitch \mnodefequiv \prepeat{\Big( \bigcup_{p \in \sigfam}{\D{x}=f_p{(x)}} \Big)}$ models arbitrary switching analogously to a computer simulation~\cite[Proposition 1]{DBLP:conf/adhs/TanP21}: on each loop iteration, the program makes a (discrete) nondeterministic choice of switching decision $\bigcup_{p \in \sigfam}{\big(\cdot\big)}$ to select an ODE $\D{x}=f_p{(x)}$ which it then follows continuously for an arbitrarily chosen duration before repeating the simulation loop.

The hybrid programs language can be used to model various other classes of switching mechanisms~\cite{DBLP:books/sp/Liberzon03,DBLP:conf/adhs/TanP21}, including general \emph{controlled switching}, as illustrated in~\rref{sec:introduction}, where a (discrete) control law $\pumod{u}{ctrl(x)}$ decides the ODE $\D{x}=f_u{(x)}$ to switch to on each loop iteration.
Stability for these models is explained next.

\subsection{Stability as Quantified Loop Safety}
\label{subsec:backgroundstab}

This paper studies \emph{uniform global pre-asymptotic stability} (UGpAS) for switched systems~\cite{DBLP:books/sp/Liberzon03,4806347,10.2307/j.ctt7s02z}, defined as follows:

\begin{definition}[UGpAS~\cite{4806347,10.2307/j.ctt7s02z}]
Let $\Phi(x)$ denote the set of all (domain-obeying) solutions\footnote{A formal construction of the (right-maximal) solution $\solvar$ for a given switching signal $\sigma$ is available elsewhere~\cite[Appendix A]{DBLP:conf/adhs/TanP21}.} $\solvar : [0,T_\solvar] \to \reals^n$ for a switched system from state $x \in \reals^n$. %
The origin $0 \in \reals^n$ is:
\begin{itemize}
\item \textbf{uniformly stable} if, for all $\varepsilon > 0$, there exists $\delta > 0$ such that from all initial states $x \in \reals^n$ with $\norm{x} < \delta$, all solutions $\solvar \in \Phi(x)$ satisfy $\norm{\solvar(t)} < \varepsilon$ for all times $0 \leq \tvar \leq T_\solvar$,

\item \textbf{uniformly globally pre-attractive} if, for all $\varepsilon > 0, \delta  > 0$, there exists $T \geq 0$ such that from all initial states $x \in \reals^n$ with $\norm{x} < \delta$, all solutions $\solvar \in \Phi(x)$ satisfy $\norm{\solvar(t)} < \varepsilon$ for all times $T \leq \tvar \leq T_\solvar$, and

\item \textbf{uniformly globally pre-asymptotically stable} if the system is uniformly stable and uniformly globally pre-attractive.
\end{itemize}
\label{def:ugpas}
\end{definition}

The UGpAS definition can be understood intuitively for a system with a given switching control mechanism:
\begin{itemize}
\item \emph{stability} means the mechanism keeps the system close to the origin if the system is initially perturbed close to the origin,
\item \emph{global pre-attractivity} means the mechanism drives the system to the origin asymptotically as $t \to \infty$, and
\item \emph{uniform} means the stability and pre-attractivity properties are independent of both the nondeterminism in the switching mechanism (e.g., arbitrary switching) and the choice of initial states satisfying $\norm{x} < \delta$; for brevity in subsequent sections, ``uniform'' is elided when describing stability properties.
\end{itemize}

\begin{remark}
\label{rem:ugpas}
Switched systems whose solutions are all uniformly bounded in time, i.e., there exists $T_m$ such that for all solutions $\solvar$, $T_\solvar \leq T_m$, are trivially pre-attractive.
Goebel et al.~\cite{4806347,10.2307/j.ctt7s02z} introduce the notion of \emph{pre-attractivity} as opposed to \emph{attractivity} for hybrid systems because it separates considerations about whether a hybrid system's solutions are \emph{complete}, i.e., solutions exist for all (forward) time, from conditions for stability and attractivity.
Pre-attractivity also sidesteps the difficult question of whether a switched system exhibits \emph{Zeno} behavior, i.e., where infinitely many discrete switches occur in finite time~\cite{DBLP:books/sp/Liberzon03,doi:10.1002/rnc.592}.
Indeed, it is common in the hybrid and switched systems literature to either \emph{ignore} incomplete solutions or \emph{assume} the models under consideration only have complete solutions~\cite{DBLP:conf/hybrid/MohlmannT13,DBLP:books/sp/Liberzon03,doi:10.1002/rnc.592}. %
Instead of predicating proofs on these hypotheses, this paper formalizes the (weaker) notion of UGpAS for switched systems, leaving proofs of completeness of solutions out of scope.
\end{remark}

The definition of UGpAS nests alternating quantification over real numbers with temporal quantification over the solutions $\solvar$ of switched systems.
This combination of quantifiers can be expressed formally using the formula language of \dL~\citep{DBLP:journals/jar/Platzer17,Platzer18}, whose grammar is shown below, $\sim {\in}~\{=,\neq,\geq,>,\leq,<\}$ is a comparison operator between \dL terms $\etermA, \etermB$ and $\alpha$ is a hybrid program:
\begin{align*}
  \fvarA,\fvarB~\bebecomes&~\etermA \sim \etermB \alternative \fvarA \land \fvarB \alternative \fvarA \lor \fvarB \alternative \lnot{\fvarA} \alternative \lforall{v}{\fvarA} \alternative \lexists{v}{\fvarA} \alternative \dbox{\alpha}{\fvarA} \alternative \ddiamond{\alpha}{\fvarA}
\end{align*}

This grammar extends the first-order language of real arithmetic (\folr) with the box ($\dbox{\alpha}{\fvarA}$) and diamond ($\ddiamond{\alpha}{\fvarA}$) modality formulas which express that all or some runs of hybrid program $\alpha$ satisfy postcondition $\fvarA$, respectively.
Real arithmetic \folr is decidable by quantifier elimination~\cite{Tarski} and serves as a useful base specification language.
Various specifications are equivalently definable in \folr, e.g., Euclidean norm bounds $\norm{x} \sim \varepsilon \mdefequiv (\sum_{i=1}^n x_i^2) \sim \varepsilon^2$ (for $\varepsilon \geq 0$) and topological operations such as the boundary $\bdr{\fvarA}$ and closure $\closure{\fvarA}$ of the set characterized by formula $\fvarA$~\cite{Bochnak1998}.

The box modality formula $\dbox{\alpha}{\fvarA}$ expresses \emph{safety} properties $\fvarA$ of program $\alpha$ that must hold along all of its executions~\cite{Platzer18}.
When $\alpha$ models a switched system, the box modality quantifies (uniformly) over all times for all solutions arising from the switching mechanism.
Accordingly, UGpAS for switched systems is formally specified by nesting the box modality with the first-order quantifiers. %

\begin{lemma}[UGpAS in differential dynamic logic]
The origin $0 \in \reals^n$ for a switched system modeled by program $\alpha$ is UGpAS iff the \dL formula $\astabhp{\alpha}$ is valid. Variables $\varepsilon, \delta,T,\tvar$ are fresh in $\alpha$:
\begingroup
\allowdisplaybreaks
\begin{align*}
\stabhp{\alpha} &\mnodefequiv \lforall{\varepsilon {>} 0} { \lexists{\delta {>} 0}{ \lforall{x}{\big( \norm{x}<\delta \limply \dbox{\alpha}{\,\norm{x}<\varepsilon}\big)}}}\\
\attrhp{\alpha} &\mnodefequiv \lforall{\varepsilon {>} 0} { \lforall{\delta {>} 0}{ \lexists{T {\geq} 0}{ \lforall{x}{\big( \norm{x}<\delta \limply}}}}\\
&\qquad\qquad\qquad\dbox{\pumod{\tvar}{0};\taug{\alpha}}{\,(\tvar \geq T \limply \norm{x}<\varepsilon)}\big) \\
\astabhp{\alpha} & \mnodefequiv \stabhp{\alpha} \land \attrhp{\alpha}
\end{align*}
\endgroup

Here, $\stabhp{\alpha}$ and $\attrhp{\alpha}$ characterize stability and global pre-attractivity of $\alpha$, respectively. In $\attrhp{\alpha}$, $\taug{\alpha}$ denotes the hybrid program obtained from $\alpha$ by augmenting its continuous dynamics so that variable $\tvar$ tracks the progression of time.
\label{lem:asymstabdl}
\end{lemma}

Formulas $\stabhp{\alpha}$ and $\attrhp{\alpha}$ syntactically formalize in \dL the corresponding quantifiers in~\rref{def:ugpas}.
In $\attrhp{\alpha}$, the fresh clock variable $\tvar$ is initialized to $0$ and syntactically tracks the progression of time along switched system solutions.
The program $\taug{\alpha}$ can, e.g., be constructed by adding a clock ODE $\D{\tvar}=1$ to all ODEs in the switched system model $\alpha$.
Accordingly, the postcondition $\tvar \geq T \limply \norm{x} < \varepsilon$ expresses that the system state norm is bounded by $\varepsilon$ after $T$ time units along any switching trajectory, as required in~\rref{def:ugpas}.
Various other stability notions are of interest in the continuous and hybrid systems literature~\cite{DBLP:conf/tacas/TanP21,DBLP:books/sp/Liberzon03,DBLP:conf/hybrid/PodelskiW06,DBLP:conf/cav/GaoKDRSAK19,SunGe,10.2307/j.ctt7s02z,DBLP:phd/dnb/Oehlerking11}.
These variations can also be formally specified in \dL~\cite{DBLP:conf/tacas/TanP21} but are left out of scope for this paper.

\subsection{Proof Calculus}
\irlabel{qear|\usebox{\Rval}}
\irlabel{qearexp|\usebox{\Rvalexp}}

The \dL proof calculus enables formal, deductive verification of UGpAS stability specifications through compositional reasoning principles for hybrid programs~\citep{DBLP:journals/jar/Platzer17,Platzer18} and a complete axiomatization for ODE invariants~\citep{DBLP:journals/jacm/PlatzerT20}.
For example, an important syntactic tool for differential equations reasoning is the \emph{Lie derivative} of term $\etermA$ along ODE $\D{x}=\genDE{x}$, defined as $\lie[]{f}{\etermA} \mdefeq \nabla \etermA \cdot f$.
The sound calculation and manipulation of Lie derivatives is enabled in \dL through the use of syntactic differentials~\citep{DBLP:journals/jar/Platzer17}.

All proofs are presented in a classical sequent calculus with the usual rules for manipulating logical connectives and sequents.
The semantics of \emph{sequent} \(\lsequent{\Gamma}{\fvarA}\) is equivalent to the formula \((\landfold_{\fvarB \in\Gamma} \fvarB) \limply \fvarA\) and a sequent is \emph{valid} iff its corresponding formula is valid.
The key (derived) \dL proof rule used in this paper is:
\[
\dinferenceRule[loop|loop]{}
{\linferenceRule
  {\lsequent{\Gamma}{\invariant} & \lsequent{\invariant}{\dbox{\alpha}{\,\invariant}} & \lsequent{\invariant}{\fvarA}}
  {\lsequent{\Gamma}{\dbox{\prepeat{\alpha}}{\fvarA}}}
}{}
\]

The~\irref{loop} rule says that, in order to prove validity of the conclusion (below the rule bar), it suffices to prove the three premises (above the rule bar), respectively from left to right:
\begin{enumerate*}[label=\roman*),font=\itshape]
\item the initial assumptions $\Gamma$ imply $\invariant$,
\item $\invariant$ is preserved across the loop body $\alpha$, i.e., $\invariant$ is a \emph{loop invariant} for $\prepeat{\alpha}$, and
\item $\invariant$ implies the postcondition $\fvarA$.
\end{enumerate*}
The identification of loop invariants $\invariant$ is crucial for formal proofs of UGpAS, as illustrated by the following deductive proof skeleton for stability (a similar skeleton is used for pre-attractivity):

\noindent\begin{minipage}[c]{0.01\textwidth}%
$\overset{\substack{\scalebox{1}{\textbf{Deduction}\hidewidth}\mathstrut}}{\Vast\uparrow}$
\end{minipage}\quad
\begin{minipage}[c]{0.45\textwidth}
{\begin{sequentdeduction}
  \linfer[]{
    \linfer[]{
    \linfer[loop]{
    \linfer[]{
      \vdots
    }
    {\lsequent{\Gamma}{\invariant}} \quad
    \linfer[]{
    \linfer[]{
    \lsequent{\Gamma_1}{\fvarA_1} \quad \cdots \quad \lsequent{\Gamma_k}{\fvarA_k}
    }
    {\qquad \vdots \quad \Big(~\parbox{5em}{\footnotesize\selectfont hybrid program reasoning for $\alpha$}~\Big)}
    }
      {\redc{\lsequent{\invariant}{\dbox{\alpha}{\,\invariant}}}} \quad
    \linfer[]{
      \vdots
    }
    {\lsequent{\invariant}{\norm{x}<\varepsilon}}
    }
      {\lsequent{\Gamma}{\dbox{\prepeat{\alpha}}{\,\norm{x}<\varepsilon}}}
    }
    {\qquad \vdots \quad \Big(~\parbox{5em}{\footnotesize\selectfont logic/arithmetic reasoning for $\Gamma$} \Big)}
  }
  {\lsequent{}{\stabhp{\prepeat{\alpha}}}}
\end{sequentdeduction}}

\end{minipage}

Proofs proceed upwards by deduction, where each reasoning step is justified by sound \dL axioms and rules of inference, e.g., the~\irref{loop} rule.
The proof skeleton above syntactically \emph{derives} a proof rule that reduces a stability proof for $\prepeat{\alpha}$ to proofs of its top-most premises, $\lsequent{\Gamma_1}{\fvarA_1} \; \cdots \; \lsequent{\Gamma_k}{\fvarA_k}$.
These correspond to required logical and arithmetical conditions on Lyapunov functions for various switching mechanisms.
The loop invariant step (highlighted in \redc{red}) crucially ties together these conditions on Lyapunov functions and hybrid program reasoning for switched systems.

\section{Loop Invariants for Switched System Stability}
\label{sec:loopinv}
This section identifies loop invariants for proving UGpAS under various classes of switching mechanisms with Lyapunov functions~\cite{MR1201326,DBLP:books/sp/Liberzon03,DBLP:journals/tac/Branicky98}; relevant mathematical arguments are presented briefly%
\iflongversion%
, see~\rref{app:proofs} for more details.
\else
~(see supplement~\rref{app:}).
\fi
Throughout the section, loop invariants are progressively tweaked to account for new design insights behind increasingly complex switching mechanisms.

\subsection{Arbitrary and State-Dependent Switching}
\label{subsec:arbstatestab}

\subsubsection{Arbitrary Switching}
\label{subsec:arbstab}
Stability for the arbitrary switching model $\arbswitch$ from \rref{sec:background} can be verified by finding a so-called \emph{common Lyapunov function} $\lterm$ for all of the ODEs $\D{x}=f_p{(x)}, p \in \sigfam$ satisfying the following arithmetical conditions~\cite{DBLP:books/sp/Liberzon03,SunGe}:
\begin{enumerate}[label=\roman*),font=\itshape]
\item \label{itm:clfone} $\lterm(0) = 0$ and $\lterm(x) > 0$ for all $\norm{x} > 0$,
\item \label{itm:clftwo} $\lterm$ is \emph{radially unbounded}, i.e., for all $b$, there exists $\gamma > 0$ such that $\norm{x} < \gamma$ for all $\lterm(x) \leq b$, and
\item \label{itm:clfthree} for each ODE $\D{x}=f_p{(x)}, p \in \sigfam$, the Lie derivative $\lie[]{f_p}{\lterm}$ satisfies:
  $\lie[]{f_p}{\lterm}(0) = 0$ and $\lie[]{f_p}{\lterm}(x) < 0$ for all $\norm{x} > 0$.
\end{enumerate}

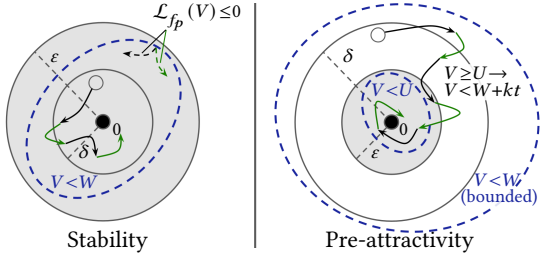
\begin{figure}
\centering
\begin{overpic}[width=0.42\textwidth,clip,trim=400 330 380 330,tics=10]{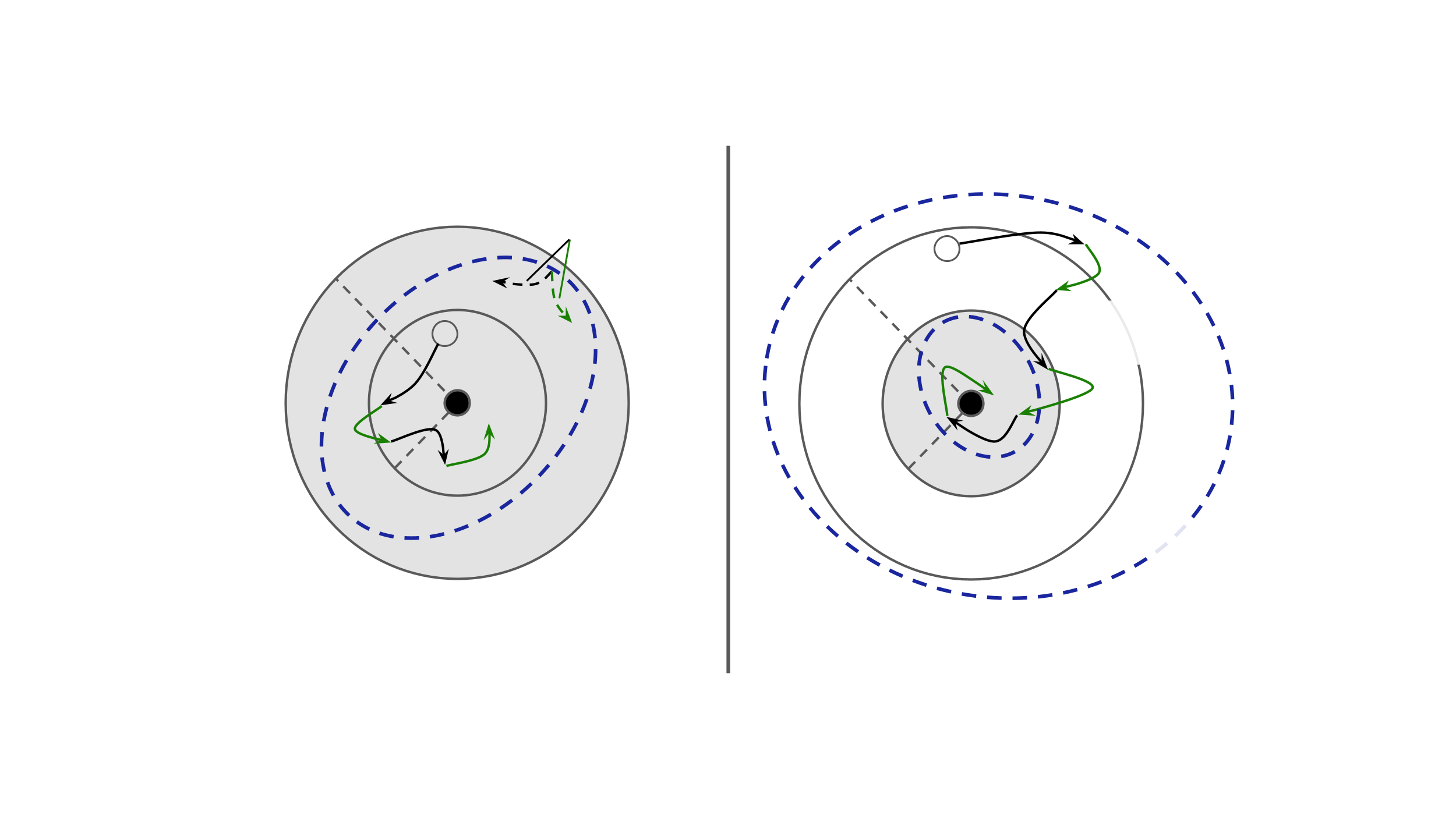}
 \put (13,33) {$\scriptstyle\varepsilon$}
 \put (18,16) {$\scriptstyle\delta$}
 \put (24,20) {$\scriptstyle0$}
 \put (13,10.5)  {\bluec{$\scriptstyle\lterm < \ltermUpper$}}
 \put (31,41) {$\scriptstyle\lie[]{f_p}{\lterm} \leq 0$}
 \put (16,0)  {Stability}
 \put (70,16) {$\scriptstyle\varepsilon$}
 \put (65,33) {$\scriptstyle\delta$}
 \put (75,20) {$\scriptstyle0$}
 \put (88.3,11)  {\bluec{$\scriptstyle\lterm < \ltermUpper$}}
 \put (86,8)  {\bluec{\footnotesize (bounded)}}
 \put (83,30) {$\scriptstyle\lterm \geq \ltermLower \limply$}
 \put (83,27) {$\scriptstyle\lterm < \ltermUpper + k \tvar$}
 \put (70,27) {\bluec{$\scriptstyle \lterm < \ltermLower$}}
 \put (62,0)  {Pre-attractivity}
\end{overpic}

\caption{Loop invariants for UGpAS (arbitrary switching), stability (left) and pre-attractivity (right).
Switching trajectories are illustrated by alternating black and \greenc{green} arrows.}
\label{fig:arbstab}
\end{figure}

Conditions~\rref{itm:clfone}--\rref{itm:clfthree} are generalizations of well-known conditions for stability of ODEs~\cite{Chicone2006,MR1201326} to arbitrary switching.
Intuitively, conditions~\rref{itm:clfone} and~\rref{itm:clfthree} ensure that $\lterm$ acts as an auxiliary energy function whose value decreases asymptotically to zero (at the origin) along all switching trajectories of the system; the radial unboundedness condition~\rref{itm:clftwo} ensures that this argument applies to all system states for \emph{global} pre-attractivity~\cite{MR1201326}.
Correctness of these conditions can be proved in \dL using loop invariants, see~\rref{fig:arbstab} (explained below).

\paragraph{Stability}
The specification $\stabhp{\arbswitch}$ requires that all trajectories of $\arbswitch$ stay in the grey ball $\norm{x}<\varepsilon$, starting from a chosen ball $\norm{x}<\delta$, see~\rref{fig:arbstab} (left).
Condition~\rref{itm:clfone} guarantees that the ball $\norm{x}<\varepsilon$ contains (a connected component of) the sublevel set $\lterm {<} \ltermUpper$ for some $\ltermUpper {>} 0$ (dashed \bluec{blue} curve) and this sublevel set contains a smaller ball $\norm{x}<\delta$~\cite{Chicone2006,MR1201326}.
Condition~\rref{itm:clfthree} shows that this sublevel set is invariant for each ODE $\D{x}=f_p{(x)}, p \in \sigfam$ because $\lie[]{f_p}{\lterm}(x) \leq 0$, as illustrated by the dashed black and \greenc{green} arrows for two different switching choices $p \in \sigfam$ both locally pointing inwards on the boundary of the sublevel set.
Thus, the formula $\invariants \mnodefequiv \norm{x} < \varepsilon \land \lterm < \ltermUpper$, which characterizes the blue sublevel set, is an invariant for all possible switching choices in the loop body of $\arbswitch$, which makes $\invariants$ a suitable loop invariant for $\stabhp{\arbswitch}$.

\paragraph{Pre-attractivity}
The specification $\attrhp{\arbswitch}$ requires that all trajectories of $\arbswitch$ stay in the grey ball $\norm{x}<\varepsilon$ after a chosen time $T$, starting from the initial ball $\norm{x}<\delta$, see~\rref{fig:arbstab} (right).
The ball $\norm{x} < \delta$ is bounded, so it is contained in a sublevel set satisfying $\lterm < \ltermUpper$ for some $\ltermUpper > 0$ (outer dashed \bluec{blue} curve); this sublevel set is bounded by condition~\rref{itm:clftwo}.
Like the stability argument, condition~\rref{itm:clfone} guarantees that there is a sublevel set $\lterm < \ltermLower$ for some $\ltermLower > 0$ (inner dashed \bluec{blue} curve) contained in the ball $\norm{x} < \varepsilon$, and condition~\rref{itm:clfthree} shows that the sublevel sets characterized by $\lterm < \ltermUpper$ and $\lterm < \ltermLower$ are both invariants for every ODE in the loop body of $\arbswitch$.
The set characterized by formula $\lterm \geq \ltermLower \land \lterm \leq \ltermUpper$ is compact and bounded away from the origin, which implies by condition~\rref{itm:clfthree} that there is a uniform bound $k < 0$ on this set, where for each ODE $\D{x}=f_p{(x)}, p \in \sigfam$, $\lie[]{f_p}{\lterm}(x) \leq k$.
Thus, the value of Lyapunov function $\lterm$ decreases at rate $k$, regardless of switching choices in the loop body of $\arbswitch$, \emph{as long as} it has not entered $\lterm < \ltermLower$.
The loop invariant for $\attrhp{\arbswitch}$ syntactically expresses this intuition: $\invarianta \mnodefequiv \lterm < \ltermUpper \land (\lterm \geq \ltermLower \limply \lterm < \ltermUpper + k \tvar)$.
For a sufficiently large choice of $T$ with $\ltermUpper + k T \leq \ltermLower$, trajectories at time $t \geq T$ satisfy $\lterm < \ltermLower$ so they are contained in the $\norm{x} < \varepsilon$ ball.

The loop invariants identified above enable derivation of a formal \dL stability proof rule for $\arbswitch$ (deferred to a more general version in~\rref{cor:ugpasstateswitchclf} below).
In fact, since arbitrary switching is the most permissive form of switching~\cite{DBLP:books/sp/Liberzon03}, UGpAS for any switching mechanism can be soundly justified using the loop invariants above in case a suitable common Lyapunov function can be found.

\subsubsection{State-dependent Switching}
\label{subsec:statestab}
The state-dependent switching mechanism~\cite{DBLP:books/sp/Liberzon03} constrains arbitrary switching by allowing execution of (and switching to) an ODE $\D{x}=f_p{(x)}, p \in \sigfam$ only when the system state is in domain $\ivr_p$.
This is modeled by the hybrid program $\stateswitch\mnodefequiv \prepeat{\Big( \bigcup_{p \in \sigfam}{ \pevolvein{\D{x}=f_p{(x)}}{\ivr_p}} \Big)}$~\cite[Proposition 2]{DBLP:conf/adhs/TanP21}, where arbitrary switching $\arbswitch$ corresponds to the special case with $\ivr_p \mnodefequiv \ltrue$ for all $p \in \sigfam$.

The same loop invariants for $\arbswitch$ are used for $\stateswitch$ to derive the following proof rule.
For brevity, premises of all derived stability proof rules are implicitly conjunctively quantified over $p \in \sigfam$.

\begin{corollary}[UGpAS for state-dependent switching, CLF]
\label{cor:ugpasstateswitchclf}
The following proof rule for common Lyapunov function $\lterm$ with three stacked premises is syntactically derivable in \dL.

\noindent
\begin{calculuscollection}
\begin{calculus}
\dinferenceRule[UGpASst|CLF]{}
{\linferenceRule
  {     \begin{array}{l}
    \lsequent{}{\lterm(0)=0 \land \lforall{x}{(\norm{x} > 0 \limply \lterm(x) > 0)}} \\
    \lsequent{}{\lforall{b}{\lexists{\gamma}{\lforall{x}{(\lterm(x) \leq b \limply \norm{x} \leq \gamma)}}}} \\
    \lsequent{}{\lie[]{f_p}{\lterm}(0) = 0 \land \lforall{x}{(\norm{x} > 0 \land \closure{\ivr_p} \limply \lie[]{f_p}{\lterm}(x) < 0)}}
    \end{array} }
  {\lsequent{}{\astabhp{\stateswitch}}}
}{}
\end{calculus}
\end{calculuscollection}%
\end{corollary}

\rref{cor:ugpasstateswitchclf} syntactically derives a slight generalization of conditions~\rref{itm:clfone}--\rref{itm:clfthree} from~\rref{subsec:arbstab} for $\stateswitch$, where the Lie derivatives $\lie[]{f_p}{\lterm}(x)$ for each $p \in \sigfam$ are required to be negative on their respective domain closures\footnote{The topological closure $\closure{\ivr}$ of domain $\ivr$ is needed for soundness of a technical compactness argument used in the pre-attractivity proof%
\iflongversion%
, see~\rref{app:proofs}.
\else
~(see supplement~\rref{app:}).
\fi
} $\closure{\ivr_p}$.
This generalization is justified by the same loop invariants in~\rref{subsec:arbstab} because the ODE invariance properties are only required to hold in their respective domains.

\begin{figure}
\begin{minipage}[t]{0.25\textwidth}
\begin{align*}
&\scriptscriptstyle p:\; \pevolvein{\D{x_1}=-4.6x_1+5.5x_2, \D{x_2}=-5.5x_1+4.4x_2}{x_1x_2\geq0} \\[-0.3em]
&\greenc{\scriptscriptstyle q:\; \pevolvein{\D{x_1}=4.4x_1+5.5x_2, \D{x_2}=-5.5x_1-4.6x_2}{x_1x_2\leq0}}
\end{align*}
\includegraphics[width=0.91\textwidth,clip,trim=0 0 0 0]{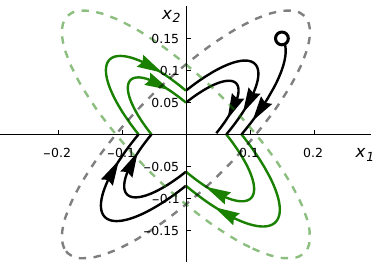}
\end{minipage}\quad
\begin{minipage}[t]{0.20\textwidth}
\begin{align*}
&\scriptscriptstyle \lterm_p = x_1^2-1.65x_1x_2+x_2^2\\[-0.3em]
&\greenc{\scriptscriptstyle \lterm_q = x_1^2+1.65x_1x_2+x_2^2}
\end{align*}
\includegraphics[width=0.91\textwidth,clip,trim=0 0 0 0]{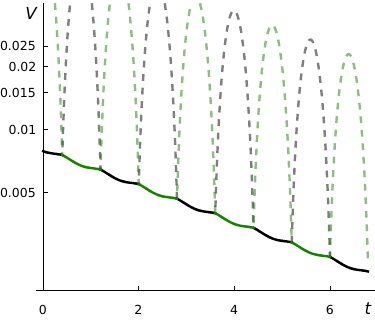}
\end{minipage}
\caption{A switching trajectory for Example 7 from~\rref{subsec:examples} with state-dependent switching (left) and the value of two Lyapunov functions along that trajectory (right, log-scale on vertical axis). Solid lines indicate the active Lyapunov function at time $t$.
Two sublevel sets $\lterm_p, \greenc{\lterm_q} < \ltermUpper = 0.012$ are shown dashed on the left within which the switching trajectory is respectively trapped at any given time.}
\label{fig:switchstab}
\end{figure}

The domain asymmetry in $\stateswitch$ suggests another way of generalizing the stability arguments, namely, through the use of \emph{multiple Lyapunov functions}, where a (possibly) different Lyapunov function $\lterm_p$ is associated to each $p \in \sigfam$~\cite{DBLP:journals/tac/Branicky98}.
Here, the function $\lterm_p$ is responsible for justifying stability within domain $\ivr_p$, i.e., its value decreases along system trajectories whenever the system is within $\ivr_p$, as illustrated in~\rref{fig:switchstab}.
Constraints on these functions are obtained by modifying the loop invariants to account for this intuition.

\paragraph{Stability} The stability loop invariant is modified by case splitting disjunctively on the domains $\ivr_p, p\in \sigfam$, and requiring that the sublevel set characterized by $\lterm_p < \ltermUpper$ is invariant within its respective domain $\ivr_p$:
$\invariants \mnodefequiv  \norm{x} < \varepsilon \land \lorfold_{p \in \sigfam}\big( \ivr_p \land \lterm_p < \ltermUpper\big)$.
Similar to~\rref{subsec:arbstab}, the bound $\ltermUpper$ is chosen so that each sublevel set characterized by $\lterm_p < W$ is contained in the ball $\norm{x} < \varepsilon$.

\paragraph{Pre-attractivity} The pre-attractivity loop invariant is similarly modified by disjunctively requiring that each $\lterm_p$ decreases along system trajectories when the system is in their respective domains $\ivr_p$: $\invarianta \mnodefequiv \lorfold_{p \in \sigfam}\big( \ivr_p \land \lterm_p < \ltermUpper \land (\lterm_p \geq \ltermLower \limply \lterm_p < \ltermUpper + k \tvar) \big)$.
The constants $\ltermLower, \ltermUpper, k, T$ are chosen as appropriate lower or upper bounds for all the Lyapunov functions (see proof of~\rref{cor:ugpasstateswitchmlf}).

Arithmetical conditions for the Lyapunov functions $\lterm_p, p\in \sigfam$ are derived from the modified invariants in the following rule.

\begin{corollary}[UGpAS for state-dependent switching, MLF]
\label{cor:ugpasstateswitchmlf}
The following proof rule for multiple Lyapunov functions $\lterm_p, p \in \sigfam$ with four stacked premises is syntactically derivable in \dL.

\noindent
\begin{calculuscollection}
\begin{calculus}
\dinferenceRule[UGpASstmlf|MLF]{}
{\linferenceRule
  {     \begin{array}{l}
    \lsequent{}{\lterm_p(0)=0 \land \lforall{x}{(\norm{x} > 0 \limply \lterm_p(x) > 0)}} \\
    \lsequent{}{\lforall{b}{\lexists{\gamma}{\lforall{x}{(\lterm_p(x) \leq b \limply \norm{x} \leq \gamma)}}}} \\
    \lsequent{}{\lie[]{f_p}{\lterm_p}(0) {=} 0 \land \lforall{x}{(\norm{x} {>} 0 \land \closure{\ivr_p} \limply \lie[]{f_p}{\lterm_p}(x) {<} 0)}}\\
    \lsequent{}{\landfold_{q \in \sigfam}{\Big(\ivr_p \land \ivr_q \limply \lterm_p = \lterm_q\Big)}}
    \end{array} }
  {\lsequent{}{\astabhp{\stateswitch}}}
}{}
\end{calculus}
\end{calculuscollection}%
\end{corollary}

The top three premises of~\rref{cor:ugpasstateswitchmlf} are similar to those of~\rref{cor:ugpasstateswitchclf}, but are now required to hold for each Lyapunov function $\lterm_p, p \in \sigfam$ separately.
The (new) bottom premise corresponds to a compatibility condition between the Lyapunov functions arising from the loop invariants.
For example, consider the stability loop invariant (similarly for pre-attractivity) and suppose the system currently satisfies disjunct $\ivr_p \land \lterm_p < \ltermUpper$ with $\lterm_p$ justifying stability in domain $\ivr_p$.
If the system switches to the ODE $\D{x}=f_q(x)$ within domain $\ivr_q$, then Lyapunov function $\lterm_q$ becomes the active Lyapunov function which must satisfy $\lterm_q < \ltermUpper$ to preserve the stability loop invariant.
The premise $\ivr_p \land \ivr_q \limply \lterm_p = \lterm_q$ says that the Lyapunov functions $\lterm_p, \lterm_q$ are equal whenever such a switch is possible (in either direction), i.e., when their domains overlap.

\subsection{Controlled Switching}
\label{subsec:controlstab}
This section turns to \emph{controlled switching} models~\cite{DBLP:conf/adhs/TanP21}, where an explicit controller program is responsible for making logical switching decisions between the ODEs $\D{x}=f_p(x), p \in \sigfam$.
This is in contrast to earlier models $\arbswitch, \stateswitch$ which exhibit \emph{autonomous switching}, i.e., without an explicit control logic~\cite{DBLP:books/sp/Liberzon03,DBLP:books/sp/necs2005/Branicky05}.
General controlled switching is modeled by the hybrid program $\ctrlswitch$:
\begin{align*}
\ctrlswitch \mnodefequiv \underset{\hidewidth\text{initialization}\hidewidth}{\underset{\downarrow}{\alpha_i}} ; \Big(
 \overset{\hidewidth\text{switching controller}\hidewidth}{\overset{\uparrow}{\vphantom{\Big(\Big)}\alpha_u}} ; \overbrace{\bigcup_{p \in \sigfam}{\big( \ptest{u = p} ; \pevolvein{\D{x}=f_p{(x,y)},\D{y}=g_p{(x,y)}}{\ivr_p} \big)}}^{\hidewidth\alpha_p \text{ (plant, actuate decision)}\hidewidth}
\Big)^*
\end{align*}

The model $\ctrlswitch$ uses three subprograms: $\alpha_i$ initializes the system, then $\alpha_u$ (modeling the switching controller) and $\alpha_p$ (modeling the continuous plant dynamics) are run in a switching loop.
The discrete programs $\alpha_i, \alpha_u$ decide on values for the control output $u=p, p \in \sigfam$ and the program $\alpha_p$ responds to this output by evolving the corresponding ODE $\pevolvein{\D{x}=f_p{(x,y)},\D{y}=g_p{(x,y)}}{\ivr_p}$.
The programs $\alpha_i, \alpha_u$ must not modify the system state variables $x$, but they may modify other auxiliaries, including \emph{auxiliary continuous state} variables $y$ used to model timers or integral terms used in controllers, see~\rref{subsec:acc}.
This control-plant loop is a typical structure for hybrid systems modeled in \dL~\cite{DBLP:books/daglib/0025392,Platzer18}, e.g., the controller $\alpha_u$ below models the discrete switching logic present in hybrid automata~\cite{DBLP:books/daglib/0025392,DBLP:conf/lics/Henzinger96,DBLP:books/sp/necs2005/Branicky05} (without jumps in the system state):
\begin{align}
\label{eqn:ctrller}
\begin{split}
\alpha_u &\mnodefequiv \bigcup_{p\in\sigfam}{\Big( \ptest{u=p}; \bigcup_{q \in \sigfam}{ \big( \ptest{G_{p,q}}; R_{p,q}; \pumod{u}{q} \big)} \Big)}\\
R_{p,q} &\mnodefequiv \pumod{y_1}{e_1};\pumod{y_2}{e_2};\dots;\pumod{y_k}{e_k}
\end{split}
\end{align}

For each mode $p \in \sigfam$, the switching controller may nondeterministically switch to mode $q \in \sigfam$ if the \emph{guard} formula $G_{p,q}$ is true in the current state ($G_{p,p} \mnodefequiv \ltrue$ for self-transitions); %
if the transition is taken, the \emph{reset map} $R_{p,q}$ sets the values of auxiliary state variables $y_1,\dots,y_k$ respectively to the value of terms $e_1,\dots,e_k$.

Stability analysis for controlled switching proceeds by identifying suitable loop invariants $\invariant$ for $\ctrlswitch$.
A powerful proof technique applied here is \emph{compositional reasoning}~\cite{DBLP:books/daglib/0025392,Platzer18} which separately analyses the discrete ($\alpha_i, \alpha_u$) and continuous ($\alpha_p$) dynamics, and then lifts those results to the full hybrid dynamics.
This idea is exemplified by the following derived variation of the~\irref{loop} rule:
\[
\dinferenceRule[loopm|loopT]{}
{\linferenceRule
  {\lsequent{\Gamma}{\dbox{\alpha_i}{\,\invariant}} & \lsequent{\invariant}{\dbox{\alpha_u}{\,\invariant}} & \lsequent{\invariant}{\dbox{\alpha_p}{\,\invariant}} & \lsequent{\invariant}{\fvarA} }
  {\lsequent{\Gamma}{\dbox{\alpha_i;\prepeat{(\alpha_u;\alpha_p)}}{\fvarA}}}
}{}
\]

The premises of rule~\irref{loopm} say that system initialization $\alpha_i$ puts the system in a state satisfying the invariant $\invariant$, and that $\invariant$ is compositionally preserved by \emph{both} the discrete switching logic $\alpha_u$ and the continuous dynamics $\alpha_p$.
This rule is applied to analyze stability for two important special instances of $\ctrlswitch$ next.

\subsubsection{Guarded State-dependent Switching}
The instance $\guardswitch$ corresponds to the automata controller from (\rref{eqn:ctrller}) with $\alpha_i \mnodefequiv \bigcup_{p \in \sigfam}{\pumod{u}{p}}$ and guard formulas $G_{p,q}$.
It does not use auxiliaries $y$ nor the reset map $R_{p,q}$.
This model adds \emph{hysteresis}~\cite{DBLP:journals/tac/JohanssonR98} to the state-dependent switching model from~\rref{subsec:statestab}, so that switching decisions at each $G_{p,q}$ depend explicitly on the current discrete mode $u$ in addition to the continuous state.
This design change is reflected in the loop invariants and in the corresponding proof rule below.

\paragraph{Stability}
The stability loop invariant is modified (cf.~\rref{subsec:statestab}) to case split on the possible discrete modes $u=p$ rather than the ODE domains:
$\invariants \mnodefequiv \norm{x} < \varepsilon \land \lorfold_{p \in \sigfam}\big( u=p \land \lterm_p < \ltermUpper\big)$.

\paragraph{Pre-attractivity} The pre-attractivity loop invariant is modified similarly:
$\invarianta \mnodefequiv \lorfold_{p \in \sigfam}\big( u{=}p \land \lterm_p {<} \ltermUpper \land (\lterm_p \geq \ltermLower \limply \lterm_p < \ltermUpper + k \tvar) \big)$.

\begin{corollary}[UGpAS for guarded state-dependent switching, MLF]
\label{cor:ugpasgstateswitchmlf}
The following proof rule for multiple Lyapunov functions $\lterm_p, p \in \sigfam$ with four stacked premises is syntactically derivable in \dL.

\noindent
\begin{calculuscollection}
\begin{calculus}
\dinferenceRule[UGpASgstmlf|MLF${_{G}}$]{}
{\linferenceRule
  {     \begin{array}{l}
    \lsequent{}{\lterm_p(0)=0 \land \lforall{x}{(\norm{x} > 0 \limply \lterm_p(x) > 0)}} \\
    \lsequent{}{\lforall{b}{\lexists{\gamma}{\lforall{x}{(\lterm_p(x) \leq b \limply \norm{x} \leq \gamma)}}}} \\
    \lsequent{}{\lie[]{f_p}{\lterm_p}(0) {=} 0 \land \lforall{x}{(\norm{x} {>} 0 \land \closure{\ivr_p} \limply \lie[]{f_p}{\lterm_p}(x) {<} 0)}}\\
    \lsequent{}{\landfold_{q \in \sigfam}{\Big(G_{p,q} \limply \lterm_q \leq \lterm_p\Big)}}
    \end{array} }
  {\lsequent{}{\astabhp{\guardswitch}}}
}{}
\end{calculus}
\end{calculuscollection}%
\end{corollary}

The premises of rule~\irref{UGpASgstmlf} are identical to those from~\irref{UGpASstmlf} except the bottom premise, which derives from~\irref{loopm} and unfolding the controller $\alpha_u$ with \dL's hybrid program axioms, e.g., the following proof skeleton shows the unfolding for the stability loop invariant $\invariants$ corresponding to a switch from mode $p$ to mode $q$:

\noindent\begin{minipage}[c]{0.01\textwidth}%
~\\~\\~\\
$\underset{\substack{\scalebox{1}{\textbf{Unfold}\hidewidth}\mathstrut}}{\qquad\Big\uparrow}$
\end{minipage}
\begin{minipage}[c]{0.45\textwidth}
{\begin{sequentdeduction}[array]
  \linfer[]{
    \linfer[]{
    \linfer[]{
      \lsequent{}{ G_{p,q} \limply \lterm_q \leq \lterm_p}
    }
      {\lsequent{\lterm_p < \ltermUpper}{ G_{p,q} \limply \lterm_q < \ltermUpper}}
    }
    {\lsequent{u=p \land \lterm_p < \ltermUpper}{\dbox{\ptest{G_{p,q}} ; \pumod{u}{q}}{(u=q \land \lterm_q < \ltermUpper)}}}
    }
  {\lsequent{\invariants}{\dbox{\alpha_u}{\invariants}}}
\end{sequentdeduction}}%
\end{minipage}
\begin{minipage}[c]{0.01\textwidth}%
$\overset{\hidewidth\substack{\scalebox{1}{\textbf{Arithmetic}}\mathstrut}}{\big\uparrow}$
~\\
~\\
~\\
\end{minipage}

\noindent
Unlike rule~\irref{UGpASstmlf}, the bottom premise of rule~\irref{UGpASgstmlf} only uses an inequality, because the guards $G_{p,q}$ determine permissible switching.

\subsubsection{Time-dependent Switching}
The instance $\timeswitch$ shown below models \emph{time-dependent switching}, where the controller $\alpha_u$ makes switching decisions based on the time $\tau$ elapsed in each mode.
\begin{align*}
\timeswitch \mnodefequiv \left\{
\begin{aligned}
\alpha_i &\mnodefequiv \pumod{\tau}{0};\bigcup_{p \in \sigfam}{\pumod{u}{p}} \\
\alpha_u &\mnodefequiv \bigcup_{p\in\sigfam}{\Big( \ptest{u=p}; \bigcup_{q \in \sigfam}{\big(\ptest{\theta_{p,q} \leq \tau}; \pumod{\tau}{0} ; \pumod{u}{q}\big)} \Big)} \\
\alpha_p &\mnodefequiv \bigcup_{p \in \sigfam}{\big( \ptest{u = p} ; \pevolvein{\D{x}=f_p{(x)},\D{\tau}=1}{\tau \leq \Theta_p} \big)}
\end{aligned}\right.
\end{align*}

The controller $\alpha_u$ enables switching from mode $p$ to $q$ when a \emph{minimum} dwell time $0 \leq \theta_{p,q} \leq \tau$ has elapsed and resets the timer whenever such a switch occurs.
Conversely, the plant $\alpha_p$ restricts modes with a \emph{maximum} dwell time $\tau \leq \Theta_p, \Theta_p > 0$; an unbounded dwell time $\Theta_p = \infty$ is represented by the domain constraint $\ltrue$.
Dwell time restrictions can be used to stabilize systems that switch between stable \emph{and unstable} modes \cite{DBLP:journals/ijsysc/ZhaiHYM01}.
Intuitively, the system should stay in stable modes for sufficient duration ($\theta_{p,q} \leq \tau$) while it should avoid staying in unstable modes for too long ($\tau \leq \Theta_p$).

To reason about stability for $\timeswitch$, consider Lyapunov function conditions $\lie[]{f_p}{\lterm_p}(x) \leq -\lambda_p \lterm_p$, where $\lambda_p$ is a constant associated with each mode $p \in \sigfam$.
This condition bounds the value of $\lterm_p$ along the solution of $\D{x}=f_p(x)$ by either a decaying exponential for stable modes ($\lambda_p > 0$) or a growing exponential for unstable modes ($\lambda_p \leq 0$).
Let $ \sigfams \mnodefeq \{ p \in \sigfam , \lambda_p > 0\}$ and $\sigfamu \mnodefeq \{p \in \sigfam, \lambda_p \leq 0\}$ be the indexes of the stable and unstable modes in the loop invariants below, and let $\expo{(\cdot)}$ denote the real exponential function, which is definable in \dL by differential axiomatization~\cite{DBLP:journals/jacm/PlatzerT20,DBLP:books/daglib/0025392}.

\paragraph{Stability} The stability loop invariant expresses the required exponential bounds with a case split depending if $p \in \sigfams$ or $p \in \sigfamu$:
\begin{align*}
\invariants &\mnodefequiv \tau \geq 0 \land \norm{x} < \varepsilon \, \land \\
&\quad\left(
\begin{aligned}
&\lorfold_{p \in \sigfams}\big( u=p \land \lterm_p < \ltermUpper\expo{-\lambda_p \tau} \big) \lor\\
&\lorfold_{p \in \sigfamu}\big( u=p \land \lterm_p < \ltermUpper\expo{-\lambda_p(\tau-\Theta_p)} \land \tau \leq \Theta_p \big)
\end{aligned}\right)
\end{align*}

For $p \in \sigfams$, $\expo{-\lambda_p \tau}$ is the accumulated decay factor for $\lterm_p$ after staying in the stable mode for time $\tau$.
For $p \in \sigfamu$, $\expo{-\lambda_p(\tau-\Theta_p)}$ is a buffer factor for the growth of $\lterm_p$ in the unstable mode so that $\lterm_p < W$ still holds at the maximum dwell time $\tau = \Theta_p$.
In both cases, the internal timer variable is non-negative ($\tau \geq 0$).

\paragraph{Pre-attractivity} The pre-attractivity loop invariant has similar exponential decay and growth bounds for each $p \in \sigfam$ in the current mode.
In addition, it has an overall exponential decay term $\expo{-\sigma(\tvar-\tau)}$ for some $\sigma > 0$, which ensures that the value of $\lterm_p$ tends to $0$ as $\tvar \to \infty$ for all switching trajectories; recall $\tvar$ is the global clock introduced in the specification of pre-attractivity in~\rref{lem:asymstabdl}.
\begin{align*}
\invarianta &\mnodefequiv \tau \geq 0 \land t \geq \tau \, \land \\
&\quad \left(
\begin{aligned}
&\lorfold_{p \in \sigfams}\big( u=p \land \lterm_p < \ltermUpper\expo{-\sigma(\tvar-\tau)}\expo{-\lambda_p \tau}   \big) \lor\\
&\lorfold_{p \in \sigfamu}\big( u=p \land \lterm_p < \ltermUpper\expo{-\sigma(\tvar-\tau)}\expo{-\lambda_p(\tau-\Theta_p)} \land \tau \leq \Theta_p\big)
\end{aligned}\right)
\end{align*}

Intuitively, $\expo{-\sigma(\tvar-\tau)}$ is the accumulated \emph{overall} decay factor for $\lterm_p$ \emph{until} the switch to mode $p$ which occurred at time $\tvar-\tau$, while $\expo{-\lambda_p \tau}$ (resp. $\expo{-\lambda_p(\tau-\Theta_p)}$) is the \emph{current} decay (resp. growth) factor \emph{since} the switch to mode $p$.

\begin{corollary}[UGpAS for time-dependent switching, MLF]
\label{cor:ugpastimeswitchmlf}
The following proof rule for multiple Lyapunov functions $\lterm_p, p \in \sigfam$ with five stacked premises is syntactically derivable in \dL.

\noindent
\begin{calculuscollection}
\begin{calculus}
\dinferenceRule[UGpAStimemlf|MLF${_{\tau}}$]{}
{\linferenceRule
  {     \begin{array}{l}
    \lsequent{}{\lterm_p(0)=0 \land \lforall{x}{(\norm{x} > 0 \limply \lterm_p(x) > 0)}} \\
    \lsequent{}{\lforall{b}{\lexists{\gamma}{\lforall{x}{(\lterm_p(x) \leq b \limply \norm{x} \leq \gamma)}}}} \\
    \lsequent{}{\lie[]{f_p}{\lterm_p} \leq -\lambda_p \lterm_p} \\
    \redc{\lsequent{\invariants}{\dbox{\alpha_u}{\invariants}} \qquad \lsequent{\invarianta}{\dbox{\alpha_u}{\invarianta}}}
    \end{array} }
  {\lsequent{}{\astabhp{\timeswitch}}}
}{}
\end{calculus}
\end{calculuscollection}%

The two \redc{red} premises on the bottom row are expanded to arithmetical conditions on $\lterm_p$ by unfolding the program structure of $\alpha_u$ with \dL axioms in\iflongversion%
~\rref{app:proofs}.
\else
~the supplement~\rref{app:}.
\fi
\end{corollary}

The bottom premises of~\irref{UGpAStimemlf} and~\irref{UGpASgstmlf} exemplify a key benefit of \dL stability reasoning: conditions on $\lterm_p$ that arise from $\invariants, \invarianta$ are derived by systematically unfolding the discrete dynamics of $\alpha_u$ with sound \dL axioms.
This enables automatic, \emph{correct-by-construction} derivation of those conditions, which is especially important for controlled switching because the number of possible transitions scales quadratically $|\sigfam|^2$ with the number of modes $|\sigfam|$.

\section{\KeYmaeraX Implementation}
\label{sec:impl}
This section presents a prototype implementation of switched systems support in the \KeYmaeraX prover based on \dL~\cite{DBLP:conf/cade/FultonMQVP15}.
The implementation consists of ${\approx}2700$ lines and, crucially, does not require any extension to \KeYmaeraX's existing soundness-critical core.
Accordingly, verification results for switched systems obtained through this implementation directly inherit the strong correctness properties guaranteed by the design of \KeYmaeraX~\cite{DBLP:conf/cade/FultonMQVP15,DBLP:series/lncs/MitschP20}.

\subsection{Modeling and Proof Interface}
\label{subsec:interface}
\begin{figure}[htb]
  \centering
  \includegraphics[width=\columnwidth]{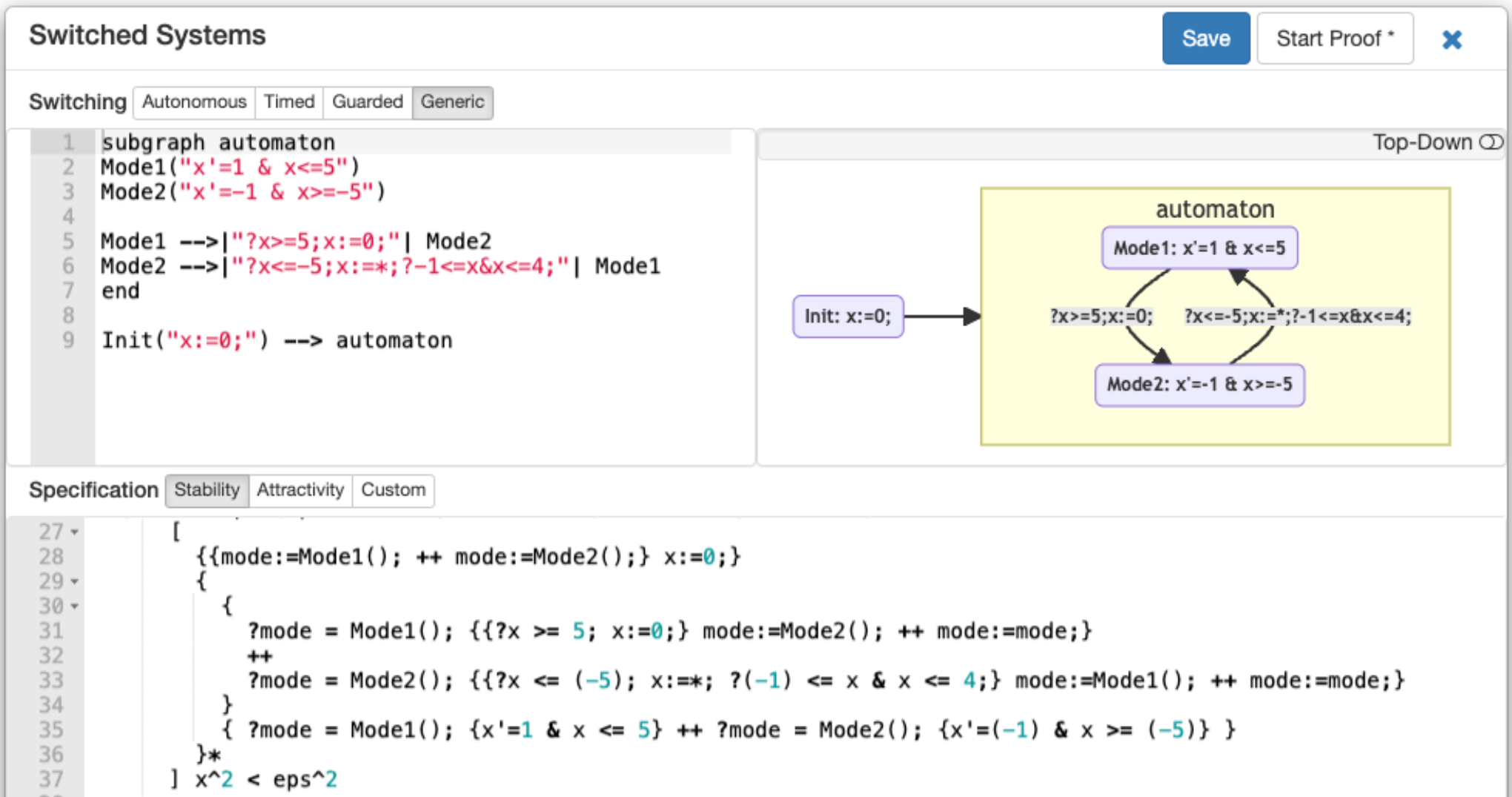}
  \Description{The \KeYmaeraX switched systems modeling editor provides a hybrid automata-style input language with graphical rendering. It automatically encodes hybrid automata as hybrid programs and specifications in differential dynamic logic. The generated programs and specifications can be tweaked manually if necessary.}
  \caption{Screenshot of the \KeYmaeraX switched systems modeling editor: automata input on top-left, rendered automaton top-right, generated hybrid program and specification(s) in \dL at the bottom}
  \label{fig:modeling}
\end{figure}

The implementation builds on \KeYmaeraX's proof IDE~\cite{DBLP:conf/fide/MitschP16} to provide a convenient interface for modeling switching mechanisms, as shown in~\rref{fig:modeling}.
The interface allows users to express switching mechanisms intuitively by rendering automaton plots while abstracting away the underlying hybrid programs.
It provides templates for switched systems following the switching mechanisms of \rref{sec:loopinv}: state-dependent, guarded, timed, and general controlled switching (tabs ``Autonomous'', ``Guarded'', ``Timed'', ``Generic'' in \rref{fig:modeling}).
From these templates, \KeYmaeraX automatically generates programs and stability specifications, ensuring that they have the correct \dL hybrid program and formula structure.

Switched systems are represented internally with a common interface \texttt{SwitchedSystem} which is currently implemented by four classes: \texttt{StateDependent} $\stateswitch$, \texttt{Guarded} $\guardswitch$, \texttt{Timed} $\timeswitch$, and \texttt{Controlled} $\ctrlswitch$.
The \texttt{SwitchedSystem} interface provides default stability and pre-attractivity specifications, which can be adapted by users on the UI if needed.
Corollaries~\ref{cor:ugpasstateswitchclf}--\ref{cor:ugpastimeswitchmlf} are implemented as UGpAS \emph{proof tactics} in \KeYmaeraX's Bellerophon tactic language~\cite{DBLP:conf/itp/FultonMBP17}.
These tactics automate all of the reasoning steps underlying stability proofs for their respective switching mechanisms, so that users only need to input candidate Lyapunov functions for \KeYmaeraX to (attempt to) complete their proofs.
Additionally, when candidates are not provided by the user, the implementation uses sum-of-squares programming~\cite{1243743,sostools} to automatically generate candidate Lyapunov functions for a subset of switching designs.
The generated candidates are checked for correctness by \KeYmaeraX so the generator does not need to be trusted for correctness of the resulting proofs.
\rref{tab:tactics} summarizes the available proof tactics and Lyapunov function generation for classes of switching mechanisms.

\begin{table}[]
\caption{Available tactics in \KeYmaeraX for switched systems stability proofs and Lyapunov function generation.}
\label{tab:tactics}
\begin{tabular}{@{}lcccc@{}}
\toprule
\multirow{2}{*}{\texttt{SwitchedSystem}} & \multicolumn{2}{c}{Common Lyap.} & \multicolumn{2}{c}{Multiple Lyap.} \\
                                         & Proof       & Gen.      & Proof       & Gen. \\ \midrule
\texttt{StateDependent} $\stateswitch$   & \checkmark  & \checkmark   & \checkmark  & \checkmark  \\
\texttt{Guarded} $\guardswitch$     & \checkmark  & \checkmark   & \checkmark  & \checkmark  \\
\texttt{Timed} $\timeswitch$             & \checkmark  & \checkmark   & \checkmark  & --- \\
\texttt{Controlled} $\ctrlswitch$        & \checkmark  & \checkmark   & ---         & --- \\ \bottomrule
\end{tabular}
\end{table}

\subsection{Examples}
\label{subsec:examples}
The implementation is tested on a suite of examples drawn from the literature~\cite{DBLP:journals/tac/Branicky98,DBLP:journals/tac/JohanssonR98,1243743,SunGe} featuring various switching mechanisms, with results summarized in~\rref{tab:examples}.
These examples have a $2$ dimensional state space and switch between $2$ modes except Example 6 ($3$ dimensions, $2$ modes) and Example 4 ($2$ dimensions, $4$ modes).

\begin{table}[]
\caption{Stability proofs for examples drawn from the literature. The ``Time'' columns indicate time (in seconds) to run the \KeYmaeraX proofs, $\times$ indicates incomplete proof. A \checkmark~in the ``Gen.'' column indicates successful Lyapunov function(s) generation, ? indicates that a candidate was generated but with numerical issues, and --- indicates inapplicability.
In the latter two cases (?, ---) known Lyapunov functions from the literature were used for the proofs (if available).}
\label{tab:examples}
\begin{tabular}{@{}llrrc@{}}
\toprule
Example & Model & Time (Stab.) & Time (Attr.) & Gen. \\ \midrule
1 \cite[Ex. 2.1]{DBLP:journals/tac/Branicky98}       & $\stateswitch$ & 2.6  & 3.0      & \checkmark \\ %
2 \cite[Motiv. ex.]{DBLP:journals/tac/JohanssonR98}  & $\stateswitch$ & 2.2  & 2.3      & \checkmark \\
3 \cite[Ex. 1]{DBLP:journals/tac/JohanssonR98}       & $\stateswitch$ & 3.3  & 4.1      & \checkmark \\
4 \cite[Ex. 2 \& 3]{DBLP:journals/tac/JohanssonR98}  & $\guardswitch$ & 2.8  & 3.8      & ?          \\
5 \cite[Ex. 6]{1243743}                              & $\guardswitch$ & $\times$    & $\times$        & ?          \\
6 \cite[Ex. 2.45]{SunGe}                             & $\arbswitch$   & 19.4 & 11.1     & \checkmark \\
7 \cite[Ex. 3.25]{SunGe}                             & $\stateswitch$ & 2.4  & 2.9      & \checkmark \\
8 \cite[Ex. 3.49]{SunGe}                             & $\timeswitch$  & 4.4  & 5.6      & ---        \\
9 \cite[Ex. 1]{DBLP:journals/ijsysc/ZhaiHYM01}       & $\timeswitch$  & 4.7  & 5.3      & ---        \\
10 \cite[Ex. 2]{DBLP:journals/ijsysc/ZhaiHYM01}      & $\timeswitch$  & 256.9& $\times$ & ---        \\\bottomrule
\end{tabular}
\end{table}

The proof tactics successfully prove most of the examples across various switching mechanisms.
For Example 5, a suitable Lyapunov function (without numerical errors) could not be found.
For the time-dependent switching models (Examples 8--10), \KeYmaeraX internally uses verified polynomial Taylor approximations to the exponential function for decidability of arithmetic~\cite{Bochnak1998,Tarski}; Example 10 needs a high degree approximation ($15$ terms in the polynomial) for sufficient accuracy and its attractivity proof could not be completed in reasonable time.
\section{Case Studies}
\label{sec:casestudies}
This section presents three case studies applying the deductive verification approach to justify various non-standard stability arguments in \KeYmaeraX.\footnote{See \url{https://github.com/LS-Lab/KeYmaeraX-projects/blob/master/stability/UGpAS}}

\subsection{Canonical Max System}

Branicky~\cite{735143} investigates the longitudinal dynamics of an aircraft with an elevator controller that mediates between two control objectives: \begin{enumerate*}[label=\roman*),font=\itshape] \item tracking potentially unsafe pilot input and \item respecting safety constraints on the aircraft's angle of attack\end{enumerate*}.
Assuming a state feedback control law, the model is transformed to the following \emph{canonical max system}~\cite[Remark 5]{735143}, with state variables $x,y$ and parameters $a,b,f,g,\gamma$ satisfying $a,b,a-f,b-g > 0$ and $\gamma \leq 0$.
\begin{equation}
\D{x} = y, \D{y}=-ax-by+\max(fx+gy+\gamma,0)
\label{eq:canonmax}
\end{equation}

The right-hand side of system~\rref{eq:canonmax} is non-differentiable but the equations can be equivalently rewritten as a family of two ODEs corresponding to either possibility for the $\max(fx+gy+\gamma,0)$ term in the equation for $\D{y}$ as follows, where the system follows ODE $\circled{A}$ in domain $fx+gy+\gamma\leq0$ and ODE $\circled{B}$ in domain $fx+gy+\gamma \geq 0$.
\begin{align*}
\circled{A} &\mnodefequiv\pevolve{\D{x}=y,\D{y}=-ax-by} \\
\circled{B} &\mnodefequiv\pevolve{\D{x}=y,\D{y}=-(a-f)x-(b-g)y+\gamma}
\end{align*}

Stability of this parametric system is \emph{not} directly provable using standard techniques for state-dependent switching presented in~\rref{subsec:statestab}.
For example, the ODE \circled{A} stabilizes the system to the origin but the ODE \circled{B} stabilizes to the point $(-\frac{\gamma}{a-f},0)$, away from the origin for $\gamma < 0$.
Branicky proves global asymptotic stability of~\rref{eq:canonmax} with the following ``noncustomary''~\cite{871309} Lyapunov function involving a nondifferentiable integrand:
\begin{align}
\lterm = \frac{1}{2}y^2 + \int_{0}^{x} a\xi - \max(f\xi+\gamma,0) d\xi
\label{eq:branicky}
\end{align}

The key idea used to deductively prove stability here instead is \emph{ghost switching}: analogous to ghost variables in program verification which are added for the sake of program proofs~\cite{DBLP:journals/cacm/OwickiG76,Platzer18,DBLP:journals/jacm/PlatzerT20}, ghost switching modes do not change the physical dynamics of the system but are introduced for the purposes of the stability analysis.
Here, ghost switching between $fx+\gamma\leq 0$ and $f x+\gamma \geq0$ is used to obtain closed form representations for the integral in~\rref{eq:branicky}.
This yields an instance of state-dependent switching $\stateswitch$ with 4 switching modes and the corresponding stability specification $\exmaxfml$:
\begingroup
\allowdisplaybreaks
\begin{align*}
\exmax &\mnodefequiv \Big( \pchoice{\pchoice{\circled{A}_1}{\circled{A}_2}}{\pchoice{\circled{B}_1}{\circled{B}_2}} \Big)^* \quad \ptermA\mnodefequiv fx+gy+\gamma \quad \ptermB \mnodefequiv fx+\gamma \\
\circled{A}_1&\mnodefequiv \pevolvein{\circled{A}}{\ptermA\leq0 \land \ptermB \leq 0}\qquad \circled{A}_2\mnodefequiv \pevolvein{\circled{A}}{\ptermA\leq0 \land \ptermB \geq 0}\\
\circled{B}_1&\mnodefequiv \pevolvein{\circled{B}}{\ptermA\geq0 \land \ptermB \leq 0}\qquad\, \circled{B}_2\mnodefequiv \pevolvein{\circled{B}}{\ptermA\geq0 \land \ptermB \geq 0}\\
\exmaxfml &\mnodefequiv a {>} 0 {\land} b {>} 0 {\land} a{-}f {>} 0 {\land} b{-}g {>} 0 {\land} f {\neq} 0 {\land} \gamma {\leq} 0 \limply \astabhp{\exmax}
\end{align*}
\endgroup

The ghost switching modes enable a multiple Lyapunov function argument for stability using the following modified closed-form representations of Branicky's Lyapunov function~\rref{eq:branicky}, with $\lterm_1 \mnodefeq \frac{1}{2}(b c x^2 + 2c x y + y^2) + \redc{\frac{a}{2}x^2}$ for $\circled{A}_1, \circled{B}_1$ and $\lterm_2 \mnodefeq \frac{1}{2}(b c x^2 + 2c x y + y^2) + \redc{\frac{a}{2}x^2 - \frac{(fx + \gamma)^2}{2f}}$ for $\circled{A}_2, \circled{B}_2$.\footnote{An important technical requirement for $\lterm_2$ to be well-defined is $f \neq 0$.
The case with $f=0$ is also verified in \KeYmaeraX but the details are omitted here for brevity. It does not require ghost switching and uses only $\lterm_1$ as its common Lyapunov function.}
The sub-terms highlighted in \redc{red} for $\lterm_1, \lterm_2$ are closed form expressions for $\int_{0}^{x} a\xi - \max(f\xi+\gamma,0) d\xi$ where $f\xi+\gamma \leq 0$ and $f\xi+\gamma \geq 0$ respectively.
The Lyapunov functions $\lterm_1, \lterm_2$ are modified from~\rref{eq:branicky} to use a quadratic form with an additional constant $c$ satisfying constraints $0 < c < b, c < b-g, c < \frac{(a - f) (b - g)}{a - f + g^2}, c < \frac{a (b - g)}{a + g^2}$ (such a constant always exists under the assumptions on $a,b,f,g$).
This technical modification is required to prove UGpAS for $\exmax$ directly with the Lyapunov functions.
Branicky's earlier proof requires LaSalle's principle~\cite{735143}.

Another challenging aspect of this case study is verification of the \emph{parametric} arithmetical conditions for $\lterm_1, \lterm_2$, i.e., stability is verified for \emph{all} possible parameter values $a,b,f,g,\gamma$ that satisfy the assumptions in $\exmaxfml$.
Such questions are decidable in theory~\cite{Bochnak1998,Tarski}, but are difficult for automated solvers in practice (even out of reach of solvers that require numerically bounded parameters~\cite{DBLP:conf/cade/GaoKC13}).
\KeYmaeraX enables a user-aided proof of the required arithmetic conditions.
For example, the Lie derivative of the Lyapunov function $\lterm_1$ for $\circled{B}_1$ is given by $\D{\lterm_1}= -(b-c)y^2-acx^2+(cx+y)(fx+gy+\gamma)$, where $\D{\lterm_1}$ is required to be strictly negative away from the origin for stability. %
The arithmetical argument is as follows: if $c x + y \leq 0$, then by constraint $f x + g y + \gamma \geq 0$, $\D{\lterm_1}$ satisfies $\D{\lterm_1} \leq -(b-c)y^2-acx^2$.
Otherwise, $c x + y > 0$, then by constraint $f x + \gamma \leq 0$, $\D{\lterm_1}$ satisfies $\D{\lterm_1} \leq -(b-g-c)y^2-acx^2+gcxy$.
In either case, the RHS bound is a negative definite quadratic form by the earlier choice of parameter $c$ and therefore, $\D{\lterm_1}$ is negative away from the origin.

\subsection{Automated Cruise Control}
\label{subsec:acc}
\begin{figure*}
\begin{minipage}[t]{.45\textwidth}
\begin{Verbatim}[frame=single,fontsize=\small]
normalPI("v' = -0.001*x-0.052*v, x' = v, t' = 0
        & -15 <= v & v <= 15 & -500 <= x & x <= 500")
normalPI -->|"?(13 <= v & v <= 15 &
            -500 <= x & x <= 500); t := 0;"| sbrakeact
normalPI -->|"?(-15 <= v & v <= -14 &
            -500 <= x & x <= 500);"| accelerate
... // Other modes
\end{Verbatim}
\centering\includegraphics[width=0.8\textwidth,clip,trim=0 0 0 0]{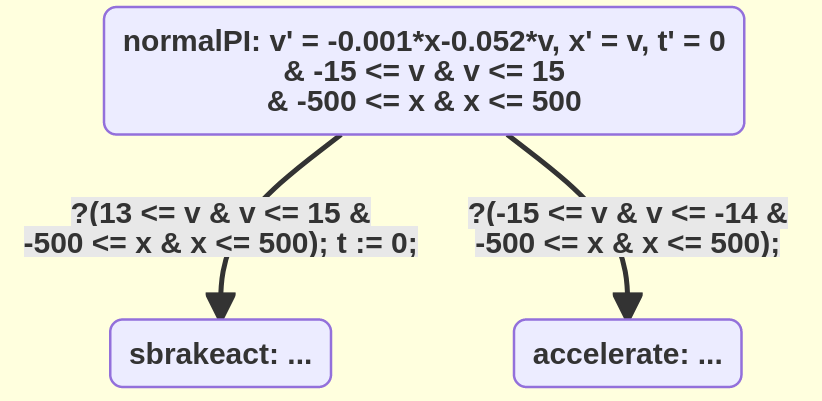}
\end{minipage}\qquad
\begin{minipage}[t]{.50\textwidth}
\begin{Verbatim}[frame=single,fontsize=\small]
\forall eps ( eps > 0 -> // Abridged stability specification
  ...
  [ ... // Initialize
  {  { ... ++ // Transitions for other modes
       ?mode = normalPI();
       { {?13 <= v & v <= 15 & -500 <= x & x <= 500; t := 0;}
         mode := sbrakeact(); ++
         ?-15 <= v & v <= -14 & -500 <= x & x <= 500;
         mode := accelerate(); ++
         mode := mode; } }
     { ...  ++ // Plant ODEs for other modes
       ?mode = normalPI();
       { v' = -0.001*x-0.052*v, x' = v, t' = 0 &
         -15 <= v & v <= 15 & -500 <= x & x <= 500 } }
  }*  // Switching loop
  ] v^2 < eps^2
\end{Verbatim}
\end{minipage}

\caption{Snippets of an automated cruise controller~\cite{DBLP:phd/dnb/Oehlerking11} modeled as a (switching) hybrid automaton. Users express the automaton within the description language (top left) and \KeYmaeraX visualizes the automaton on-the-fly (bottom left).
The implementation automatically generates the appropriate hybrid program representation and UGpAS specification (right); \texttt{++,\&,()} denote choice, conjunction, and constants in \KeYmaeraX's ASCII syntax respectively.}
\label{fig:automaton}
\end{figure*}

Oehlerking~\cite[Sect.\,4.6]{DBLP:phd/dnb/Oehlerking11} verifies the stability of an automatic cruise controller modeled as a hybrid automaton with 6 operating modes and 11 transitions between them: normal proportional-integral (PI) control, acceleration, service braking (2 modes), and emergency braking (2 modes).
Figure~\ref{fig:automaton} shows an abridged version of the corresponding \KeYmaeraX model (using $\ctrlswitch$) with the PI control mode, where $v$ is the relative velocity to be controlled to $v=0$ and $x,t$ are auxiliary integral and timer variables used in the controller. %
Briefly, this controller is designed to use the PI controller near $v=0$ for stability, while its other control modes drive the system toward $v=0$ by accelerating or braking.

Lyapunov function candidates for this model can be successfully generated using the Stabhyli~\cite{DBLP:conf/hybrid/MohlmannT13} stability tool for hybrid automata.
However, Stabhyli (with default configurations) outputs a Lyapunov function candidate for the PI control mode that is numerically unsound,
\iflongversion%
see~\rref{app:casestudies} for the output and a counterexample;
\else
see the supplement~\rref{app:casestudies};
\fi
this is a known issue with Stabhyli for control modes at the origin~\cite{DBLP:conf/hybrid/MohlmannT13}.
For this case study, the issue is manually resolved by truncating terms with very small magnitude coefficients in the generated output and then checking in \KeYmaeraX that the arithmetical conditions for the PI mode are satisfied \emph{exactly} for the truncated candidate.

Further insights from the controller design are used in the UGpAS proof in \KeYmaeraX.
Briefly, stability only concerns states and modes that are active near the origin so the stability argument and loop invariant only need to mention a single Lyapunov function for the PI control mode, while choosing $\delta$ (in~\rref{def:ugpas}) sufficiently small so that none of the other modes can be entered.\footnote{In fact, the PI controller equations are exactly those of a linearized pendulum, which has known Lyapunov functions~\cite{DBLP:conf/tacas/TanP21,MR1201326}. It could be interesting to modify Stabhyli to accept user-provided Lyapunov function hints for certain modes.}
Similarly, pre-attractivity only requires reasoning about \emph{asymptotic} convergence to the origin for the PI control mode so it suffices to show that the system leaves all other modes in finite time.

\subsection{Brockett's Nonholonomic Integrator}

Verification of stabilizing control laws for Brockett's nonholonomic integrator~\cite{Brockett83asymptoticstability} is of significant interest because stability for a large class of models can be reduced to that of the integrator via coordinate transformations, e.g., Liberzon~\cite{DBLP:books/sp/Liberzon03} transforms a unicycle model to the integrator and provides a stabilizing switching control law corresponding to parking of the unicycle.
The nonholonomic integrator is described by the system of differential equations $\D{x} = u, \D{y}= v, \D{z} = xv-yu$, with state variables $x,y,z$ and state feedback control inputs $u=u(x,y,z), v=v(x,y,z)$ (to be determined below).
Notably, this is a classical example of a system that is not stabilizable by purely continuous feedback control.
Intuitively, no choice of controls $u,v$ can produce motion along the $z$-axis ($x=y=0$).
Thus, to stabilize the system to the origin, the controller must first drive the system away from the $z$-axis before switching to a control law that stabilizes the system from states away from the $z$-axis.
This intuition can be realized using two different switching strategies that are analogous to the event-triggered and time-triggered CPS design paradigms respectively~\cite{Platzer18}.

\subsubsection{Event-triggered Controller}
Bloch and Drakunov~\cite{BLOCH199691} use the switching controller $u=-x+ay \sign(z), v=-y-ax \sign(z)$ to asymptotically stabilize the integrator in the region $\frac{a}{2}(x^2+y^2) \geq \abs{z}$ for any given constant $a > 0$.
This controller first drives the system towards the plane $z=0$ and, once it reaches the plane, \emph{slides} along the plane towards the origin.
The closed-loop system is modeled as an instance of state-dependent switching $\stateswitch$ with $3$ modes depending on the sign of $z$ and specification $\exholoeventfml$:
\begin{align*}
\circled{A}&\mnodefequiv \pevolvein{\D{x} = -x+ay, \D{y}= -y-ax, \D{z} = -a(x^2+y^2)}{z \geq 0} \\
\circled{B}&\mnodefequiv \pevolvein{\D{x} = -x-ay, \D{y}= -y+ax, \D{z} = a(x^2+y^2)}{z \leq 0}\\
\circled{C}&\mnodefequiv \pevolvein{\D{x} = -x   , \D{y}= -y   , \D{z} = 0}{z = 0} \quad \exholoevent\mnodefequiv \prepeat{\Big(\pchoice{\pchoice{\circled{A}}{\circled{B}}}{\circled{C}}\Big)} \\
\exholoeventfml &\mnodefequiv a> 0 \limply \stabhp{\alpha} \land \\
& \lforall{\delta{>}0}{\lforall{\varepsilon{>}0}{\lexists{T{\geq}0}{\lforall{x,y,z}{\Big(\norm{x,y,z} < \delta  \land \redc{\frac{a}{2}(x^2+y^2) \geq \abs{z}} \limply}}}} \\
&\quad\qquad\qquad\qquad\qquad \dbox{\pumod{\tvar}{0};\taug{\exholoevent}}{(t \geq T \limply \norm{x,y,z} < \varepsilon\Big)}
\end{align*}

The specification $\exholoeventfml$ is identical to UGpAS except it restricts pre-attractivity to the applicable region $\redc{\frac{a}{2}(x^2+y^2) \geq \abs{z}}$ for the controller.\footnote{The applicable region is equivalently characterized by the real arithmetic formula $\redc{(z {\geq} 0 \limply \frac{a}{2}(x^2+y^2) {\geq} z) \land (z {\leq} 0 \limply \frac{a}{2}(x^2+y^2) {\geq} {-}z)}$, omitted for brevity.}
Its verification uses the squared norm $\lterm = x^2+y^2+z^2$ as a common Lyapunov function.
The key modification to the pre-attractivity proof, cf.~\rref{subsec:arbstatestab}, is to use (and verify) the fact that $\redc{\frac{a}{2} (x^2+y^2) \geq \abs{z}}$ is a loop invariant of $\exholoevent$.
This additional invariant corresponds to the fact that the controller keeps the system within its applicable region (if the system is initially within that region).

In fact, $\exholoevent$ can be extended to a globally stabilizing controller, as modeled by $\exholoeventfull$ below ($\keyword{if},\keyword{else}$ branching is supported as an abbreviation in \KeYmaeraX~\cite{Platzer18}):
\begin{align*}
\circled{D}&\mnodefequiv \pevolvein{\D{x}=u,\D{y}=v,\D{z}=xv-yu}{\frac{a}{2}(x^2+y^2)\leq \abs{z}} \\
\circled{E}&\mnodefequiv \pevolvein{\D{x}=u,\D{y}=v,\D{z}=xv-yu}{\frac{a}{2}(x^2+y^2)\geq \abs{z}} \\
\exholoeventfull&\mnodefequiv \Big(\keyword{if}\big(\redc{\frac{a}{2}(x^2+y^2) \geq \abs{z}}\big)~\big\{\pchoice{\pchoice{\circled{A}}{\circled{B}}}{\circled{C}}\big\} \\
&~\quad\;\;\keyword{else}~\Big\{\quad \keyword{if}((x-y)z \leq 0) \{ \pumod{u}{c}; \pumod{v}{c} \}\\
&\;\;\quad\qquad\qquad \keyword{else}\{ \pumod{u}{-c}; \pumod{v}{-c} \}; \\
&\;\;\quad\qquad\qquad \big\{ \pchoice{\circled{D}}{\circled{E}} \big\} \quad \Big\}\Big)^*
\end{align*}

If the system is in the applicable region (outer \keyword{if} branch), then the previous controller from $\exholoevent$ is used.
Otherwise, outside the applicable region (outer \keyword{else} branch), the system applies a constant control $c > 0$ chosen to drive the system into the applicable region.
The pair of ODEs \circled{D} and \circled{E} model an event-trigger in \dL~\cite{Platzer18}, where the switching controller is triggered to make its next decision when the system reaches the switching surface $\frac{a}{2}(x^2+y^2) = \abs{z}$.

The specification $\exholoeventfullfml \mnodefequiv a {>} 0 \land c {>} 0 \limply \astabhp{\exholoeventfull}$ is proved by modifying the loop invariants to account for an initial period where the system is outside the applicable region.
For example, the stability loop invariant
$\invariants \mnodefequiv (\lnot{\redc{\frac{a}{2}(x^2+y^2) {\geq} \abs{z}}} \limply |z| {<} \delta) \land (\redc{\frac{a}{2}(x^2+y^2) {\geq} \abs{z}} \limply \norm{x,y,z} {<} \varepsilon)$ expresses that the controller keeps $\abs{z}$ sufficiently small with $|z| {<} \delta$ to preserve stability outside the applicable region.
The pre-attractivity loop invariant is similarly split between the two cases, with an explicit time estimate on the time it takes for the system to enter the applicable region.

\subsubsection{Time-triggered Controller}

The time-triggered switching strategy~\cite{Platzer18}, modeled by $\exholotime$ below, is similar to that proposed by Liberzon~\cite[Section 4.2]{DBLP:books/sp/Liberzon03}.
If the system is on the $z$-axis and away from the origin \circled{A}, the controller sets an internal stopwatch $\tau$ and drives the system away from the axis for maximum duration $T_0 > 0$ with $u=z, v=z$.
Otherwise \circled{B}, the controller drives the system towards the origin along a parabolic curve of the form $\frac{a}{2}(x^2+y^2) = z$.
\begin{align*}
\exholotime&\mnodefequiv \Big(\keyword{if}(x=0\land y=0 \land z\not=0)~\big\{\\
\circled{A}&\quad\qquad \tau:=0;\pevolvein{x'=z,y'=z,z'=xz-yz}{\tau \leq T_0}\quad\big\}\\
&\quad\;\;\;\keyword{else}~\big\{~~~ a:=\frac{2z}{x^2+y^2};\\
\circled{B}&
\quad\qquad\qquad
\pevolve{x'=-x+ay,y'=-y-ax,z'=-a(x^2+y^2)}\quad\big\}\Big)^*
\end{align*}

The specification $\exholotimefml \mnodefequiv T_0 > 0 \limply \astabhp{\exholotime}$ is again proved by analyzing both cases of the controller in the loop invariants, e.g., with the pre-attractivity invariant $\invarianta$:
\begin{align*}
&\big(x=0 \land y=0 \land z\not=0 \limply \abs{z} < \delta \land \tvar=0\big)~\land\\
&\big(\lnot{(x=0 \land y=0 \land z\not=0)} \limply \\
&\qquad\qquad\norm{x,y,z}>\varepsilon \limply \norm{x,y,z}^2 < \delta^2(2T_0^2+1)-\varepsilon^2(t-T_0)\big)
\end{align*}

The top conjunct says the system may start transiently on the $z$-axis (away from $z=0$) at time $t=0$.
The bottom conjunct gives explicit bounds on $\norm{x,y,z}$, which, for sufficiently large $t \geq T$, implies that the system enters $\norm{x,y,z} < \varepsilon$ as required for pre-attractivity.
The transient term $\delta^2(2T_0^2+1)$ upper bounds the (squared) norm of the system state after starting on the $z$-axis in ball $\norm{x,y,z} < \delta$ and following mode \circled{A} for the maximum stopwatch duration $\tau = T_0$.

\section{Related Work}

\paragraph{Switched Systems} Comprehensive introductions to the analysis and design of switching control can be found in the literature~\cite{DBLP:books/sp/Liberzon03,SunGe,871309}.
An important design consideration (which this paper sidesteps, cf.~\rref{rem:ugpas}) is whether a given switched or hybrid system has complete solutions~\cite{4806347,10.2307/j.ctt7s02z,doi:10.1002/rnc.592,DBLP:journals/tac/LygerosJSZS03}.
Justification of such design considerations, and other stability notions of interest for switching designs, e.g., quadratic, region, or set-based stability~\cite{DBLP:books/sp/Liberzon03,SunGe,4806347,10.2307/j.ctt7s02z,DBLP:conf/hybrid/PodelskiW06}, can be done in \dL with appropriate formal specifications of the desired properties from the literature~\cite{DBLP:conf/tacas/TanP21,DBLP:conf/adhs/TanP21,Platzer18,DBLP:books/daglib/0025392}.
Another complementary question is how to design a switching control law that \emph{stabilizes} a given system.
Switching design approaches are often guided by underlying stability arguments~\cite{DBLP:conf/cdc/RavanbakhshS15,DBLP:books/sp/Liberzon03,SunGe}; the loop invariants from~\rref{sec:loopinv} are expected to help guide correct-by-construction synthesis of such controllers.

\paragraph{Stability Analysis and Verification} Corollaries~\ref{cor:ugpasstateswitchclf}--\ref{cor:ugpastimeswitchmlf} formalize various Lyapunov function-based stability arguments from the literature~\cite{DBLP:journals/ijsysc/ZhaiHYM01,DBLP:journals/tac/Branicky98} using loop invariants, yielding trustworthy, computer-checked stability proofs in \KeYmaeraX~\cite{DBLP:conf/cade/FultonMQVP15,DBLP:conf/itp/FultonMBP17}.
Other computer-aided approaches for switched system stability analysis are based on finding Lyapunov functions that satisfy the requisite arithmetical conditions~\cite{DBLP:journals/siamco/SheX14,DBLP:phd/dnb/Oehlerking11,DBLP:conf/hybrid/MohlmannT13,DBLP:conf/hybrid/KapinskiDSA14,1243743,DBLP:conf/nolcos/Sankaranarayanan0A13}.
Although the search for such functions can often be done efficiently with numerical techniques~\cite{1243743,DBLP:conf/hybrid/MohlmannT13,sostools}, various authors have emphasized the need to check that their outputs satisfy the arithmetical conditions \emph{exactly}, i.e., without numerical errors compromising the resulting stability claims~\cite{DBLP:conf/hybrid/KapinskiDSA14,DBLP:journals/fmsd/RouxVS18,DBLP:conf/tacas/AhmedPA20} (see, e.g.,~\rref{subsec:acc}).
This paper's deductive approach goes further as it comprehensively verifies \emph{all} steps of the stability argument down to its underlying discrete and continuous reasoning steps~\cite{Platzer18,DBLP:journals/jar/Platzer17}.
The generality of this approach is precisely what enables verification of various classes of switching mechanisms all within a common logical framework (\rref{sec:loopinv}) and verification of non-standard stability arguments (\rref{sec:casestudies}).
Alternative approaches to stability verification are based on abstraction~\cite{GARCIASOTO2020100856,DBLP:conf/hybrid/SotoP18} and model checking~\cite{DBLP:conf/hybrid/PodelskiW06}.

\section{Conclusion}

This paper shows how to deductively verify switched system stability, using \dL's nested quantification over hybrid programs to specify stability, and \dL's axiomatics to prove those specifications.
Loop invariants---a classical technique from verification---are used to succinctly capture the desired properties of a given switching design; through deductive proofs, these invariants yield systematic, correct-by-construction derivation of the requisite arithmetical conditions on Lyapunov functions for stability arguments in implementations.
An interesting direction for future work is to use other Lyapunov function generation techniques~\cite{DBLP:conf/hybrid/MohlmannT13,DBLP:phd/dnb/Oehlerking11,DBLP:journals/siamco/SheX14,DBLP:conf/hybrid/KapinskiDSA14}, which---thanks to the presented approach---do not have to be trusted since their results can be checked independently by \KeYmaeraX.
This would enable fully automated, yet sound and trustworthy verification of switched system stability based on \dL's parsimonious hybrid program reasoning principles.

\begin{acks}
We thank Thomas Baar and the anonymous reviewers for their helpful feedback on this paper.
This material is based upon work supported by the National Science Foundation under Grant No. CNS-1739629.
This research was sponsored by the AFOSR under grant number FA9550-16-1-0288.
\iflongversion
The views and conclusions contained in this document are those of the author and should not be interpreted as representing the official policies, either expressed or implied, of any sponsoring institution, the U.S. government or any other entity.
\fi
\end{acks}

\bibliographystyle{ACM-Reference-Format}
\iflongversion
\else
\fi
\bibliography{paper}

\iflongversion
\newpage
\appendix

\section{Proofs}
\label{app:proofs}
This appendix provides proofs for the results presented in the main paper.
Relevant background for \dL's semantics and axiomatics is given, expanding on the material in~\rref{sec:background}.
Full definitions are available in the literature~\citep{DBLP:journals/jar/Platzer17,Platzer18}.

A \dL state $\iget[state]{\I} : \allvars \to \reals$ assigns a real value to each variable in the set of all variables $\allvars$.
The set $\allvars$ consists of the continuously evolving state variables $x=(x_1,\dots,x_n)$ of a switched system model and additional variables $\allvars \setminus \{x\}$ used as program auxiliaries for those models.
Following Tan and Platzer~\cite{DBLP:conf/adhs/TanP21}, \dL states are projected on the state variables $x$ and the (projected) \dL states $\iget[state]{\I}$ are equivalently treated as points in $\reals^n$.
The semantics of program auxiliaries is as usual~\citep{Platzer18}.
The derivations use a classical sequent calculus with the usual propositional and first-order sequent rules, e.g.,~\irref{andr+cut} and~\irref{allr+existsr}~\citep{DBLP:journals/jar/Platzer17,Platzer18}.
Rule~\irref{qear} denotes the use of an underlying decision procedure for first-order real arithmetic \folr~\cite{Tarski}.
The axioms and proof rules of \dL used in the proofs are as follows.
\irlabel{qear|\usebox{\Rval}}
\irlabel{notr|$\lnot$\rightrule}
\irlabel{notl|$\lnot$\leftrule}
\irlabel{orr|$\lor$\rightrule}
\irlabel{orl|$\lor$\leftrule}
\irlabel{andr|$\land$\rightrule}
\irlabel{andl|$\land$\leftrule}
\irlabel{implyr|$\limply$\rightrule}
\irlabel{implyl|$\limply$\leftrule}
\irlabel{equivr|$\lbisubjunct$\rightrule}
\irlabel{equivl|$\lbisubjunct$\leftrule}
\irlabel{id|id}
\irlabel{cut|cut}
\irlabel{weakenr|W\rightrule}
\irlabel{weakenl|W\leftrule}
\irlabel{existsr|$\exists$\rightrule}
\irlabel{existsrinst|$\exists$\rightrule}
\irlabel{alll|$\forall$\leftrule}
\irlabel{alllinst|$\forall$\leftrule}
\irlabel{allr|$\forall$\rightrule}
\irlabel{existsl|$\exists$\leftrule}
\irlabel{iallr|i$\forall$}
\irlabel{iexistsr|i$\exists$}

\noindent
\begin{calculuscollection}
\begin{calculus}
\cinferenceRule[assignb|$\dibox{:=}$]{assignment / substitution axiom}
      {\linferenceRule[equiv]
        {\rfvar(e)}
        {\dbox{\pupdate{\umod{x}{\etermA}}}{\rfvar(x)}}
      }
      {\text{$\etermA$ free for $x$ in $\rfvar$}}
\end{calculus} \\
\begin{calculus}
\cinferenceRule[testb|$\dibox{?}$]{test}
{\linferenceRule[equiv]
  {(\ivr \limply \rfvar)}
  {\dbox{\ptest{\ivr}}{\rfvar}}
}{}
\cinferenceRule[composeb|$\dibox{{;}}$]{composition} %
      {\linferenceRule[equiv]
        {\dbox{\alpha}{\dbox{\beta}{\rfvar}}}
        {\dbox{\alpha;\beta}{\rfvar}}
      }{}
\end{calculus} \quad
\begin{calculus}
\cinferenceRule[choiceb|$\dibox{\cup}$]{axiom of nondeterministic choice}
{\linferenceRule[equiv]
  {\dbox{\alpha}{\rfvar} \land \dbox{\beta}{\rfvar}}
  {\dbox{\pchoice{\alpha}{\beta}}{\rfvar}}
}{}
\cinferenceRule[iterateb|$\dibox{{}^*}$]{iteration/repeat unwind} %
{\linferenceRule[equiv]
  {\rfvar \land \dbox{\alpha}{\dbox{\prepeat{\alpha}}{\rfvar}}}
  {\dbox{\prepeat{\alpha}}{\rfvar}}
}{}
\end{calculus}\\
\begin{calculus}
\dinferenceRuleQuote{loop}
\dinferenceRuleQuote{loopm}
\end{calculus}\\
\begin{calculus}
\quad\;\;\cinferenceRule[G|G]{$\dbox{}{}$ generalization} %
{\linferenceRule[formula]
  {\lsequent{}{\rfvar}}
  {\lsequent{\Gamma}{\dbox{\alpha}{\rfvar}}}
}{}
\end{calculus}\qquad
\begin{calculus}
\dinferenceRule[Mb|M${\dibox{\cdot}}$]{}
{\linferenceRule
  {\lsequent{\rrfvar}{\rfvar} \qquad \lsequent{\Gamma}{\dbox{\alpha}{\rrfvar}}}
  {\lsequent{\Gamma}{\dbox{\alpha}{\rfvar}}}
}{}
\end{calculus}\\
\begin{calculus}
\dinferenceRule[dIcmp|dI$_\cmp$]{}
{\linferenceRule
  {
    \lsequent{\Gamma,\ivr}{\ptermA {\cmp} \ptermB} \quad
    \lsequent{\ivr}{\lie[]{\genDE{x}}{\ptermA}{\geq}\lie[]{\genDE{x}}{\ptermB}}
  }
  {\lsequent{\Gamma}{\dbox{\pevolvein{\D{x}=\genDE{x}}{\ivr}}{\ptermA {\cmp} \ptermB}} }
  ~~\hfill
}{$\cmp$ is either $\geq$ or $>$}
\dinferenceRule[dC|dC]{}
{\linferenceRule
  {\lsequent{\Gamma}{\dbox{\pevolvein{\D{x}=\genDE{x}}{\ivr}}{\rcfvar}}
  &\lsequent{\Gamma}{\dbox{\pevolvein{\D{x}=\genDE{x}}{\ivr \land \rcfvar}}{\rfvar}}
  }
  {\lsequent{\Gamma}{\dbox{\pevolvein{\D{x}=\genDE{x}}{\ivr}}{\rfvar}}}
}{}
\dinferenceRule[dW|dW]{}
{\linferenceRule
  {\lsequent{\ivr}{\rfvar}}
  {\lsequent{\Gamma}{\dbox{\pevolvein{\D{x}=\genDE{x}}{\ivr}}{\rfvar}}}
}{}
\dinferenceRule[dbxineq|dbx${_\cmp}$]{Darboux inequality}
{\linferenceRule
  {\lsequent{\ivr} {\lie[]{\genDE{x}}{\ptermA}\geq \cofterm\ptermA}}
  {\lsequent{\ptermA\cmp0} {\dbox{\pevolvein{\D{x}=\genDE{x}}{\ivr}}{\ptermA\cmp0}}}
}{\text{$\cmp$ is either $\geq$ or $>$}}

\dinferenceRule[BC|Barr]{}
{\linferenceRule
  { \lsequent{\ivr, \ptermA=0}{\lie[]{\genDE{x}}{p} > 0}
  }
  {\lsequent{\Gamma, \ptermA \cmp 0}{\dbox{\pevolvein{\D{x}=\genDE{x}}{\ivr}}{\ptermA \cmp 0}} }  \quad\hfill
}{$\cmp$ is either $\geq$ or $>$}

\dinferenceRule[DCC|DCC]{}
{
\linferenceRule[impll]
  {\dbox{\pevolvein{\D{x}{=}\genDE{x}}{\ivr {\land} \rfvar}}{\rrfvar} \land \dbox{\pevolvein{\D{x}{=}\genDE{x}}{\ivr}}{( \lnot{\rfvar} {\limply} \dbox{\pevolvein{\D{x}{=}\genDE{x}}{\ivr}}{\lnot{\rfvar}})}}
  {\dbox{\pevolvein{\D{x}{=}\genDE{x}}{\ivr}}{(\rfvar \limply \rrfvar)}}
}{}

\cinferenceRule[DX|DX]{}
{\linferenceRule[equiv]
  {(\ivr \limply \rfvar \land \dbox{\pevolvein{\D{x}{=}\genDE{x}}{\ivr}}{\rfvar})}
  {\dbox{\pevolvein{\D{x}{=}\genDE{x}}{\ivr}}{\rfvar}}
}{\text{$\D{x} \not\in \rfvar,\ivr$}}

\end{calculus}
\end{calculuscollection}

Axioms~\irref{assignb+testb+composeb+choiceb+iterateb} unfold box modalities of their respective hybrid programs according to their semantics~\citep{DBLP:journals/jar/Platzer17,Platzer18}.
These equivalences are especially useful for obtaining correct-by-construction arithmetical conditions on Lyapunov functions in derivations and implementations (see Corollaries~\ref{cor:ugpasgstateswitchmlf} and~\ref{cor:ugpastimeswitchmlf}).
The derived loop induction rules~\irref{loop+loopm} are used to prove stability properties of switched system models with suitably chosen loop invariants $\invariant$ (see~\rref{sec:loopinv}).
Rule~\irref{G} is G\"odel generalization, and rule~\irref{Mb} is the derived monotonicity rule for box modality postconditions; antecedents that have no free variables bound in $\alpha$ are soundly kept across uses of rules~\irref{loop+loopm+G+Mb}~\citep{DBLP:journals/jar/Platzer17,Platzer18}.

The remaining axioms and proof rules are used in \dL to reason about differential equations $\pevolvein{\D{x}=\genDE{x}}{\ivr}$~\citep{DBLP:journals/jar/Platzer17,Platzer18,DBLP:journals/jacm/PlatzerT20,DBLP:conf/tacas/TanP21}.
Differential invariants~\irref{dIcmp} prove ODE invariance for inequalities $\ptermA\cmp \ptermB$ whenever their Lie derivatives satisfy $\lie[]{\genDE{x}}{\ptermA} \geq \lie[]{\genDE{x}}{\ptermB}$.
Differential cuts~\irref{dC} say that if one can separately prove that formula $\rcfvar$ is always satisfied along the solution, then $\rcfvar$ may be assumed in the domain constraint when proving the same for formula $\rfvar$.
Differential weakening~\irref{dW} says that postcondition $\rfvar$ is always satisfied along solutions if it is already implied by the domain constraint $\ivr$.
Rule~\irref{dbxineq} is the Darboux inequality proof rule for the invariance of $\ptermA \cmp 0$, where $\cofterm$ is an arbitrary cofactor term~\cite{DBLP:journals/jacm/PlatzerT20}.
Rule~\irref{BC} is a \dL rendition of the strict barrier certificates proof rule~\cite{DBLP:journals/tac/PrajnaJP07} for invariance of $\ptermA \cmp 0$.
Axiom~\irref{DCC} says to prove that an implication $\rfvar \limply \rrfvar$ is always true along an ODE, it suffices to prove that $\rrfvar$ is always true along that ODE assuming $\rfvar$ in the domain if $\lnot{\rfvar}$ is invariant along the ODE~\cite{DBLP:conf/tacas/TanP21}.
Differential skip~\irref{DX} unfolds the effect of a differential equation on the initial state in the box modality.

To improve readability in the proofs below, formula and premises are often abbreviated, e.g., with \circled{a}, \circled{1}.
To avoid confusion, the scope of these abbreviations always extend to the end of each {\it paragraph} label, i.e., the abbreviations used in the {\it Stability} proofs should not be confused with those used in the {\it Pre-attractivity} proofs.

\begin{proof}[Proof of~\rref{lem:asymstabdl}]
Let $\Phi(x)$ be the set of all domain-obeying solutions $\solvar : [0,T_\solvar] \to \reals^n$ for a given switched system from state $x \in \reals^n$ as in~\rref{def:ugpas}.
Hybrid program $\alpha$ \emph{models} the given switched system if, for any initial state $\iget[state]{\I} \in \reals^n$, the state $\iget[state]{\It}$ is reachable from state $\iget[state]{\I}$, i.e., $\iaccessible[\alpha]{\I}{\It}$ by \dL's hybrid program semantics~\citep{DBLP:journals/jar/Platzer17,Platzer18}, iff $\iget[state]{\It}=\solvar(\tau)$ for some $\solvar \in \Phi(\iget[state]{\I})$ and $\tau \in [0,T_\solvar]$.
For the augmented program $\taug{\alpha}$, in particular, $\tvar$ syntactically tracks the progression of time so that $\iaccessible[\taug{\alpha}]{\I}{\It}$ iff $\iget[state]{\It}=\solvar(\tau)$ for some $\solvar \in \Phi(\iget[state]{\I})$ and $\tau = \iget[state]{\It}(\tvar) - \iget[state]{\I}(\tvar)$.
Tan and Platzer~\cite{DBLP:conf/adhs/TanP21} prove the adequacy of looping hybrid program models for several switching mechanisms.

The formulas $\stabhp{\alpha}$ and $\attrhp{\alpha}$ syntactically express their respective quantifiers from~\rref{def:ugpas}, where the box modality $\dibox{\cdot}$ is used in both formulas to quantify over all reachable states of $\alpha$ (and $\taug{\alpha}$), i.e., all times $\tau \in [0,T_\solvar]$ along all solutions $\solvar \in \Phi$.
Thus, the correctness of these specifications follows directly from the definition of \dL's formula semantics~\cite{Platzer18,DBLP:journals/jar/Platzer17}.
In $\attrhp{\alpha}$, the clock variable $\tvar$ is set to $0$ initially and has ODE $\D{\tvar}=1$ so it tracks the progression of time along the continuous evolution of program $\alpha$.
Thus, the implication $\tvar \geq T \limply \dots$ in the postcondition of the box modality restricts temporal quantification to all times $\tau$ with $\iget[state]{\I}(T) \leq \tau \leq T_\solvar$ for all solutions $\solvar \in \Phi(\iget[state]{\I})$, as required in the definition of uniform pre-attractivity.
\end{proof}

\begin{proof}[Proof of~\rref{cor:ugpasstateswitchclf}]
The proof rule~\irref{UGpASst} is an instance of rule~\irref{UGpASstmlf} from~\rref{cor:ugpasstateswitchmlf} where the Lyapunov functions for all modes $p \in \sigfam$ are chosen identically with $\lterm_p = \lterm$.
Nevertheless, a full derivation of~\irref{UGpASst} is given here because it provides the building blocks used in later derivations.
The stability and pre-attractivity conjuncts of $\astabhp{\stateswitch}$ are proved separately with~\irref{andr}:

{\small\begin{sequentdeduction}[array]
  \linfer[andr]{
    \lsequent{}{\stabhp{\stateswitch}} !
    \lsequent{}{\attrhp{\stateswitch}}
  }
  {\lsequent{}{\astabhp{\stateswitch}}}
\end{sequentdeduction}
}%

\paragraph{Stability} The derivation for stability begins by Skolemizing the succedent with~\irref{allr+implyr}, followed by two arithmetic cuts which are justified as follows.
For any $\varepsilon > 0$, the Lyapunov function $\lterm$ attains a minimum value on the compact set characterized by $\norm{x}=\varepsilon$.
From the first (topmost) premise of rule~\irref{UGpASst}, this minimum is attained away from the origin so it is positive, which proves the first~\irref{cut} of formula $\lexists{\ltermUpper{>}0}{\circled{a}}$ where $\circled{a}\mnodefequiv \lforall{x}{(\norm{x}=\varepsilon \limply \lterm \geq \ltermUpper)}$.
After Skolemizing $\ltermUpper$ with~\irref{existsl}, the premise $V(0)=0$ implies, by continuity of \dL term semantics~\cite{DBLP:journals/jar/Platzer17}, that the sublevel set characterized by $\lterm < \ltermUpper$ with $\ltermUpper > 0$ (see~\rref{fig:arbstab}) contains a sufficiently small $\delta$ ball around the origin (with $\delta \leq \varepsilon)$.
This proves the second arithmetic~\irref{cut} with the formula $ \lexists{\delta}{(0 < \delta \leq \varepsilon \land \circled{b})}$ where $\circled{b} \mnodefequiv \lforall{x}{(\norm{x}< \delta \limply \lterm < \ltermUpper)}$.
After both cuts, the Skolemized $\delta$ from the antecedent is used to witness the succedent by~\irref{existsr}.

{\small\begin{sequentdeduction}[array]
  \linfer[allr+implyr]{
  \linfer[cut+qear+existsl]{
  \linfer[cut+qear+existsl]{
  \linfer[existsr]{
    \lsequent{\circled{a}, \delta \leq \varepsilon, \circled{b}} {\lforall{x}{\big( \norm{x}<\delta \limply \dbox{\stateswitch}{\,\norm{x}<\varepsilon}\big)}}
  }
    {\lsequent{\circled{a}, 0 < \delta \leq \varepsilon, \circled{b}} { \lexists{\delta {>} 0}{ \lforall{x}{\big( \norm{x}<\delta \limply \dbox{\stateswitch}{\,\norm{x}<\varepsilon}\big)}}}}
  }
  {\lsequent{\varepsilon {>} 0, \ltermUpper{>}0, \circled{a}} { \lexists{\delta {>} 0}{ \lforall{x}{\big( \norm{x}<\delta \limply \dbox{\stateswitch}{\,\norm{x}<\varepsilon}\big)}}}}
  }
  {\lsequent{\varepsilon {>} 0} { \lexists{\delta {>} 0}{ \lforall{x}{\big( \norm{x}<\delta \limply \dbox{\stateswitch}{\,\norm{x}<\varepsilon}\big)}}}}
  }
  {\lsequent{}{\stabhp{\stateswitch}}}
\end{sequentdeduction}
}%

The derivation continues from the open premise by Skolemizing the succedent with~\irref{allr+implyr} and proving the LHS of the implication in \circled{b} with~\irref{alll+implyl}.
Then, the~\irref{loop} rule is used with the stability loop invariant $\invariants \mnodefequiv \norm{x} < \varepsilon \land \lterm < \ltermUpper$.
This results in three premises:
\circled{1} which shows that the invariant is implied by the initial antecedent assumptions;
\circled{2} the crucial premise, which shows that the invariant $\invariants$ is preserved across the loop body of $\stateswitch$; and
\circled{3} which shows that the invariant implies the postcondition.
These premises are shown and proved further below.

{\small\begin{sequentdeduction}[array]
  \linfer[allr+implyr]{
  \linfer[alll+implyl]{
  \linfer[loop]{
    \circled{1} !
    \circled{2} !
    \circled{3}
  }
    {\lsequent{\circled{a}, \delta \leq \varepsilon, \norm{x}<\delta, \lterm < \ltermUpper}{\dbox{\stateswitch}{\,\norm{x}<\varepsilon}}}
  }
    {\lsequent{\circled{a}, \delta \leq \varepsilon, \circled{b}, \norm{x}<\delta}{\dbox{\stateswitch}{\,\norm{x}<\varepsilon}}}
  }
  {\lsequent{\circled{a}, \delta \leq \varepsilon, \circled{b}} {\lforall{x}{\big( \norm{x}{<}\delta \limply \dbox{\stateswitch}{\,\norm{x}<\varepsilon}\big)}}}
\end{sequentdeduction}
}%

Premise \circled{1} proves by~\irref{qear} from the antecedents using the inequalities $\norm{x} < \delta$ and $\delta \leq \varepsilon$.

{\small\begin{sequentdeduction}[array]
  \linfer[qear]{
    \lclose
  }
  {\lsequent{\delta \leq \varepsilon, \norm{x}<\delta, \lterm < \ltermUpper}{\invariants}}
\end{sequentdeduction}
}%

Premise \circled{3} proves trivially since the postcondition $\norm{x}<\varepsilon$ is part of the loop invariant:

{\small\begin{sequentdeduction}[array]
  \linfer[qear]{
    \lclose
  }
  {\lsequent{\invariants}{\norm{x}<\varepsilon}}
\end{sequentdeduction}
}%

The derivation continues from premise \circled{2} by unfolding the loop body of $\stateswitch$ with~\irref{choiceb+andr}.
This results in one premise for each switching choice $p \in \sigfam$, indexed below by $p$.

{\small\begin{sequentdeduction}[array]
  \linfer[choiceb+andr]{
    \lsequent{\circled{a}, \invariants}{\dbox{\pevolvein{\D{x}=f_p{(x)}}{\ivr_p}}{\invariants}} \qquad (p \in \sigfam)
  }
  {\lsequent{\circled{a}, \invariants}{\dbox{\bigcup_{p \in \sigfam}{ \pevolvein{\D{x}=f_p{(x)}}{\ivr_p}} }{\invariants}}}
\end{sequentdeduction}
}%

Each of these $p \in \sigfam$ premises is an ODE invariance question, which completely reduces to an arithmetic question by proof in \dL~\cite{DBLP:journals/jacm/PlatzerT20}.
The derivation below shows how to derive arithmetical conditions on $\lterm$ from these premises.
The right conjunct of $\invariants$, $\lterm < \ltermUpper$, is added to the domain constraint with a~\irref{dC} step; the cut premise is labeled \circled{4} and proved below.
A subsequent~\irref{dC} step adds $\norm{x}\not=\varepsilon$ to the domain constraint using the contrapositive of antecedent \circled{a} and the derivation is completed with rule~\irref{BC} since the resulting $\norm{x}=\varepsilon$ assumption in its premise contradicts the domain constraint $\norm{x}\not{=}\varepsilon$.

{\small\begin{sequentdeduction}[array]
  \linfer[dC]{
  \linfer[dC]{
  \linfer[BC]{
  \linfer[qear]{
    \lclose
  }
    {\lsequent{\norm{x}\not{=}\varepsilon, \norm{x}{=}\varepsilon}{\lfalse}}
  }
    {\lsequent{\norm{x}<\varepsilon}{\dbox{\pevolvein{\D{x}{=}f_p{(x)}}{\ivr_p \land \lterm < \ltermUpper \land \norm{x}\not{=}\varepsilon}}{\,\norm{x}<\varepsilon}}}
  }
    {\lsequent{\circled{a}, \norm{x}<\varepsilon}{\dbox{\pevolvein{\D{x}{=}f_p{(x)}}{\ivr_p \land \lterm < \ltermUpper}}{\,\norm{x}<\varepsilon}}} \qquad \circled{4}
  }
  {\lsequent{\circled{a}, \invariants}{\dbox{\pevolvein{\D{x}{=}f_p{(x)}}{\ivr_p}}{\invariants}}}
\end{sequentdeduction}
}%

The derivation from \circled{4} is completed with a~\irref{dIcmp} step whose resulting arithmetic is implied by the bottom premise of rule~\irref{UGpASst}.

{\small\begin{sequentdeduction}[array]
  \linfer[dIcmp]{
  \linfer[qear]{
    \lclose
  }
    {\lsequent{\ivr_p}{ \lie[]{f_p}{\lterm} \leq 0}}
  }
  {\lsequent{\lterm < \ltermUpper}{\dbox{\pevolvein{\D{x}=f_p{(x)}}{\ivr_p}}{\lterm < \ltermUpper}}}
\end{sequentdeduction}
}%

\paragraph{Pre-attractivity}  The derivation for pre-attractivity begins by Skolemizing the succedent $\delta, \varepsilon$ with~\irref{allr+implyr}, followed by a series of arithmetic cuts which are justified stepwise.
First, the Lyapunov function $\lterm$ is bounded above on the ball characterized by $\norm{x} < \delta$, which justifies a~\irref{cut} of the formula $\lexists{\ltermUpper{>}0}{\circled{a}}$ with $\circled{a} \mnodefequiv \lforall{x}{\big(\norm{x}<\delta \limply \lterm < \ltermUpper\big)}$.
After Skolemizing the upper bound $\ltermUpper$, note that the set characterized by formula $\lterm \leq \ltermUpper$ is compact by radial unboundedness (middle premise of rule~\irref{UGpASst}).
Therefore, the set characterized by formula $\lterm \leq \ltermUpper \land \norm{x} \geq \varepsilon$ is an intersection of a compact and closed set, which is itself compact.
Thus, $\lterm$ attains a minimum $\ltermLower$ on that set which is positive by the first (topmost) premise.
This justifies the next arithmetic~\irref{cut} of the formula $\lexists{\ltermLower{>}0}{\circled{b}}$ with $\circled{b} \mnodefequiv \lforall{x}{(\lterm \leq \ltermUpper \land \norm{x} \geq \varepsilon \limply \lterm \geq \ltermLower)}$, where $\ltermLower$ is subsequently Skolemized with~\irref{existsl}.
The steps are shown below, with the box modality in $\attrhp{\stateswitch}$ temporarily hidden with $\dots$ as it is not relevant for this part of the derivation.

{\small\begin{sequentdeduction}[array]
  \linfer[allr+implyr]{
  \linfer[cut+qear+existsl]{
  \linfer[cut+qear+existsl]{
    \lsequent{\varepsilon {>} 0, \ltermUpper{>}0, \circled{a}, \ltermLower{>}0, \circled{b}} {\exists{T {\geq} 0}{ \lforall{x}{\big( \norm{x}<\delta \limply \dots \big)}}}
  }
  {\lsequent{\varepsilon {>} 0, \ltermUpper{>}0, \circled{a}} {\exists{T {\geq} 0}{ \lforall{x}{\big( \norm{x}<\delta \limply \dots \big)}}}}
  }
  {\lsequent{\varepsilon {>} 0} {\exists{T {\geq} 0}{ \lforall{x}{\big( \norm{x}<\delta \limply \dots \big)}}}}
  }
  {\lsequent{}{\attrhp{\stateswitch}}}
\end{sequentdeduction}
}%

Intuitively (see~\rref{fig:arbstab}) the next arithmetic steps syntactically determine $T \geq 0$ such that the value of $\lterm$ decreases from $\ltermUpper$ to $\ltermLower$ along all switching trajectories within time $T$.
Consider the set characterized by formula $\ivr_p \land \ltermLower \leq \lterm \leq \ltermUpper$, which is the set of states (before reaching $\lterm < \ltermLower$) where switching to ODE $\pevolvein{\D{x}=f_p(x)}{\ivr_p} (p \in \sigfam)$ is possible.
From the third (bottom) premise of rule~\irref{UGpASst}, $\lie[]{f_p}{\lterm}$ is negative on the set characterized by the formula $\closure{\ivr_p} \land \ltermLower \leq \lterm \leq \ltermUpper$ because conjunct $\ltermLower \leq \lterm$ bounds the set away from the origin as $\ltermLower > 0$.
Using radial unboundedness again, $\lterm \leq \ltermUpper$ is compact, so the set characterized by $\closure{\ivr_p} \land \ltermLower \leq \lterm \leq \ltermUpper$ is an intersection of closed sets and compact sets which is therefore compact.
Accordingly, $\lie[]{f_p}{\lterm}$ attains a maximum value $k_p < 0$ on that set, which justifies the following arithmetic cut, where the bound $k < 0$ is chosen uniformly across all choices of $p$, e.g., as the maximum over all $k_p$ for $p \in \sigfam$:
\[ \lexists{k {<} 0}{\underbrace{\landfold_{p \in \sigfam}{\lforall{x}{\big( \closure{\ivr_p} \land \ltermLower \leq \lterm \leq \ltermUpper \limply \lie[]{f_p}{\lterm} \leq k \big)}}}_{\circled{c}}} \]

After Skolemizing $k$, it suffices to pick $T \geq 0$ for the succedent such that $\ltermUpper + k T \leq \ltermLower$. Such a $T$ always exists since $k < 0$.

{\small\begin{sequentdeduction}[array]
  \linfer[cut+qear+existsl]{
  \linfer[existsr]{
    \lsequent{\circled{a}, \circled{b}, k {<} 0, \circled{c}, \ltermUpper + k T \leq \ltermLower} {\lforall{x}{\big( \norm{x}<\delta \limply \dots \big)}}
  }
    {\lsequent{\varepsilon {>} 0, \ltermUpper{>}0, \circled{a}, \ltermLower{>}0, \circled{b}, k {<} 0, \circled{c} } {\exists{T {\geq} 0}{ \lforall{x}{\big( \norm{x}<\delta \limply \dots \big)}}}}
  }
  {\lsequent{\varepsilon {>} 0, \ltermUpper{>}0, \circled{a}, \ltermLower{>}0, \circled{b}} {\exists{T {\geq} 0}{ \lforall{x}{\big( \norm{x}<\delta \limply \dots \big)}}}}
\end{sequentdeduction}
}%

The derivation continues by Skolemizing with~\irref{allr+implyr} and proving the LHS of the implication in \circled{a} with~\irref{alll+implyl}.
The assignment $\pumod{t}{0}$ is unfolded with axioms~\irref{composeb+assignb}, then the~\irref{loop} rule is used with the pre-attractivity loop invariant $\invarianta \mnodefequiv \lterm < \ltermUpper \land (\lterm \geq \ltermLower \limply \lterm < \ltermUpper + k \tvar)$.
Similar to the stability derivation, this results in three premises, where the crucial premise \circled{2} requires showing that $\invarianta$ is preserved across the loop body, while the other premises are labeled \circled{1} and \circled{3} (all three premises are shown further below).

{\small\begin{sequentdeduction}[array]
  \linfer[allr+implyr]{
  \linfer[alll+implyl]{
  \linfer[composeb+assignb]{
  \linfer[loop]{
    \circled{1} !
    \circled{2} !
    \circled{3}
  }
    {\lsequent{ \lterm {<} \ltermUpper, \circled{b}, k {<} 0, \circled{c}, \ltermUpper + k T \leq \ltermLower, \tvar {=} 0} {\dbox{\taug{\stateswitch}}{\dots}}}
  }
    {\lsequent{ \lterm {<} \ltermUpper, \circled{b}, k {<} 0, \circled{c}, \ltermUpper + k T \leq \ltermLower} {\dbox{\pumod{\tvar}{0};\taug{\stateswitch}}{\dots}}}
  }
    {\lsequent{\circled{a}, \circled{b}, k {<} 0, \circled{c}, \ltermUpper + k T \leq \ltermLower, \norm{x}{<}\delta} {\dbox{\pumod{\tvar}{0};\taug{\stateswitch}}{\dots}}}
  }
  {\lsequent{\circled{a}, \circled{b}, k {<} 0, \circled{c}, \ltermUpper + k T \leq \ltermLower} {\lforall{x}{\big( \norm{x}{<}\delta \limply \dots \big)}}}
\end{sequentdeduction}
}%

Premise \circled{1} proves by~\irref{qear} from the antecedents, using assumption $t=0$ to simplify the term $\ltermUpper + k \tvar$ in $\invarianta$.

{\small\begin{sequentdeduction}[array]
  \linfer[qear]{
    \lclose
  }
  {\lsequent{ \lterm {<} \ltermUpper, \tvar = 0}{\invarianta}}
\end{sequentdeduction}
}%

Premise \circled{3} proves by~\irref{qear} from the loop invariant using the following arithmetic argument.
Suppose for contradiction that there is a state satisfying the negation of the postcondition, i.e., assume the negation $t \geq T \land \norm{x} \geq \varepsilon$.
Then, using the left conjunct of $\invarianta$ together with $\norm{x}\geq \varepsilon$ to prove the LHS of the implication in \circled{b} gives assumption $\lterm \geq \ltermLower$.
The right conjunct of $\invarianta$ then yields the chain of inequalities $\lterm < \ltermUpper + k \tvar \leq \ltermUpper + k T \leq \ltermLower$, which is a contradiction to assumption $\lterm \geq \ltermLower$.
The steps are outlined below.

{\small\begin{sequentdeduction}[array]
  \linfer[qear]{
  \linfer[qear]{
  \linfer[qear]{
  \linfer[qear]{
    \lclose
  }
    {\lsequent{\lterm \geq \ltermLower,  k {<} 0, \ltermUpper + k T \leq \ltermLower, \lterm < \ltermUpper + k \tvar , \tvar \geq T}{\lfalse}}
  }
    {\lsequent{\lterm \geq \ltermLower,  k {<} 0, \ltermUpper + k T \leq \ltermLower, \invarianta, \tvar \geq T}{\lfalse}}
  }
    {\lsequent{\circled{b},  k {<} 0, \ltermUpper + k T \leq \ltermLower, \invarianta, \tvar \geq T, \norm{x}\geq \varepsilon }{\lfalse}}
  }
  {\lsequent{\circled{b},  k {<} 0, \ltermUpper + k T \leq \ltermLower, \invarianta}{\tvar \geq T \limply \norm{x}<\varepsilon}}
\end{sequentdeduction}
}%

The proof for premise \circled{2} proceeds by unfolding the loop body with~\irref{choiceb+andr}, yielding one premise for each switching choice $p \in \sigfam$.
A~\irref{dC} step proves the invariance of the left conjunct $\lterm < \ltermUpper$ of $\invarianta$ with~\irref{dIcmp} (see the stability proof, sublevel sets of $\lterm$ are invariant).
The right conjunct of $\invarianta$ is abbreviated $I \mnodefequiv \lterm {\geq} \ltermLower \limply \lterm {<} \ltermUpper + k \tvar$ and it is proved below using axiom~\irref{DCC}, which results in premises \circled{4} and \circled{5} (shown and proved further below).

{\small\begin{sequentdeduction}[array]
  \linfer[choiceb+andr]{
  \linfer[dC+dIcmp]{
  \linfer[DCC+andr]{
    \circled{4} ! \circled{5}
  }
    {\lsequent{\circled{c}, I} {\dbox{\pevolvein{\D{x}=f_p{(x)},\D{\tvar}=1}{\ivr_p \land \lterm < \ltermUpper}}{ I } }}
  }
    {\lsequent{\circled{c}, \invarianta} {\dbox{\pevolvein{\D{x}=f_p{(x)},\D{\tvar}=1}{\ivr_p}}{\invarianta} }\qquad (p \in \sigfam)}
  }
  {\lsequent{\circled{c}, \invarianta} {\dbox{\bigcup_{p \in \sigfam}{ \pevolvein{\D{x}=f_p{(x)},\D{\tvar}=1}{\ivr_p}} }{\invarianta} }}
\end{sequentdeduction}
}%

From premise \circled{4}, the proof is completed with a~\irref{dIcmp} step using the quantified assumption \circled{c} because the domain constraint $\ivr$ implies its closure formula $\closure{\ivr}$ and the strict inequality $\lterm < \ltermUpper$ implies the nonstrict inequality $\lterm \leq \ltermUpper$ which is needed for the LHS of the nested implication in \circled{c}.
The Lie derivative of RHS $W + k \tvar$ is $k$ using $\D{\tvar}=1$.
{\small\begin{sequentdeduction}
  \linfer[dIcmp]{
  \linfer[qear]{
    \lclose
  }
    {\lsequent{\circled{c},\ivr_p \land \lterm < \ltermUpper \land  \lterm \geq \ltermLower } {\lie[]{f_p}{\lterm} \leq k }}
  }
  {\lsequent{\circled{c}, I} {\dbox{\pevolvein{\D{x}=f_p{(x)},\D{\tvar}=1}{\ivr_p \land \lterm < \ltermUpper \land  \lterm \geq \ltermLower }}{ \lterm < \ltermUpper + k \tvar} }}
\end{sequentdeduction}
}%

From premise \circled{5}, the proof is completed with a generalization~\irref{G} step followed by~\irref{dIcmp} to prove the invariance of formula $\lterm < \ltermLower$ (see the stability proof above, sublevel sets of $\lterm$ are invariant).
The ODE in the outer box modality is elided with $\dots$ here.

{\small\begin{sequentdeduction}[array]
  \linfer[G+implyr]{
  \linfer[dIcmp]{
    \lclose
  }
    {\lsequent{\lterm {<} \ltermLower}{\dbox{\pevolvein{\D{x}=f_p{(x)},\D{\tvar}=1}{\ivr_p \land \lterm {<} \ltermUpper}}{\lterm {<} \ltermLower }}}
  }
  {\lsequent{}{\dbox{\dots}{(\lterm {<} \ltermLower \limply \dbox{\pevolvein{\D{x}=f_p{(x)},\D{\tvar}=1}{\ivr_p \land \lterm {<} \ltermUpper}}{\lterm {<} \ltermLower})}} \;\;\qedhere}
\end{sequentdeduction}
}%
\end{proof}

\begin{proof}[Proof of~\rref{cor:ugpasstateswitchmlf}]
The derivation of rule~\irref{UGpASstmlf} builds on the ideas of the derivation of rule~\irref{UGpASst} from~\rref{cor:ugpasstateswitchclf} so similar proof steps are explained in less detail here.
The derivation starts with an~\irref{andr} step for the stability and pre-attractivity conjuncts which are proved separately below.

{\small\begin{sequentdeduction}[array]
  \linfer[andr]{
    \lsequent{}{\stabhp{\stateswitch}} !
    \lsequent{}{\attrhp{\stateswitch}}
  }
  {\lsequent{}{\astabhp{\stateswitch}}}
\end{sequentdeduction}
}%

\paragraph{Stability} The derivation for stability similarly begins with~\irref{cut} and Skolemization steps.
The difference compared to the derivation of rule~\irref{UGpASst} is the cut formulas are now conjunctions over all possible modes $p \in \sigfam$ for the Lyapunov functions $\lterm_p$.
The first cut is $\lexists{\ltermUpper{>}0}{\circled{a}}$ with $\circled{a}\mnodefequiv \landfold_{p \in \sigfam}{\lforall{x}{(\norm{x}=\varepsilon \limply \lterm_p \geq \ltermUpper)}}$, where the upper bound $\ltermUpper{>}0$ is chosen to be the maximum of the respective bounds for each $\lterm_p$ on the compact set characterized by $\norm{x}=\varepsilon$.
After Skolemizing $\ltermUpper$, the second arithmetic~\irref{cut} is the formula $ \lexists{\delta}{(0 < \delta \leq \varepsilon \land \circled{b})}$ with $\circled{b} \mnodefequiv \landfold_{p \in \sigfam}{\lforall{x}{(\norm{x}< \delta \limply \lterm_p < \ltermUpper)}}$.
Such a $\delta$ exists by continuity for each $\lterm_p, p \in \sigfam$ since $\lterm_p(0)=0$ from the first (topmost) premise of rule~\irref{UGpASstmlf}.
After both cuts, the Skolemized $\delta$ from the antecedent is used to witness the succedent by~\irref{existsr}.

{\small\begin{sequentdeduction}[array]
  \linfer[allr+implyr]{
  \linfer[cut+qear+existsl]{
  \linfer[cut+qear+existsl]{
  \linfer[existsr]{
    \lsequent{\circled{a}, \delta \leq \varepsilon, \circled{b}} {\lforall{x}{\big( \norm{x}<\delta \limply \dbox{\stateswitch}{\,\norm{x}<\varepsilon}\big)}}
  }
    {\lsequent{\circled{a}, 0 < \delta \leq \varepsilon, \circled{b}} { \lexists{\delta {>} 0}{ \lforall{x}{\big( \norm{x}<\delta \limply \dbox{\stateswitch}{\,\norm{x}<\varepsilon}\big)}}}}
  }
  {\lsequent{\varepsilon {>} 0, \ltermUpper{>}0, \circled{a}} { \lexists{\delta {>} 0}{ \lforall{x}{\big( \norm{x}<\delta \limply \dbox{\stateswitch}{\,\norm{x}<\varepsilon}\big)}}}}
  }
  {\lsequent{\varepsilon {>} 0} { \lexists{\delta {>} 0}{ \lforall{x}{\big( \norm{x}<\delta \limply \dbox{\stateswitch}{\,\norm{x}<\varepsilon}\big)}}}}
  }
  {\lsequent{}{\stabhp{\stateswitch}}}
\end{sequentdeduction}
}%

The derivation continues with logical simplification steps, Skolemizing the succedent and then proving the LHS of the implications in antecedent \circled{b}.

{\small\begin{sequentdeduction}
  \linfer[allr+implyr]{
  \linfer[alll+implyl]{
    \lsequent{\circled{a}, \delta \leq \varepsilon, \norm{x}<\delta, \landfold_{p \in \sigfam}{\lterm_p < \ltermUpper}}{\dbox{\stateswitch}{\,\norm{x}<\varepsilon}}
  }
  {\lsequent{\circled{a}, \delta \leq \varepsilon, \circled{b}, \norm{x}<\delta} {\dbox{\stateswitch}{\,\norm{x}<\varepsilon}}}
  }
  {\lsequent{\circled{a}, \delta \leq \varepsilon, \circled{b}} {\lforall{x}{\big( \norm{x}{<}\delta \limply \dbox{\stateswitch}{\,\norm{x}<\varepsilon}\big)}}}
\end{sequentdeduction}
}%

Next, a~\irref{cut+orl} step case splits on whether the switched system is initially in its domain of definition characterized by formula $\ivr \mnodefequiv \lorfold_{p \in \sigfam}{\ivr_p}$.
The case where the system is \emph{not} in its domain is labeled \circled{0} and the proof for this case is deferred to the end.
In case the system is in its domain, the~\irref{loop} rule is used with stability loop invariant $\invariants \mnodefequiv \norm{x} < \varepsilon \land \lorfold_{p \in \sigfam}\big( \ivr_p \land \lterm_p < \ltermUpper\big)$.
This yields three premises labeled \circled{1}--\circled{3} shown and proved further below.

{\small\begin{sequentdeduction}[array]
  \linfer[cut+orl]{
  \linfer[loop]{
    \circled{1} !
    \circled{2} !
    \circled{3}
  }
    {\lsequent{\circled{a}, \delta \leq \varepsilon, \norm{x}<\delta, \landfold_{p \in \sigfam}{\lterm_p < \ltermUpper}, \ivr}{\dbox{\stateswitch}{\,\norm{x}<\varepsilon}} \qquad \circled{0}}
  }
    {\lsequent{\circled{a}, \delta \leq \varepsilon, \norm{x}<\delta, \landfold_{p \in \sigfam}{\lterm_p < \ltermUpper}}{\dbox{\stateswitch}{\,\norm{x}<\varepsilon}}}
\end{sequentdeduction}
}%

Premise \circled{1} proves by~\irref{qear} from the antecedents using the inequalities $\norm{x} < \delta$ and $\delta \leq \varepsilon$ for the left conjunct and propositionally from antecedents $\ivr$ and $\landfold_{p \in \sigfam}{\lterm_p < \ltermUpper}$ for the right conjunct.

{\small\begin{sequentdeduction}[array]
  \linfer[qear]{
    \lclose
  }
  {\lsequent{\delta \leq \varepsilon, \norm{x}<\delta, \landfold_{p \in \sigfam}{\lterm_p < \ltermUpper}, \ivr}{\invariants}}
\end{sequentdeduction}
}%

Premise \circled{3} proves trivially since the postcondition $\norm{x}<\varepsilon$ is part of the loop invariant:

{\small\begin{sequentdeduction}[array]
  \linfer[qear]{
    \lclose
  }
  {\lsequent{\invariants}{\norm{x}<\varepsilon}}
\end{sequentdeduction}
}%

The derivation continues from premise \circled{2} by unfolding the loop body of $\stateswitch$ with~\irref{choiceb+andr}.
Premises are indexed by $p \in \sigfam$ in the derivation.
The~\irref{Mb} step propositionally strengthens the postcondition to its constituent disjunct $\norm{x} < \varepsilon \land \lterm_p < \ltermUpper$ for the chosen mode $p$.
Then,~\irref{DX} assumes domain $\ivr_p$ in the antecedent and a~\irref{cut} step adds the assumption $\norm{x} < \varepsilon \land \lterm_p < \ltermUpper$.
This cut corresponds to the last (bottom) premise of rule~\irref{UGpASstmlf}. It is labeled \circled{4} and explained below.
The rest of the proof after the cut proceeds identically to the corresponding derivation for rule~\irref{UGpASst} using the respective conjunct for $p \in \sigfam$ from \circled{a}. The steps are omitted here.

{\small\begin{sequentdeduction}[array]
  \linfer[choiceb+andr]{
  \linfer[Mb]{
  \linfer[DX]{
  \linfer[cut]{
  \linfer[]{\lclose}{\lsequent{\circled{a}, \norm{x} {<} \varepsilon \land \lterm_p {<} \ltermUpper} {\dbox{\pevolvein{\D{x}=f_p{(x)}}{\ivr_p}}{(\norm{x} {<} \varepsilon \land \lterm_p {<} \ltermUpper)}}  \qquad \circled{4}}
  }
    {\lsequent{\circled{a}, \invariants, \ivr_p} {\dbox{\pevolvein{\D{x}=f_p{(x)}}{\ivr_p}}{(\norm{x} {<} \varepsilon \land \lterm_p {<} \ltermUpper)}}}
  }
    {\lsequent{\circled{a}, \invariants} {\dbox{\pevolvein{\D{x}=f_p{(x)}}{\ivr_p}}{(\norm{x} {<} \varepsilon \land \lterm_p {<} \ltermUpper)}}}
  }
    {\lsequent{\circled{a}, \invariants} {\dbox{\pevolvein{\D{x}=f_p{(x)}}{\ivr_p}}{\invariants} }\qquad (p \in \sigfam)}
  }
  {\lsequent{\circled{a}, \invariants} {\dbox{\bigcup_{p \in \sigfam}{ \pevolvein{\D{x}=f_p{(x)}}{\ivr_p}} }{\invariants} }}
\end{sequentdeduction}
}%

The cut premise~\circled{4} is proved by splitting the disjunction in $\invariants$ with~\irref{orl} (indexed by $q \in \sigfam$ below).
The disjunct corresponding to mode $p$ proves trivially.
For modes $q \not= p$, the derivation yields a compatibility condition for switching from mode $q$ to mode $p$ which is proved using the last (bottom) premise of rule~\irref{UGpASstmlf}.
Note that the rule uses succedent $\lterm_p = \lterm_q$ since a symmetric condition ($\lterm_q \leq \lterm_p$) is obtained when the roles of modes $p,q \in \sigfam$ are swapped.

{\small\begin{sequentdeduction}[array]
  \linfer[]{
  \linfer[orl]{
  \linfer[qear]{
  \linfer[qear]{
    \lclose
  }
    {\lsequent{\ivr_q , \ivr_p} {\lterm_p \leq \lterm_q}}
  }
    {\lsequent{p \not= q, \ivr_q , \lterm_q < \ltermUpper, \ivr_p} {\lterm_p {<} \ltermUpper}\qquad (q \in \sigfam)}
  }
    {\lsequent{\lorfold_{q \in \sigfam}\big( \ivr_q \land \lterm_q < \ltermUpper\big), \ivr_p} {\lterm_p {<} \ltermUpper}}
  }
  {\lsequent{\invariants, \ivr_p} {\norm{x} {<} \varepsilon \land \lterm_p {<} \ltermUpper}}
\end{sequentdeduction}
}%

Returning to premise \circled{0}, for initial states not in the switched system's domain, i.e., satisfying $\lnot{\ivr}$, no continuous motion is possible within the model.
This is proved using the loop invariant $\invariants^0 \mnodefequiv \norm{x} < \varepsilon \land \lnot{\ivr}$.
The first and third premise resulting from the~\irref{loop} rule are proved trivially (not shown below).
For the remaining premise, $\lnot{\ivr}$ is preserved (trivially) across the loop body after unfolding it with~\irref{choiceb+andr} and using~\irref{DX} to show that the system is unable to switch to the ODE with domain $\ivr_p$ because $\lnot{\ivr}$ implies $\lnot{\ivr_p}$ propositionally.

{\small\begin{sequentdeduction}[array]
  \linfer[loop]{
  \linfer[choiceb+andr]{
  \linfer[DX]{
  \linfer[]{
    \lclose
  }
    {\lsequent{\lnot{\ivr}, \ivr_p}{\lfalse}}
  }
    {\lsequent{\lnot{\ivr}}{\dbox{\pevolvein{\D{x}=f_p{(x)}}{\ivr_p}}{\invariants^0}} \qquad (p \in \sigfam)}
  }
    {\lsequent{\invarianta^0}{\dbox{\bigcup_{p \in \sigfam}{ \pevolvein{\D{x}=f_p{(x)}}{\ivr_p}} }{\invariants^0 }}}
  }
    {\lsequent{\delta \leq \varepsilon, \norm{x}<\delta, \lnot{\ivr}}{\dbox{\stateswitch}{\,\norm{x}<\varepsilon}}}
\end{sequentdeduction}
}%

\paragraph{Pre-attractivity}  The derivation for pre-attractivity begins with logical simplification followed by a series of arithmetic cuts.
First, the multiple Lyapunov functions $\lterm_p, p \in \sigfam$ are simultaneously bounded above on the ball characterized by $\norm{x} < \delta$, with the~\irref{cut} $\lexists{\ltermUpper{>}0}{\circled{a}}$ where $\circled{a} \mnodefequiv \landfold_{p\in\sigfam}{\lforall{x}{\big(\norm{x}<\delta \limply \lterm_p < \ltermUpper\big)}}$.
The upper bound $\ltermUpper$ is Skolemized, then the next arithmetic~\irref{cut} uses $\lexists{\ltermLower{>}0}{\circled{b}}$ with $\circled{b} \mnodefequiv \landfold_{p \in\sigfam}{\lforall{x}{(\lterm_p \leq \ltermUpper \land \norm{x} \geq \varepsilon \limply \lterm_p \geq \ltermLower)}}$ (using radial unboundedness of all functions $\lterm_p$ from the second premise of~\irref{UGpASstmlf}).
Then, $\ltermLower$ is Skolemized with~\irref{existsl}.
The steps are shown below, with the box modality in $\attrhp{\stateswitch}$ temporarily hidden with $\dots$ as it is not relevant for this part of the derivation.

{\small\begin{sequentdeduction}[array]
  \linfer[allr+implyr]{
  \linfer[cut+qear+existsl]{
  \linfer[cut+qear+existsl]{
    \lsequent{\varepsilon {>} 0, \ltermUpper{>}0, \circled{a}, \ltermLower{>}0, \circled{b}} {\exists{T {\geq} 0}{ \lforall{x}{\big( \norm{x}<\delta \limply \dots \big)}}}
  }
  {\lsequent{\varepsilon {>} 0, \ltermUpper{>}0, \circled{a}} {\exists{T {\geq} 0}{ \lforall{x}{\big( \norm{x}<\delta \limply \dots \big)}}}}
  }
  {\lsequent{\varepsilon {>} 0} {\exists{T {\geq} 0}{ \lforall{x}{\big( \norm{x}<\delta \limply \dots \big)}}}}
  }
  {\lsequent{}{\attrhp{\stateswitch}}}
\end{sequentdeduction}
}%

Similar to the derivation of rule~\irref{UGpASst} from~\rref{cor:ugpasstateswitchclf}, the premises of rule~\irref{UGpASstmlf} prove that, for each $p \in \sigfam$, the respective Lie derivatives $\lie[]{f_p}{\lterm_p}$ are bounded above by some $k_p < 0$ on the compact set characterized by formula $\closure{\ivr_p} \land \ltermLower \leq \lterm_p \leq \ltermUpper$.
This justifies the following arithmetic cut, where the bound $k < 0$ is chosen to be the maximum over all $k_p$ across all switching choices $p \in \sigfam$:
\[ \lexists{k {<} 0}{\underbrace{\landfold_{p \in \sigfam}{\lforall{x}{\big( \closure{\ivr_p} \land \ltermLower \leq \lterm_p \leq \ltermUpper \limply \lie[]{f_p}{\lterm_p} \leq k \big)}}}_{\circled{c}}} \]

The derivation continues similarly to rule~\irref{UGpASst}, first picking $T {>} 0$ satisfying $\ltermUpper + k T \leq \ltermLower$, then Skolemizing and unfolding the succedent propositionally.

{\small\begin{sequentdeduction}[array]
  \linfer[cut+qear+existsl]{
  \linfer[existsr]{
  \linfer[allr+implyr]{
    \lsequent{\circled{a}, \circled{b}, k {<} 0, \circled{c}, T {>} 0, \ltermUpper {+} k T {\leq} \ltermLower, \norm{x}{<}\delta} {\dots}
  }
    {\lsequent{\circled{a}, \circled{b}, k {<} 0, \circled{c}, T {>} 0, \ltermUpper {+} k T {\leq} \ltermLower} {\lforall{x}{\big( \norm{x}{<}\delta \limply \dots \big)}}}
  }
    {\lsequent{\varepsilon {>} 0, \ltermUpper{>}0, \circled{a}, \ltermLower{>}0, \circled{b}, k {<} 0, \circled{c} } {\exists{T {\geq} 0}{\dots}}}
  }
  {\lsequent{\varepsilon {>} 0, \ltermUpper{>}0, \circled{a}, \ltermLower{>}0, \circled{b}} {\exists{T {\geq} 0}{\dots}}}
\end{sequentdeduction}
}%

The LHS in antecedent \circled{a} is proved and the succedent is further unfolded with~\irref{composeb+assignb}.
The antecedents are abbreviated with $\Gamma \mnodefequiv \circled{b}, k {<} 0, \circled{c}, T > 0, \ltermUpper + k T \leq \ltermLower$ below.
Similar to the stability proof, the derivation continues with a~\irref{cut+orl} step that case splits on whether the switched system is initially in its domain of definition $\ivr \mnodefequiv \lorfold_{p \in \sigfam}{\ivr_p}$.
The case where the system is \emph{not} in its domain is labeled \circled{0} and its proof is deferred to the end.
In case the system is in domain $\ivr$, the~\irref{loop} rule is used with pre-attractivity loop invariant $\invarianta \mnodefequiv \lorfold_{p \in \sigfam}\big( \ivr_p \land \lterm_p < \ltermUpper \land (\lterm_p \geq \ltermLower \limply \lterm_p < \ltermUpper + k \tvar) \big)$.
This results in three premises \circled{1}--\circled{3} which are proved below.

{\small\begin{sequentdeduction}[array]
  \linfer[alll+implyl]{
  \linfer[composeb+assignb]{
  \linfer[cut+orl]{
  \linfer[loop]{
    \circled{1} !
    \circled{2} !
    \circled{3}
  }
    {\lsequent{\Gamma, \landfold_{p\in\sigfam}{\lterm_p {<} \ltermUpper}, \tvar=0, \ivr} {\dbox{\taug{\stateswitch}}{\dots}}  \qquad \circled{0}}
  }
    {\lsequent{\Gamma, \landfold_{p\in\sigfam}{\lterm_p {<} \ltermUpper}, \tvar=0} {\dbox{\taug{\stateswitch}}{\dots}}}
  }
    {\lsequent{\Gamma, \landfold_{p\in\sigfam}{\lterm_p {<} \ltermUpper}} {\dbox{\pumod{\tvar}{0};\taug{\stateswitch}}{\dots}}}
  }
    {\lsequent{\Gamma, \circled{a}, \norm{x}{<}\delta} {\dbox{\pumod{\tvar}{0};\taug{\stateswitch}}{\dots}}}
\end{sequentdeduction}
}%

Premise \circled{1} proves propositionally from the antecedents after simplifying the term $\ltermUpper + k \tvar$ using assumption $\tvar=0$.

{\small\begin{sequentdeduction}[array]
  \linfer[qear]{
    \lclose
  }
  {\lsequent{ \landfold_{p\in\sigfam}{\lterm_p {<} \ltermUpper}, \tvar=0, \ivr}{\invarianta}}
\end{sequentdeduction}
}%

Premise \circled{3} proves by~\irref{qear} from the loop invariant after using~\irref{orl} to split the disjuncts of the loop invariant.
The disjunct for mode $p \in \sigfam$ is abbreviated $\rrfvar \mnodefequiv \lterm_p < \ltermUpper \land (\lterm_p \geq \ltermLower \limply \lterm_p < \ltermUpper + k \tvar)$.
The rest of the arithmetic argument is identical to the corresponding premise for~\irref{UGpASst} using the conjunct for $p$ in \circled{b} (summarized below).

{\small\begin{sequentdeduction}[array]
  \linfer[orl]{
  \linfer[qear]{
  \linfer[qear]{
  \linfer[qear]{
  \linfer[qear]{
    \lclose
  }
    {\lsequent{\lterm_p \geq \ltermLower,  k {<} 0, \ltermUpper + k T \leq \ltermLower, \lterm_p < \ltermUpper + k \tvar , \tvar \geq T}{\lfalse}}
  }
    {\lsequent{\lterm_p \geq \ltermLower,  k {<} 0, \ltermUpper + k T \leq \ltermLower, \rrfvar, \tvar \geq T}{\lfalse}}
  }
    {\lsequent{\circled{b},  k {<} 0, \ltermUpper + k T \leq \ltermLower, \rrfvar, \tvar \geq T, \norm{x}\geq \varepsilon }{\lfalse}}
  }
  {\lsequent{\circled{b},  k {<} 0, \ltermUpper + k T \leq \ltermLower, \rrfvar}{\tvar \geq T \limply \norm{x}<\varepsilon} }
  }
  {\lsequent{\circled{b},  k {<} 0, \ltermUpper + k T \leq \ltermLower, \invarianta}{\tvar \geq T \limply \norm{x}<\varepsilon}}
\end{sequentdeduction}
}%

The derivation from premise \circled{2} proceeds by unfolding the loop body with~\irref{choiceb+andr+DX}, yielding one premise for each switching choice $p \in \sigfam$.
The~\irref{Mb} step selects the disjunct $\rrfvar$ (as defined above for premise \circled{3}) in the postcondition corresponding to mode $p$ and the~\irref{cut} adds this disjunct to the antecedents (the cut premise \circled{4} is shown and proved below).
The rest of the proof after the cut is omitted here as it is identical to the corresponding derivation for rule~\irref{UGpASst} using the respective conjunct for mode $p$ in \circled{c}.

{\small\begin{sequentdeduction}[array]
  \linfer[choiceb+andr+DX]{
  \linfer[Mb]{
  \linfer[cut]{
  \linfer[]{
    \lclose
  }
    {\circled{4} \qquad \lsequent{\circled{c}, \rrfvar} {\dbox{\pevolvein{\D{x}=f_p{(x)},\D{\tvar}=1}{\ivr_p}}{\rrfvar} }}
  }
    {\lsequent{\circled{c}, \invarianta, \ivr_p} {\dbox{\pevolvein{\D{x}=f_p{(x)},\D{\tvar}=1}{\ivr_p}}{\rrfvar} }}
  }
    {\lsequent{\circled{c}, \invarianta, \ivr_p} {\dbox{\pevolvein{\D{x}=f_p{(x)},\D{\tvar}=1}{\ivr_p}}{\invarianta} } \qquad (p \in \sigfam)}
  }
  {\lsequent{\circled{c}, \invarianta} {\dbox{\bigcup_{p \in \sigfam}{ \pevolvein{\D{x}=f_p{(x)},\D{\tvar}=1}{\ivr_p}} }{\invarianta} }}
\end{sequentdeduction}
}%

The cut premise~\circled{4} is proved by splitting the disjunction in $\invarianta$ with~\irref{orl} (indexed by $q \in \sigfam$ below).
For modes $q \not= p$, the derivation needs a compatibility condition which proves using the last (bottom) premise of rule~\irref{UGpASstmlf}, similar to the stability proof.

{\small\begin{sequentdeduction}[array]
  \linfer[]{
  \linfer[orl]{
  \linfer[qear]{
  \linfer[qear]{
    \lclose
  }
    {\lsequent{\ivr_q , \ivr_p} {\lterm_p \leq \lterm_q}}
  }
    {\lsequent{p \not= q,  \ivr_q \land \lterm_q < \ltermUpper \land (\lterm_q \geq \ltermLower \limply \lterm_q < \ltermUpper + k \tvar) , \ivr_p} {\rrfvar} \quad (q \in \sigfam)}
  }
    {\lsequent{\lorfold_{q \in \sigfam}\big( \ivr_q \land \lterm_q < \ltermUpper \land (\lterm_q \geq \ltermLower \limply \lterm_q < \ltermUpper + k \tvar) \big), \ivr_p} {\rrfvar}}
  }
  {\lsequent{\invarianta, \ivr_p} {\rrfvar}}
\end{sequentdeduction}
}%

Returning to premise \circled{0}, similar to the case for stability, initial states satisfying $\lnot{\ivr}$ have no continuous motion possible so they are stuck at the initial state (with global clock $t=0$).
This is proved using the loop invariant $\invarianta^0 \mnodefequiv t=0 \land \lnot{\ivr}$.
The first and third premise resulting from the~\irref{loop} rule are proved trivially (not shown below).
For the remaining premise, $\lnot{\ivr}$ is preserved (trivially) across the loop body after unfolding it with~\irref{choiceb+andr} and using~\irref{DX} to show that the system is unable to switch to the ODE with domain $\ivr_p$ because $\lnot{\ivr}$ implies $\lnot{\ivr_p}$ propositionally.

{\small\begin{sequentdeduction}[array]
  \linfer[loop]{
  \linfer[choiceb+andr]{
  \linfer[DX]{
  \linfer[]{
    \lclose
  }
    {\lsequent{\lnot{\ivr}, \ivr_p}{\lfalse}}
  }
    {\lsequent{\lnot{\ivr}}{\dbox{\pevolvein{\D{x}=f_p{(x)},\D{\tvar}=1}{\ivr_p}}{\invarianta^0}} \qquad (p \in \sigfam)}
  }
    {\lsequent{\invarianta^0 }{\dbox{\bigcup_{p \in \sigfam}{ \pevolvein{\D{x}=f_p{(x), \D{\tvar}=1}}{\ivr_p}} }{\invarianta^0 }}}
  }
    {\lsequent{T > 0, \tvar=0, \lnot{\ivr}} {\dbox{\taug{\stateswitch}}{(\tvar \geq T \limply \norm{x}<\varepsilon)}} \qedhere}
\end{sequentdeduction}
}%
\end{proof}

\begin{proof}[Proof of~\rref{cor:ugpasgstateswitchmlf}]
The derivation of rule~\irref{UGpASgstmlf} is similar to the derivation rule~\irref{UGpASstmlf} from~\rref{cor:ugpasstateswitchmlf}, but adapted to the shape of the guarded state-dependent switching model $\guardswitch$ and its corresponding loop invariants.
The derivation starts with an~\irref{andr} step for the stability and pre-attractivity conjuncts which are proved separately below.

{\small\begin{sequentdeduction}[array]
  \linfer[andr]{
    \lsequent{}{\stabhp{\guardswitch}} !
    \lsequent{}{\attrhp{\guardswitch}}
  }
  {\lsequent{}{\astabhp{\guardswitch}}}
\end{sequentdeduction}
}%

\paragraph{Stability} The derivation for stability proceeds identically to the derivation for rule~\irref{UGpASstmlf} from~\rref{cor:ugpasstateswitchmlf} until the step before the stability loop invariant is used.
These steps are omitted below with $\dots$ and the resulting premise has antecedent formula abbreviated $\circled{a}\mnodefequiv \landfold_{p \in \sigfam}{\lforall{x}{(\norm{x}=\varepsilon \limply \lterm_p \geq \ltermUpper)}}$.

{\small\begin{sequentdeduction}[array]
  \linfer[]{
  \linfer[]{
    \lsequent{\circled{a}, \delta \leq \varepsilon, \norm{x}<\delta, \landfold_{p \in \sigfam}{\lterm_p < \ltermUpper}}{\dbox{\guardswitch}{\,\norm{x}<\varepsilon}}
  }
  {\dots}
  }
  {\lsequent{}{\stabhp{\guardswitch}}}
\end{sequentdeduction}
}%

The derivation continues using the~\irref{loopm} rule with stability loop invariant $\invariants \mnodefequiv \norm{x} < \varepsilon \land \lorfold_{p \in \sigfam}\big( u=p \land \lterm_p < \ltermUpper\big)$.
This yields four premises labeled \circled{1}--\circled{4}, shown and proved further below.

{\small\begin{sequentdeduction}[array]
  \linfer[loopm]{
    \circled{1} !
    \circled{2} !
    \circled{3} !
    \circled{4}
  }
    {\lsequent{\circled{a}, \delta \leq \varepsilon, \norm{x}<\delta, \landfold_{p \in \sigfam}{\lterm_p < \ltermUpper}}{\dbox{\guardswitch}{\,\norm{x}<\varepsilon}}}
\end{sequentdeduction}
}%

Premise \circled{1} shows that the system state satisfies the invariant $\invariants$ after running the initialization program $\alpha_i \mnodefequiv \bigcup_{p \in \sigfam}{\pumod{u}{p}}$.
This is proved by~\irref{qear} after unfolding $\alpha_i$ using~\irref{choiceb+assignb}.

{\small\begin{sequentdeduction}[array]
  \linfer[choiceb+assignb]{
  \linfer[qear]{
    \lclose
  }
    {\lsequent{\delta \leq \varepsilon, \norm{x}<\delta, \landfold_{p \in \sigfam}{\lterm_p < \ltermUpper}, u = p}{\invariants}  \qquad (p \in \sigfam)}
  }
  {\lsequent{\delta \leq \varepsilon, \norm{x}<\delta, \landfold_{p \in \sigfam}{\lterm_p < \ltermUpper}}{\dbox{\alpha_i}{\invariants}}}
\end{sequentdeduction}
}%

Premise \circled{4} proves trivially since the postcondition $\norm{x}<\varepsilon$ is part of the loop invariant.

{\small\begin{sequentdeduction}[array]
  \linfer[qear]{
    \lclose
  }
  {\lsequent{\invariants}{\norm{x}<\varepsilon}}
\end{sequentdeduction}
}%

The derivation from premise \circled{2} yields \emph{correct-by-construction} arithmetical conditions on the Lyapunov functions from unfolding the switching controller in $\guardswitch$, recall \[ \alpha_u \mnodefequiv \bigcup_{p\in\sigfam}{\Big( \ptest{u=p}; \bigcup_{q \in \sigfam}{ \big( \ptest{G_{p,q}}; \pumod{u}{q} \big)}\Big)}\]

Axiom~\irref{choiceb} unfolds the outer choice $\bigcup_{p\in\sigfam}{\big(\cdot\big)}$, yielding one premise for each mode $p \in \sigfam$.
Then, axioms~\irref{composeb+testb} add the current mode $u=p$ (before switching) to the assumptions.
The~\irref{cut} step propositionally unfolds antecedent loop invariant assumption $\invariants$ to the corresponding disjunct for $u=p$.
The inner choice $\bigcup_{q\in\sigfam}{\big(\cdot\big)}$ is unfolded next with axioms~\irref{choiceb+composeb+testb}, yielding one premise for each possible transition to mode $q \in \sigfam$ guarded by formula $G_{p,q}$.
The assignment $\pumod{u}{q}$ is unfolded with~\irref{assignb}, so the succedent simplifies to the disjunct for $u=q$ in $\invariants$.
An arithmetic simplification step yields the bottom premise of rule~\irref{UGpASgstmlf}.

{\small\begin{sequentdeduction}[array]
  \linfer[choiceb]{
  \linfer[composeb+testb]{
  \linfer[cut]{
  \linfer[choiceb+composeb+testb]{
  \linfer[assignb]{
  \linfer[qear]{
  \linfer[qear]{
    \lclose
  }
    {\lsequent{G_{p,q} }{\lterm_q \leq \lterm_p}}
  }
    {\lsequent{\lterm_p {<} \ltermUpper, G_{p,q} }{\lterm_q {<} \ltermUpper}}
  }
    {\lsequent{ \norm{x} {<} \varepsilon,\lterm_p {<} \ltermUpper, G_{p,q} }{\dbox{\pumod{u}{q}}}{\invariants}  \qquad (q \in \sigfam)}
  }
    {\lsequent{ \norm{x} {<} \varepsilon, \lterm_p {<} \ltermUpper}{\dbox{\bigcup_{q \in \sigfam}{ \big( \ptest{G_{p,q}}; \pumod{u}{q} \big)}}{\invariants}}}
  }
    {\lsequent{\invariants, u{=}p}{\dbox{\bigcup_{q \in \sigfam}{ \big( \ptest{G_{p,q}}; \pumod{u}{q} \big)}}{\invariants}}}
  }
    {\lsequent{\invariants}{\dbox{\ptest{u{=}p}; \bigcup_{q \in \sigfam}{ \big( \ptest{G_{p,q}}; \pumod{u}{q} \big)}}{\invariants}} \;(p \in \sigfam)}
  }
  {\lsequent{\invariants}{\dbox{\alpha_u}{\invariants}}}
\end{sequentdeduction}
}%

The derivation from premise \circled{3} unfolds the plant model $\alpha_p \mnodefequiv \bigcup_{p \in \sigfam}{\big( \ptest{u = p} ; \pevolvein{\D{x}=f_p{(x,y)}}{\ivr_p} \big)}$.
The choice $\bigcup_{p\in\sigfam}{\big(\cdot\big)}$ is unfolded with axiom~\irref{choiceb}, yielding one premise for each mode $p \in \sigfam$.
Then, axioms~\irref{composeb+testb} add the mode selected by $\alpha_u$ to the antecedent, where the antecedent loop invariant assumption $\invariants$ is simplified by~\irref{cut} to the disjunct for $u=p$.
Similarly~\irref{Mb} strengthens the postcondition to the disjunct for $u=p$.
The rest of the proof proceeds identically to the corresponding derivation for rule~\irref{UGpASst} in~\rref{cor:ugpasstateswitchclf} so it is omitted here.

{\small\begin{sequentdeduction}[array]
  \linfer[choiceb]{
  \linfer[composeb+testb]{
  \linfer[cut]{
  \linfer[Mb]{
  \linfer[]{
    \lclose
  }
    {\lsequent{\circled{a}, \norm{x} {<} \varepsilon , \lterm_p {<} \ltermUpper} {\dbox{\pevolvein{\D{x}{=}f_p{(x)}}{\ivr_p}}{(\norm{x} {<} \varepsilon \land \lterm_p {<} \ltermUpper)}}}
  }
    {\lsequent{\circled{a}, \norm{x} {<} \varepsilon , \lterm_p {<} \ltermUpper, u{=}p} {\dbox{\pevolvein{\D{x}{=}f_p{(x)}}{\ivr_p}}{\invariants}}}
  }
    {\lsequent{\circled{a}, \invariants, u{=}p} {\dbox{\pevolvein{\D{x}{=}f_p{(x)}}{\ivr_p}}{\invariants}}}
  }
    {\lsequent{\circled{a}, \invariants} {\dbox{\ptest{u {=} p} ; \pevolvein{\D{x}{=}f_p{(x,y)}}{\ivr_p}}{\invariants}}  \;(p \in \sigfam)}
  }
  {\lsequent{\circled{a}, \invariants} {\dbox{\alpha_p}{\invariants} }}
\end{sequentdeduction}
}%

\paragraph{Pre-attractivity}  The derivation for pre-attractivity is also identical to~\irref{UGpASstmlf} until the step before the pre-attractivity loop invariant is used.
These steps are omitted below with $\dots$ and the resulting premise has antecedent formulas abbreviated with:
\begin{align*}
\circled{b} &\mnodefequiv \landfold_{p \in\sigfam}{\lforall{x}{(\lterm_p \leq \ltermUpper \land \norm{x} \geq \varepsilon \limply \lterm_p \geq \ltermLower)}} \\
\circled{c} &\mnodefequiv \landfold_{p \in \sigfam}{\lforall{x}{\big( \closure{\ivr_p} \land \ltermLower \leq \lterm_p \leq \ltermUpper \limply \lie[]{f_p}{\lterm_p} \leq k \big)}}
\end{align*}

{\small\begin{sequentdeduction}[array]
  \linfer[]{
  \linfer[]{
  \lsequent{ \landfold_{p\in\sigfam}{\lterm_p {<} \ltermUpper}, \circled{b}, k {<} 0, \circled{c}, \ltermUpper + k T \leq \ltermLower, \tvar=0} {\dbox{\taug{\guardswitch}}{\dots}}
  }
  {\dots}
  }
  {\lsequent{}{\attrhp{\guardswitch}}}
\end{sequentdeduction}
}

The derivation continues using the~\irref{loopm} rule with pre-attractivity loop invariant $\invarianta \mnodefequiv \lorfold_{p \in \sigfam}\big( u{=}p \land \lterm_p {<} \ltermUpper \land (\lterm_p {\geq} \ltermLower \limply \lterm_p {<} \ltermUpper {+} k \tvar) \big)$.
This yields four premises labeled \circled{1}--\circled{4} which are shown and proved further below.

{\small\begin{sequentdeduction}[array]
  \linfer[loopm]{
    \circled{1}!
    \circled{2}!
    \circled{3}!
    \circled{4}
  }
  {\lsequent{ \landfold_{p\in\sigfam}{\lterm_p {<} \ltermUpper}, \circled{b}, k {<} 0, \circled{c}, \ltermUpper {+} k T {\leq} \ltermLower, \tvar{=}0} {\dbox{\taug{\guardswitch}}{\dots}}
}
\end{sequentdeduction}
}%

Premise \circled{1} proves the invariant $\invarianta$ after unfolding the initialization program $\alpha_i$ using~\irref{choiceb+assignb}.

{\small\begin{sequentdeduction}[array]
  \linfer[choiceb+assignb]{
  \linfer[qear]{
    \lclose
  }
    {\lsequent{ \landfold_{p\in\sigfam}{\lterm_p {<} \ltermUpper}, \tvar{=}0 , u = p}{\invarianta}}
  }
  {\lsequent{ \landfold_{p\in\sigfam}{\lterm_p {<} \ltermUpper}, \tvar{=}0 }{\dbox{\alpha_i}{\invarianta}}}
\end{sequentdeduction}
}%

Premise \circled{4} is proved by~\irref{qear} after unfolding the disjuncts of the loop invariant with~\irref{orl} (the arithmetical argument is identical to earlier proofs).
The selected disjunct of $\invarianta$ (indexed by $p$) is abbreviated $\rrfvar \mnodefequiv  u{=}p \land \lterm_p {<} \ltermUpper \land (\lterm_p {\geq} \ltermLower \limply \lterm_p {<} \ltermUpper {+} k \tvar)$.

{\small\begin{sequentdeduction}[array]
  \linfer[orl]{
  \linfer[qear]{
    \lclose
  }
  {\lsequent{\circled{b},  k {<} 0, \ltermUpper + k T \leq \ltermLower, \rrfvar}{\tvar \geq T \limply \norm{x}<\varepsilon} \qquad (p \in \sigfam) }
  }
  {\lsequent{\circled{b},  k {<} 0, \ltermUpper + k T \leq \ltermLower, \invarianta}{\tvar \geq T \limply \norm{x}<\varepsilon}}
\end{sequentdeduction}
}%

The derivation from premise \circled{2} unfolds $\alpha_u$ using \dL's hybrid program axioms similar to the stability proof, and an arithmetic simplification step yields the premises of~\irref{UGpASgstmlf} for guarded mode switches from $p$ to $q$, for $p,q \in \sigfam$.

{\small\begin{sequentdeduction}[array]
  \linfer[choiceb]{
  \linfer[composeb+testb]{
  \linfer[cut+orl]{
  \linfer[choiceb+composeb+testb]{
  \linfer[assignb]{
  \linfer[qear]{
  \linfer[qear]{
    \lclose
  }
    {\lsequent{G_{p,q} }{\lterm_q \leq \lterm_p}}
  }
    {\lsequent{\rrfvar, G_{p,q} }{\lterm_q {<} \ltermUpper \land (\lterm_q {\geq} \ltermLower \limply \lterm_q {<} \ltermUpper {+} k \tvar)}}
  }
    {\lsequent{\rrfvar, G_{p,q} }{\dbox{\pumod{u}{q}}}{\invarianta} \qquad (q \in \sigfam)}
  }
    {\lsequent{\rrfvar}{\dbox{\bigcup_{q \in \sigfam}{ \big( \ptest{G_{p,q}}; \pumod{u}{q} \big)}}{\invarianta}}}
  }
    {\lsequent{\invarianta, u=p}{\dbox{\bigcup_{q \in \sigfam}{ \big( \ptest{G_{p,q}}; \pumod{u}{q} \big)}}{\invarianta}}}
  }
    {\lsequent{\invarianta}{\dbox{\ptest{u=p}; \bigcup_{q \in \sigfam}{ \big( \ptest{G_{p,q}}; \pumod{u}{q} \big)}}{\invarianta}} \; (p \in \sigfam)}
  }
  {\lsequent{\invarianta}{\dbox{\alpha_u}{\invarianta}}}
\end{sequentdeduction}
}%

The derivation from premise \circled{3} unfolds the plant model and then proceeds identically to the corresponding derivation for rule~\irref{UGpASst} with the same abbreviation $\rrfvar \mnodefequiv  u{=}p \land \lterm_p {<} \ltermUpper \land (\lterm_p {\geq} \ltermLower \limply \lterm_p {<} \ltermUpper {+} k \tvar)$.

{\small\begin{sequentdeduction}[array]
  \linfer[choiceb]{
  \linfer[composeb+testb]{
  \linfer[cut]{
  \linfer[Mb]{
  \linfer[]{
    \lclose
  }
    {\lsequent{\circled{c}, \rrfvar} {\dbox{\pevolvein{\D{x}=f_p{(x)},\D{\tvar}=1}{\ivr_p}}{\rrfvar}}}
  }
    {\lsequent{\circled{c}, \rrfvar} {\dbox{\pevolvein{\D{x}=f_p{(x)},\D{\tvar}=1}{\ivr_p}}{\invarianta}}}
  }
    {\lsequent{\circled{c}, \invarianta, u=p} {\dbox{\pevolvein{\D{x}=f_p{(x)},\D{\tvar}=1}{\ivr_p}}{\invarianta}}}
  }
    {\lsequent{\circled{c}, \invarianta} {\dbox{\ptest{u = p} ; \pevolvein{\D{x}{=}f_p{(x,y)},\D{\tvar}=1}{\ivr_p}}{\invarianta}}}
  }
  {\lsequent{\circled{c}, \invarianta} {\dbox{\taug{\alpha_p}}{\invarianta} }}
\\[-\normalbaselineskip]\tag*{\qedhere}
\end{sequentdeduction}
}%
\end{proof}

\begin{proof}[Proof of~\rref{cor:ugpastimeswitchmlf}]
The derivation of rule~\irref{UGpAStimemlf} departs more significantly from the derivations of rules~\irref{UGpASst+UGpASstmlf+UGpASgstmlf}.
For this proof,~\irref{qearexp} is used to indicate arithmetic steps that use properties of the real exponential function.
Although arithmetic over the exponential function is not known to be decidable, tools are available for answering specialized subsets of such questions~\cite{DBLP:conf/cade/GaoKC13}.
Additional explanation is given below for~\irref{qearexp} steps that only require elementary properties of the exponential function.

The proof also shows how to derive arithmetic conditions (arising from the time-dependent switching controller) in a correct by construction manner through the hybrid program axioms of \dL~\citep{DBLP:journals/jar/Platzer17,Platzer18}.
Recall from~\rref{cor:ugpastimeswitchmlf} that the modes $p \in \sigfam$ are partitioned into two subsets consisting of the stable $ \sigfams \mnodefeq \{ p \in \sigfam , \lambda_p > 0\}$ and unstable $\sigfamu \mnodefeq \{p \in \sigfam, \lambda_p \leq 0\}$ modes.
The derivation starts with an~\irref{andr} step for the stability and pre-attractivity conjuncts which are proved separately below.

{\small\begin{sequentdeduction}[array]
  \linfer[andr]{
    \lsequent{}{\stabhp{\timeswitch}} !
    \lsequent{}{\attrhp{\timeswitch}}
  }
  {\lsequent{}{\astabhp{\timeswitch}}}
\end{sequentdeduction}
}%

\paragraph{Stability} The stability derivation begins with~\irref{cut} and Skolemization steps.
The first cut is $\lexists{\ltermUpper{>}0}{\circled{a}}$ with the abbreviation $\circled{a}\mnodefequiv \landfold_{p \in \sigfam}{\lforall{x}{(\norm{x}=\varepsilon \limply \lterm_p \geq \ltermUpper)}}$, where the upper bound $\ltermUpper{>}0$ is chosen to be the maximum of the respective bounds for each $\lterm_p$ on the compact set characterized by $\norm{x}=\varepsilon$.
After Skolemizing $\ltermUpper$, the second arithmetic~\irref{cut} is the formula $ \lexists{\delta}{(0 < \delta \leq \varepsilon \land \circled{b})}$, where the conjuncts for $p\in\sigfamu$ need the arithmetic fact $\expo{\lambda_p\Theta_p} > 0$.
\begin{align*}
\circled{b} &\mnodefequiv \landfold_{p \in \sigfams}{\lforall{x}{(\norm{x}< \delta \limply \lterm_p < \ltermUpper)}} \\
&\land \landfold_{p \in \sigfamu}{\lforall{x}{(\norm{x}< \delta \limply \lterm_p < \ltermUpper\expo{\lambda_p\Theta_p})}}
\end{align*}

Such a $\delta$ exists by continuity for each $\lterm_p, p \in \sigfam$, $\lterm_p(0)=0$ from the premise of rule~\irref{UGpAStimemlf}.
After both cuts, the Skolemized $\delta$ from the antecedent is used to witness the succedent by~\irref{existsr}.

{\small\begin{sequentdeduction}[array]
  \linfer[allr+implyr]{
  \linfer[cut+qear+existsl]{
  \linfer[cut+qearexp+existsl]{
  \linfer[existsr]{
    \lsequent{\circled{a}, \delta \leq \varepsilon, \circled{b}} {\lforall{x}{\big( \norm{x}<\delta \limply \dbox{\timeswitch}{\,\norm{x}<\varepsilon}\big)}}
  }
    {\lsequent{\circled{a}, 0 < \delta \leq \varepsilon, \circled{b}} { \lexists{\delta {>} 0}{ \lforall{x}{\big( \norm{x}<\delta \limply \dbox{\timeswitch}{\,\norm{x}<\varepsilon}\big)}}}}
  }
  {\lsequent{\varepsilon {>} 0, \ltermUpper{>}0, \circled{a}} { \lexists{\delta {>} 0}{ \lforall{x}{\big( \norm{x}<\delta \limply \dbox{\timeswitch}{\,\norm{x}<\varepsilon}\big)}}}}
  }
  {\lsequent{\varepsilon {>} 0} { \lexists{\delta {>} 0}{ \lforall{x}{\big( \norm{x}<\delta \limply \dbox{\timeswitch}{\,\norm{x}<\varepsilon}\big)}}}}
  }
  {\lsequent{}{\stabhp{\timeswitch}}}
\end{sequentdeduction}
}%

The derivation continues after both cuts similarly to~\irref{UGpASstmlf} from~\rref{cor:ugpasstateswitchmlf} by unfolding and proving the LHS of the implications in antecedent \circled{b}.
The resulting assumption on the initial state is abbreviated $B \mnodefequiv \landfold_{p \in \sigfams}{\lterm_p {<} \ltermUpper} \land \landfold_{p \in \sigfamu}{\lterm_p {<} \ltermUpper\expo{\lambda_p\Theta_p}}$.
Then, the~\irref{loopm} rule is used with the following stability loop invariant $\invariants$, which yields premises \circled{1}--\circled{4} shown and proved further below:
\begin{align*}
\invariants &\mnodefequiv \tau \geq 0 \land \norm{x} < \varepsilon \, \land \\
&\quad\left(
\begin{aligned}
&\lorfold_{p \in \sigfams}\big( u=p \land \lterm_p < \ltermUpper\expo{-\lambda_p \tau} \big) \lor\\
&\lorfold_{p \in \sigfamu}\big( u=p \land \lterm_p < \ltermUpper\expo{-\lambda_p(\tau-\Theta_p)} \land \tau \leq \Theta_p \big)
\end{aligned}\right)
\end{align*}

{\small\begin{sequentdeduction}[array]
  \linfer[allr+implyr]{
  \linfer[alll+implyl]{
  \linfer[loopm]{
    \circled{1} !
    \circled{2} !
    \circled{3} !
    \circled{4}
  }
    {\lsequent{\circled{a}, \delta {\leq} \varepsilon, \norm{x}{<}\delta, B}{\dbox{\timeswitch}{\,\norm{x}{<}\varepsilon}}}
  }
  {\lsequent{\circled{a}, \delta {\leq} \varepsilon, \circled{b}, \norm{x}{<}\delta} {\dbox{\timeswitch}{\,\norm{x}{<}\varepsilon}}}
  }
  {\lsequent{\circled{a}, \delta {\leq} \varepsilon, \circled{b}} {\lforall{x}{\big( \norm{x}{{<}}\delta \limply \dbox{\timeswitch}{\,\norm{x}{<}\varepsilon}\big)}}}
\end{sequentdeduction}
}%

Premise \circled{1} shows that the system state satisfies the invariant $\invariants$ after initialization with program $\alpha_i \mnodefequiv \pumod{\tau}{0};\bigcup_{p \in \sigfam}{\pumod{u}{p}}$.
This is proved from $B$ after unfolding $\alpha_i$ using~\irref{choiceb+assignb} and substituting $\tau = 0$ in the loop invariant (using $e^0=1$).

{\small\begin{sequentdeduction}[array]
  \linfer[choiceb+assignb]{
  \linfer[qearexp]{
    \lclose
  }
    {\lsequent{\delta \leq \varepsilon, \norm{x}<\delta, B, \tau=0, u = p}{\invariants}}
  }
  {\lsequent{\delta \leq \varepsilon, \norm{x}<\delta, B}{\dbox{\alpha_i}{\invariants}}}
\end{sequentdeduction}
}%

Premise \circled{4} proves trivially since the postcondition $\norm{x}<\varepsilon$ is part of the loop invariant.

{\small\begin{sequentdeduction}[array]
  \linfer[qear]{
    \lclose
  }
  {\lsequent{\invariants}{\norm{x}<\varepsilon}}
\end{sequentdeduction}
}%

The derivation from premise \circled{2} unfolds the switching controller $\alpha_u$ in $\timeswitch$ with \dL's hybrid program axioms, recall:
\[ \alpha_u \mnodefequiv \bigcup_{p\in\sigfam}{\Big( \ptest{u=p}; \bigcup_{q \in \sigfam}{\big(\ptest{\theta_{p,q} \leq \tau}; \pumod{\tau}{0} ; \pumod{u}{q}\big)} \Big)} \]

This unfolding yields four possible shapes of premises (abbreviated as $\dots$ and shown immediately below) for a switch from the current mode $p$ to mode $q$.
In each case, the antecedent assumption corresponds to the disjunct of $\invariants$ for mode $p$, while the succedent assumption corresponds to the disjunct for mode $q$ with timer $\tau$ reset to $0$ by the switching controller $\alpha_u$.

{\small\begin{sequentdeduction}
  \linfer[choiceb]{
  \linfer[composeb+testb]{
  \linfer[choiceb+composeb+testb+assignb]{
    \dots
    }
    {\lsequent{\invariants, u=p}{\dbox{\bigcup_{q \in \sigfam}{\big(\ptest{\theta_{p,q} \leq \tau}; \pumod{\tau}{0} ; \pumod{u}{q}\big)}}{\invariants}}}
  }
    {\lsequent{\invariants}{\dbox{\ptest{u=p}; \bigcup_{q \in \sigfam}{\big(\ptest{\theta_{p,q} \leq \tau}; \pumod{\tau}{0} ; \pumod{u}{q}\big)} }{\invariants}}}
  }
  {\lsequent{\invariants}{\dbox{\alpha_u}{\invariants}}}
\end{sequentdeduction}
}%

The four cases correspond to whether $p \in \sigfams$ or $p\in\sigfamu$ and similarly for $q$, as labeled below.
{\small\begin{align*}
&\theta_{p,q} \leq \tau, \lterm_p < \ltermUpper\expo{-\lambda_p \tau}\vdash\lterm_q < \ltermUpper \, &&(p {\in} \sigfams, q {\in} \sigfams)\\
&\theta_{p,q} \leq \tau, \lterm_p < \ltermUpper\expo{-\lambda_p \tau}\vdash\lterm_q < \ltermUpper\expo{\lambda_q\Theta_q}  \, &&(p {\in} \sigfams, q {\in} \sigfamu)\\
&\theta_{p,q} \leq \tau, \lterm_p < \ltermUpper\expo{-\lambda_p(\tau-\Theta_p)}, \tau \leq \Theta_p\vdash\lterm_q < \ltermUpper \, &&(p {\in} \sigfamu, q {\in} \sigfams)\\
&\theta_{p,q} \leq \tau, \lterm_p < \ltermUpper\expo{-\lambda_p(\tau-\Theta_p)}, \tau \leq \Theta_p\vdash\lterm_q < \ltermUpper\expo{\lambda_q\Theta_q}  \, &&(p {\in} \sigfamu, q {\in} \sigfamu)
\end{align*}}

These premises are correct-by-construction and can be handed to an arithmetic solver directly.
They can also be simplified, e.g., for $p {\in} \sigfams, q {\in} \sigfams$, the inequalities can be rearranged to eliminate $\ltermUpper$ and $\tau$.
The first~\irref{qear} step uses transitivity of $<$ and $\leq$, while the second~\irref{qearexp} step uses monotonicity $ \expo{\lambda_p \theta_{p,q}}\leq \expo{\lambda_p \tau}$ whenever $\lambda_p > 0$ (since $p \in \sigfams$) and $\theta_{p,q} \leq \tau$.
Intuitively, the resulting (simplified) premise says that by choosing sufficiently large dwell time $\theta_{p,q}$ (for stable mode $p$), one can offset an increase in value when switching from $\lterm_p$ to $\lterm_q$.
The resulting arithmetic condition $\lterm_q \leq \lterm_p\expo{\lambda_p \theta_{p,q}}$ is a \emph{correct-by-construction} premise for rule~\irref{UGpAStimemlf}.

{\small\renewcommand{\arraystretch}{1.4}%
\begin{sequentdeduction}[array]
  \linfer[qear]{
  \linfer[qearexp]{
    \lsequent{}{\lterm_q \leq \lterm_p\expo{\lambda_p \theta_{p,q}}}
  }
    {\lsequent{\theta_{p,q} \leq \tau}{\lterm_q \leq \lterm_p\expo{\lambda_p \tau}}}
  }
  {\lsequent{\theta_{p,q} \leq \tau, \lterm_p < \ltermUpper\expo{-\lambda_p \tau}}{\lterm_q < \ltermUpper }}
\end{sequentdeduction}
}%

The derivation from premise \circled{3} unfolds the plant model $\alpha_p\mnodefequiv \bigcup_{p \in \sigfam}{\big( \ptest{u = p} ; \pevolvein{\D{x}=f_p{(x)},\D{\tau}=1}{\tau \leq \Theta_p} \big)}$ using \dL axioms.
There are two possible shapes of the premises resulting from this unfolding, depending if $p\in\sigfams$ or $p\in\sigfamu$, these are abbreviated \circled{5} and \circled{6} respectively.
In either case, the derivation shows that the appropriate upper bound on $\lterm_p$ is preserved for the invariant.

{\small\begin{sequentdeduction}[array]
  \linfer[choiceb]{
  \linfer[composeb+testb]{
  \linfer[composeb+testb]{
    \circled{5} ! \circled{6}
  }
    {\lsequent{\circled{a}, \invariants, u=p} {\dbox{\pevolvein{\D{x}=f_p{(x)},\D{\tau}=1}{\tau \leq \Theta_p}}{\invariants}}}
  }
    {\lsequent{\circled{a}, \invariants} {\dbox{\ptest{u = p} ; \pevolvein{\D{x}=f_p{(x)},\D{\tau}=1}{\tau \leq \Theta_p}}{\invariants}}}
  }
  {\lsequent{\circled{a}, \invariants} {\dbox{\alpha_p}{\invariants} }}
\end{sequentdeduction}
}%

For premise \circled{5}, the proof uses~\irref{dbxineq} with cofactor $-\lambda_p$, where the Lie derivative of subterm $\ltermUpper\expo{-\lambda_p \tau}$ is $(-\lambda_p)\ltermUpper\expo{-\lambda_p \tau}$ from $\D{\tau}=1$.
The resulting premise simplifies to the third premise of rule~\irref{UGpAStimemlf}.

{\small\renewcommand{\arraystretch}{1.4}%
\begin{sequentdeduction}[array]
  \linfer[cut+Mb]{
  \linfer[dbxineq]{
  \linfer[]{
  \linfer[]{
    \lclose
  }
    {\lsequent{} { \lie[]{f_p}{\lterm_p} {\leq} {-}\lambda_p\lterm_p }}
  }
    {\lsequent{} { \lie[]{f_p}{\lterm_p} {-} ({-}\lambda_p)\ltermUpper\expo{{-}\lambda_p \tau} {\leq} {-}\lambda_p(\lterm_p {-}\ltermUpper\expo{{-}\lambda_p \tau})}}
  }
  {\lsequent{\lterm_p {-}\ltermUpper\expo{{-}\lambda_p \tau} < 0} {\dbox{\pevolvein{\D{x}=f_p{(x)},\D{\tau}=1}{\tau {\leq} \Theta_p}}{\lterm_p {-}\ltermUpper\expo{{-}\lambda_p \tau} < 0} }}
  }
  {\lsequent{\lterm_p < \ltermUpper\expo{{-}\lambda_p \tau}} {\dbox{\pevolvein{\D{x}=f_p{(x)},\D{\tau}=1}{\tau {\leq} \Theta_p}}{\lterm_p < \ltermUpper\expo{{-}\lambda_p \tau}} }}
\end{sequentdeduction}
}%

The proof for premise \circled{6} also uses~\irref{dbxineq} with cofactor $-\lambda_p$, yielding the third premise of rule~\irref{UGpAStimemlf} again.

{\small\begin{sequentdeduction}[array]
  \linfer[dbxineq]{
  \linfer[]{
    \lclose
  }
    {\lsequent{} { \lie[]{f_p}{\lterm_p} \leq -\lambda_p\lterm_p }}
  }
  {\lsequent{\lterm_p {<} \ltermUpper\expo{-\lambda_p (\tau-\Theta_p)}} {\dbox{\pevolvein{\D{x}=f_p{(x)},\D{\tau}=1}{\tau \leq \Theta_p}}{\lterm_p {<} \ltermUpper\expo{-\lambda_p (\tau-\Theta_p)}} }}
\end{sequentdeduction}
}%

\paragraph{Pre-attractivity}  The pre-attractivity proof requires an additional input parameter $\sigma>0$ for the overall decay factor with $\sigma < \lambda_p$ for $p \in \sigfams$ ($\sigma$ must also satisfy other arithmetic properties, to be derived in a correct-by-construction manner in the proof).
The derivation begins with logical simplification followed by a series of arithmetic cuts.
First, the multiple Lyapunov functions $\lterm_p, p \in \sigfam$ are simultaneously bounded above on the ball characterized by $\norm{x} < \delta$, with the~\irref{cut} $\lexists{\ltermUpper{>}0}{\circled{a}}$ (abbreviated below) where the conjuncts for $p\in\sigfamu$ need the arithmetic fact $\expo{\lambda_p\Theta_p} > 0$ (by~\irref{qearexp}).
\begin{align*}
\circled{a} &\mnodefequiv \landfold_{p \in \sigfams}{\lforall{x}{(\norm{x}< \delta \limply \lterm_p < \ltermUpper)}} \\
&\land \landfold_{p \in \sigfamu}{\lforall{x}{(\norm{x}< \delta \limply \lterm_p < \ltermUpper\expo{\lambda_p\Theta_p})}}
\end{align*}

The upper bound $\ltermUpper$ is Skolemized, then the next arithmetic~\irref{cut} uses $\lexists{\ltermLower{>}0}{\circled{b}}$ with $\circled{b} \mnodefequiv \landfold_{p \in\sigfam}{\lforall{x}{(\lterm_p \leq \ltermUpper \land \norm{x} \geq \varepsilon \limply \lterm_p \geq \ltermLower)}}$, where $\ltermLower$ is Skolemized with~\irref{existsl}.

{\small\begin{sequentdeduction}[array]
  \linfer[allr+implyr]{
  \linfer[cut+qearexp+existsl]{
  \linfer[cut+qear+existsl]{
    \lsequent{\varepsilon {>} 0, \ltermUpper{>}0, \circled{a}, \ltermLower{>}0, \circled{b}} {\exists{T {\geq} 0}{ \lforall{x}{\big( \norm{x}<\delta \limply \dots \big)}}}
  }
  {\lsequent{\varepsilon {>} 0, \ltermUpper{>}0, \circled{a}} {\exists{T {\geq} 0}{ \lforall{x}{\big( \norm{x}<\delta \limply \dots \big)}}}}
  }
  {\lsequent{\varepsilon {>} 0} {\exists{T {\geq} 0}{ \lforall{x}{\big( \norm{x}<\delta \limply \dots \big)}}}}
  }
  {\lsequent{}{\attrhp{\timeswitch}}}
\end{sequentdeduction}
}%

The derivation continues by picking $T \geq 0$ where (abbreviated) $\rrfvar \mnodefequiv \ltermUpper \leq \ltermLower\expo{\sigma T} \land \landfold_{p \in \sigfamu}{\ltermUpper \leq \ltermLower\expo{\sigma T}\expo{-\sigma\Theta_p}}$, such a $T$ exists by~\irref{qearexp} because $\sigma > 0$ so the $\expo{\sigma T}$ term on the RHS of each inequality can be chosen arbitrarily large.
The quantifiers in the succedent are unfolded and the LHS of the implications in \circled{a} are proved.
The resulting antecedent (from \circled{a}) is abbreviated $B \mnodefequiv \landfold_{p \in \sigfams}{\lterm_p {<} \ltermUpper} \land \landfold_{p \in \sigfamu}{\lterm_p {<} \ltermUpper\expo{\lambda_p\Theta_p}}$.
The~\irref{loopm} rule is used with the following pre-attractivity loop invariant $\invarianta$, which yields premises \circled{1}--\circled{4} shown and proved further below:
\begin{align*}
\invarianta &\mnodefequiv \tau \geq 0 \land t \geq \tau \, \land \\
&\quad \left(
\begin{aligned}
&\lorfold_{p \in \sigfams}\big( u=p \land \lterm_p < \ltermUpper\expo{-\sigma(\tvar-\tau)}\expo{-\lambda_p \tau}   \big) \lor\\
&\lorfold_{p \in \sigfamu}\big( u=p \land \lterm_p < \ltermUpper\expo{-\sigma(\tvar-\tau)}\expo{-\lambda_p(\tau-\Theta_p)} \land \tau \leq \Theta_p\big)
\end{aligned}\right)
\end{align*}

{\small\begin{sequentdeduction}[array]
  \linfer[existsr+qearexp]{
  \linfer[allr+implyr]{
  \linfer[composeb+assignb]{
  \linfer[alll+implyl]{
  \linfer[loopm]{
    \circled{1} !
    \circled{2} !
    \circled{3} !
    \circled{4}
  }
    {\lsequent{\circled{b}, T \geq 0, \rrfvar, B, \tvar=0} {\dbox{\taug{\guardswitch}}{\dots}}}
  }
    {\lsequent{\circled{a}, \circled{b}, T \geq 0, \rrfvar, \norm{x}{<}\delta, \tvar=0} {\dbox{\taug{\guardswitch}}{\dots}}}
  }
    {\lsequent{\circled{a}, \circled{b}, T \geq 0, \rrfvar, \norm{x}{<}\delta} {\dbox{\pumod{\tvar}{0};\taug{\guardswitch}}{\dots}}}
  }
    {\lsequent{\circled{a}, \circled{b}, T \geq 0, \rrfvar} {\lforall{x}{\big( \norm{x}<\delta \limply \dots \big)}}}
  }
  {\lsequent{\varepsilon {>} 0, \ltermUpper{>}0, \circled{a}, \ltermLower{>}0, \circled{b}} {\exists{T {\geq} 0}{ \lforall{x}{\big( \norm{x}<\delta \limply \dots \big)}}}}
\end{sequentdeduction}
}%

Premise \circled{1} is proved by unfolding the initialization program $\alpha_i$
This is proved from $B$ after unfolding $\alpha_i$ using axioms~\irref{choiceb+assignb} and substituting $\tau = 0$ and $\tvar=0$ in the loop invariant (using $\expo{0}=1$).

{\small\begin{sequentdeduction}[array]
  \linfer[choiceb+assignb]{
  \linfer[qearexp]{
    \lclose
  }
    {\lsequent{B, \tvar=0, \tau=0, u = p}{\invarianta}}
  }
  {\lsequent{B, \tvar=0}{\dbox{\alpha_i}{\invarianta}}}
\end{sequentdeduction}
}%

Premise \circled{4} is proved by unfolding the loop invariant with~\irref{orl}.
This yields two possible premise shapes, corresponding to $p \in \sigfams$ or $p \in\sigfamu$.
In both cases, assuming the negation of the succedent proves the corresponding implication LHS in the antecedent assumption $\circled{b}$, which gives $\lterm < \ltermLower$ as an assumption.
The remaining arithmetic argument underlying these premises is proved by~\irref{qearexp} by contradicting assumption $\lterm < \ltermLower$ for each case resulting from~\irref{orl} (explained further below).

{\small\begin{sequentdeduction}[array]
  \linfer[orl+qearexp]{
    \lclose
  }
  {\lsequent{\circled{b}, \rrfvar, \invarianta}{\tvar \geq T \limply \norm{x}<\varepsilon}}
\end{sequentdeduction}
}%

For $p\in\sigfams$, the following sequence of inequalities is used:
\begin{align*}
\lterm_p &< \ltermUpper\expo{-\sigma(\tvar-\tau)}\expo{-\lambda_p \tau} &&\text{(from invariant)} \\
&= \ltermUpper\expo{-\sigma\tvar}\expo{-\tau(\lambda_p-\sigma)} &&\\
&\leq \ltermUpper\expo{-\sigma T}\expo{-\tau(\lambda_p-\sigma)} &&\text{(from $t \geq T, \sigma>0$)}\\
&\leq \ltermLower\expo{-\tau(\lambda_p-\sigma)} &&\text{(from $\rrfvar$)}\\
&\leq \ltermLower &&\text{(from $\sigma < \lambda_p, \tau \geq 0$, contradiction)}
\end{align*}

For $p\in\sigfamu$, the following sequence of inequalities is used (note that $\tau \leq \Theta_p$ is in the invariant $\invarianta$ for $p \in \sigfamu$):
\begin{align*}
\lterm_p &< \ltermUpper\expo{-\sigma(\tvar-\tau)}\expo{-\lambda_p(\tau-\Theta_p)} &&\text{(from invariant)} \\
&\leq \ltermUpper\expo{-\sigma(\tvar-\tau)} &&\text{(from $\tau \leq \Theta_p, \lambda_p \leq 0$)} \\
&= \ltermUpper\expo{-\sigma\tvar}\expo{\sigma\tau} && \\
&\leq \ltermUpper\expo{-\sigma\tvar}\expo{\sigma\Theta_p} &&\text{(from $\sigma > 0, \tau \leq \Theta_p$)} \\
&\leq \ltermUpper\expo{-\sigma T}\expo{\sigma\Theta_p} &&\text{(from $\tvar \geq T, \sigma>0$)}\\
&\leq \ltermLower &&\text{(from $\rrfvar$, contradiction)}
\end{align*}

The derivation from premise \circled{2} unfolds the switching controller $\alpha_u$ in $\timeswitch$ with \dL's hybrid program axioms.
Similar to the derivation for the stability conjunct, this unfolding yields four possible shapes of premises (abbreviated as $\dots$ and shown immediately below) for maintaining the invariant $\invarianta$ after a switch from the current mode $p$ to the next mode $q$.
{\small\begin{sequentdeduction}
  \linfer[choiceb]{
  \linfer[composeb+testb]{
  \linfer[choiceb+composeb+testb+assignb]{
    \dots
    }
    {\lsequent{\invarianta, u=p}{\dbox{\bigcup_{q \in \sigfam}{\big(\ptest{\theta_{p,q} \leq \tau}; \pumod{\tau}{0} ; \pumod{u}{q}\big)}}{\invarianta}}}
  }
    {\lsequent{\invarianta}{\dbox{\ptest{u=p}; \bigcup_{q \in \sigfam}{\big(\ptest{\theta_{p,q} \leq \tau}; \pumod{\tau}{0} ; \pumod{u}{q}\big)} }{\invarianta}}}
  }
  {\lsequent{\invarianta}{\dbox{\alpha_u}{\invarianta}}}
\end{sequentdeduction}
}%
{\small\begin{align*}
&t{\geq}\tau,\theta_{p,q} {\leq} \tau, \lterm_p {<} \ltermUpper\expo{-\sigma(\tvar-\tau)}\expo{-\lambda_p \tau}{\vdash}\lterm_q {<} \ltermUpper\expo{-\sigma\tvar}\\
&\qquad\qquad\qquad\qquad\qquad\qquad\qquad\qquad\qquad\qquad\qquad\qquad(p {\in} \sigfams, q {\in} \sigfams)\\
&t{\geq}\tau,\theta_{p,q} {\leq} \tau, \lterm_p {<} \ltermUpper\expo{-\sigma(\tvar-\tau)}\expo{-\lambda_p \tau}{\vdash}\lterm_q {<} \ltermUpper\expo{-\sigma\tvar}\expo{\lambda_q\Theta_q}\\
&\qquad\qquad\qquad\qquad\qquad\qquad\qquad\qquad\qquad\qquad\qquad\qquad(p {\in} \sigfams, q {\in} \sigfamu)\\
&t{\geq}\tau,\theta_{p,q} {\leq} \tau, \lterm_p {<} \ltermUpper\expo{-\sigma(\tvar-\tau)}\expo{-\lambda_p(\tau-\Theta_p)}, \tau {\leq} \Theta_p{\vdash}\lterm_q {<} \ltermUpper\expo{-\sigma\tvar}\\
&\qquad\qquad\qquad\qquad\qquad\qquad\qquad\qquad\qquad\qquad\qquad\qquad(p {\in} \sigfamu, q {\in} \sigfams)\\
&t{\geq}\tau,\theta_{p,q} {\leq} \tau, \lterm_p {<} \ltermUpper\expo{-\sigma(\tvar-\tau)}\expo{-\lambda_p(\tau-\Theta_p)}, \tau {\leq} \Theta_p{\vdash}\lterm_q {<} \ltermUpper\expo{-\sigma\tvar}\expo{\lambda_q\Theta_q}\\
&\qquad\qquad\qquad\qquad\qquad\qquad\qquad\qquad\qquad\qquad\qquad\qquad(p {\in} \sigfamu, q {\in} \sigfamu)
\end{align*}}
The derivation from premise \circled{3} unfolds the plant model $\alpha_p$.
This results in two possible shapes of premises, depending if $p\in\sigfams$ or $p\in\sigfamu$, which are abbreviated \circled{5} and \circled{6} respectively.
In either case, the key step is to show that the respective upper bound on $\lterm_p$ is preserved along evolution of the ODE.

{\small\begin{sequentdeduction}[array]
  \linfer[choiceb]{
  \linfer[composeb+testb]{
  \linfer[composeb+testb]{
    \circled{5} ! \circled{6}
  }
    {\lsequent{\invarianta, u=p} {\dbox{\pevolvein{\D{x}=f_p{(x)},\D{\tau}=1,\D{\tvar}=1}{\tau \leq \Theta_p}}{\invarianta}}}
  }
    {\lsequent{\invarianta} {\dbox{\ptest{u = p} ; \pevolvein{\D{x}=f_p{(x)},\D{\tau}=1,\D{\tvar}=1}{\tau \leq \Theta_p}}{\invarianta}}}
  }
  {\lsequent{\invarianta} {\dbox{\alpha_p}{\invarianta} }}
\end{sequentdeduction}
}%

For premise \circled{5}, the proof uses~\irref{dbxineq} with cofactor $-\lambda_p$, with abbreviation $\rfvar_s \mnodefeq \ltermUpper\expo{-\sigma(\tvar-\tau)}\expo{-\lambda_p \tau}$, noting that the Lie derivative of $\rfvar_s$ is $-\lambda_p\rfvar_s$.
This yields the third premise of rule~\irref{UGpAStimemlf}.

{\small\begin{sequentdeduction}[array]
  \linfer[dbxineq]{
  \linfer[]{
    \lclose
  }
    {\lsequent{} { \lie[]{f_p}{\lterm_p} \leq -\lambda_p\lterm_p }}
  }
  {\lsequent{\lterm_p {<}  \rfvar_s} {\dbox{\pevolvein{\D{x}=f_p{(x)},\D{\tau}=1,\D{\tvar}=1}{\tau \leq \Theta_p}}{\lterm_p {<}  \rfvar_s} }}
\end{sequentdeduction}
}%

The proof for premise \circled{6} also uses rule~\irref{dbxineq} with cofactor $-\lambda_p$, with abbreviation $\rfvar_u = \ltermUpper\expo{-\sigma(\tvar-\tau)}\expo{-\lambda_p(\tau-\Theta_p)} $, noting that the Lie derivative of $\rfvar_a$ is $-\lambda_p\rfvar_a$.
This yields the third premise of rule~\irref{UGpAStimemlf}.

{\small\begin{sequentdeduction}[array]
  \linfer[dbxineq]{
  \linfer[]{
    \lclose
  }
    {\lsequent{} { \lie[]{f_p}{\lterm_p} \leq -\lambda_p\lterm_p }}
  }
  {\lsequent{\lterm_p {<} \rfvar_u} {\dbox{\pevolvein{\D{x}=f_p{(x)},\D{\tau}=1,\D{\tvar}=1}{\tau \leq \Theta_p}}{\lterm_p {<}  \rfvar_u} }}
\\[-\normalbaselineskip]\tag*{\qedhere}
\end{sequentdeduction}
}%
\end{proof}

\section{Counterexamples}
\label{app:casestudies}

The cruise controller automaton from~\rref{subsec:acc} is taken from the suite of examples for the Stabhyli tool~\cite{DBLP:conf/hybrid/MohlmannT13,stabhyli}.
Using the default instructions on a Linux machine, Stabhyli generates a success message with the following output (newlines added for readability):

\begin{verbatim}
...
SOSSolution( Problem is solved. (accepted); ...
...
### Lyapunov template for mode normal_PI: \
  +V_23*relV^2+V_22*intV^2+V_21*intV*relV \
  +V_20*relV+V_19*intV
### Lyapunov function for mode normal_PI: \
  +572572089848357/144115188075855872*intV*relV \
  +256336575597239/281474976710656*relV^2 \
  +6008302119812893/4611686018427387904*intV^2 \
  +5787253314511645/618970019642690137449562112*relV \
  +5661677770976729/39614081257132168796771975168*intV
...
The hybrid system is stable
\end{verbatim}

The generated Lyapunov function candidate $\lterm$ does not satisfy all of the required arithmetical conditions for the normal PI mode~\cite{DBLP:conf/hybrid/MohlmannT13}.
For example, one requirement is that it should be non-negative in the mode invariant $-15 {\leq} relV {\leq} 15 \land -500 {\leq} intV {\leq} 500$.
It can be checked that $intV = -\frac{1}{17179869184}, relV = 0$ is a counterexample, with $\lterm = -3.90488 \times 10^{-24} < 0$.

A heuristic approach to resolve this numerical issue is to truncate terms in the candidate $\lterm$ with extremely small coefficients and then check the resulting truncated candidate.
This heuristic is applied for the case study in~\rref{subsec:acc}, where the \KeYmaeraX proof succeeded using the truncated candidate together with the rest of the Lyapunov function candidates generated by Stabhyli (for other automaton modes).

More interestingly, it is also possible for Stabhyli to declare that a system is stable because of numerical issues even though the system is unstable.
Consider the following unstable system with two modes and no switching allowed between the modes:
\begin{itemize}
\item $\pevolvein{\D{x}=-x+1}{-\frac{1}{1000000000} \leq x \leq \frac{1}{1000000000}}$ which is unstable at the origin and
\item $\pevolve{\D{x}=-x}$ which is stable.
\end{itemize}

Stabhyli always examines stability of the origin of the given hybrid system~\cite{DBLP:conf/hybrid/MohlmannT13}.
Using the default instructions as before, Stabhyli generates a success message with the following output (newlines added for readability):

\begin{verbatim}
...
SOSSolution( Problem is solved. (accepted); ...
...
### Lyapunov template for mode stable: +V_2*x^2+V_1*x
### Lyapunov function for mode stable: \
  +603702977637151/18889465931478580854784*x^2
### Lyapunov template for mode unstable: +V_4*x^2+V_3*x
### Lyapunov function for mode unstable: \
  +457363293760441/18889465931478580854784*x^2
  -224353181720881/77371252455336267181195264*x
The hybrid system is stable
\end{verbatim}

\fi

\end{document}